\documentclass[eqsecnum,secnumarabic,nofootinbib,showkeys,tightenlines,12pt]{revtex4}
\usepackage{amsfonts}
\usepackage{amssymb}
\usepackage{amsmath}
\usepackage[pdftex]{graphicx}
\usepackage[utf8]{inputenc}
\usepackage{fancyhdr}
\usepackage{slashed}
\usepackage[brazil,USenglish]{babel}
\usepackage{hyperref}

\begin{document}

\title{\textbf{Low-Energy Theorems and Linearity Breaking in
Anomalous~Amplitudes}}
\author{J. F. Thuorst$^{1}$\footnote{jfernando.th@gmail.com, corresponding author}, L. Ebani$^{1}$\footnote{
luci.ebani@gmail.com}, T. J. Girardi$^{1}$\footnote{thalisjg@gmail.com}\footnote{All authors equally contributed to this work}}
\affiliation{$^{1}$CBPF-Centro Brasileiro de Pesquisas F\'{\i}sicas, 22290-180, Rio de
Janeiro, RJ, Brazil}

\begin{abstract}
This study seeks a better comprehension of anomalies by exploring $(n+1)$%
-point perturbative amplitudes in a $2n$-dimensional framework. The involved
structures combine axial and vector vertices into odd tensors. This
configuration enables diverse expressions, considered identities at the
integrand level. However, connecting them is not automatic after loop
integration, as the divergent nature of amplitudes links to surface terms.
The background to this subject is the conflict between the linearity of
integration and the translational invariance observed in the context of
anomalies. That makes it impossible to simultaneously satisfy all symmetry
and linearity properties, constraints that arise through Ward identities and
relations among Green functions. Using the method known as Implicit
Regularization, we show that trace choices are a means to select the amount
of anomaly contributions appearing in each symmetry relation. Such an idea
appeared through recipes to take traces in recent works, but we introduce a
more complete view. We also emphasize low-energy theorems of finite
amplitudes as the source of these violations, proving that the total amount
of anomaly remains fixed regardless of any choices.
\end{abstract}

\maketitle

\section{Introduction}

Concisely, anomalies come from the quantum violation of symmetries present
in the classical theory. This subject arose when the authors \cite%
{Fukuda1949}-\cite{Rosenberg1963} attempted to build models with fermions
coupled to axial currents. Afterwards, it resurfaced in two dimensions
through the non-conservation of the axial current in two-point perturbative
corrections \cite{Johnson1963}. In four dimensions, it manifests through the
coupling of axial and vector currents in one fermionic loop, the\textit{\ }%
ABJ anomaly of the triangle graph \cite{Adler1969}-\cite{Jackiw1969}. The
presence of one anomalous term on the divergence of the axial current is
responsible for the decay rate of some mesons \cite{Veltman}, including the
experimentally observed decay of the neutral pion into two photons \cite%
{Sutherland}.

The concept of anomaly received prominence due to the breaking of Ward
Identities (WI), crucial in guaranteeing the renormalizability of gauge
models \cite{ChengLi1984}. Theories featuring spontaneous symmetry breaking,
such as the Standard Model, resort to anomalous cancellation to circumvent
this problem \cite{BenLee1972, Gross1972}. This mechanism becomes
fundamental for maintaining the consistency of the theory, also contributing
to the prediction of particles as the top quark \cite{Bertlmann1996}. Some
research lines suggest the need for a similar mechanism to establish a gauge
theory in the gravitational context. Anomalies manifest when gravitational
fields couple to fermions, with two gravitons contributing to the axial
anomaly from a triangle diagram \cite{Kimura1969, Salam1972}.

This subject remains important in modern investigations within the domain of
Kaluza-Klein theories, irrespective of renormalization \cite{Frampton1983,
Frampton21983}. We stress its relevance regarding the breaking of
diffeomorphism invariance in purely gravitational anomalies (without gauge
coupling) \cite{Gaume1983}. When interacting with photons and Weyl fermions,
one also acknowledges violations of conformal symmetry in the propagation of
gravitons \cite{DuffCapper1974, Duff1994}. Furthermore, recent contributions
have revisited the Weyl anomalies on the Pontryagin density \cite{Bonora2014}%
-\cite{Bonora2022}. Lorentz anomalies can be interchanged with Einstein
anomalies using the local Bardeen-Zumino polynomial \cite{BZumino1984},
which transforms the consistency into a covariant form for anomalies \cite%
{Bertlmann2001a, Bertlmann2001b}. Ultimately, anomalies are recognized as an
intrinsic aspect of symmetries \cite{Witten}, establishing criteria for
delimiting admissible field theories.

With this background established, we aim to elucidate some aspects relevant
to the anomalies study. For such, let us develop our investigation in a
general model coupling fermions with boson fields of even and odd parity
(spins zero and one). The $n$-vertex polygon graphs of spin-$1/2$ internal
propagators are the center of this analysis, being explored in two, four,
and six dimensions. The corresponding amplitudes exhibit Dirac traces
containing two gamma matrices beyond the space-time dimension, whose
computation yields combinations involving the metric tensor and the
Levi-Civita symbol. Hence, traces admit equivalent expressions that differ
in their index arrangements, signs, and number of monomials. That only
produces identities at first glance; however, subtle consequences emerge
since the involved amplitudes are divergent. This feature led to many works
developed in recent years, sometimes proposing rules to take these traces 
\cite{Breitenlohner1977}-\cite{Bruque2018}. Part of our task is to shed
light on this issue, and we use operations on general identities governing
the Clifford algebra for such \cite{Kreimer1990}-\cite{Schubert1993}.

This outset is intimately linked to the divergent content of the amplitudes,
especially regarding surface terms. When dealing with linearly divergent
structures, a shift in the integration variable requires compensation
through non-zero surface terms \cite{Treiman1985, Bertlmann1996}. These
objects bring coefficients depending on arbitrary routings attributed to
internal momenta\footnote{%
The same surface terms will appear within tensor integrals exhibiting
logarithmic power counting, albeit without arbitrary coefficients.}.
Although conservation sets differences between these routings as physical
momenta, internal momenta remain arbitrary and might assume non-covariant
expressions \cite{Sterman1993}. This feature represents a break in the
translational invariance, violating a crucial requirement for establishing
WIs and thus violating other symmetries. Alternatively, some regularization
techniques \cite{Bollini1972, tHooft1972} partially preserve symmetries
because they maintain translational invariance by eliminating surface terms.

Given the impossibility of satisfying all WIs in four dimensions \cite%
{JohnsonJackiw1969}, we attribute a central role to the axial triangle. That
motivates the pursuit of odd correlators involving axial and vector
vertices, the $AV^{n}$-type amplitudes in $2n$-dimensions. They are $\left(
n+1\right) $-order tensors expressed as functions of $n$ momentum variables,
which lead to low-energy theorems derived from well-defined finite functions 
\cite{AvivZee1972}. We obtain these theorems through momenta contractions
over general tensors, achieving meaningful results regarding the anomaly's
source and implications.

Such perspective is associated with relations among Green functions (RAGFs),
obtained from momenta contractions over amplitudes independently of
prescriptions to evaluate divergences. These relations embody the linearity
of the integration and are a central ingredient of the procedure adopted for
our calculations. We use the set of tools proposed by O. A. Battistel in his
Ph.D. thesis \cite{PhdBattistel1999}, later known as Implicit Regularization
(IReg). Several investigations applied this strategy in even and odd
dimensions \cite{Battistel2002a}-\cite{Ebani2018} and multi-loops
calculations \cite{Dallabona2023}. Other works also have a similar approach 
\cite{Ferreira2012}-\cite{VieiraPorto2023}.

This strategy uses an identity to expand propagators, allowing us to isolate
divergent objects without modifying expressions derived from Feynman rules.
Evaluating these objects is unnecessary in the initial steps; hence, one can
opt for a prescription at the end of the calculations. No choices are made
for internal momenta; they feature arbitrary routings used along the work.
Lastly, the organization of finite integrals is also a helpful feature \cite%
{Battistel2006, Dallabona2012}. We improved its efficiency by developing a
systematization through finite tensors and their properties.

By carefully exploring general tensor forms, we show how the kinematical
behavior of finite integrals links to anomalous contributions. Although
violations are unavoidable, different prescriptions affect how they manifest
within the calculations. Interpretations that set surface terms as zero make
results symmetric for even amplitudes. Meanwhile, they lead to the
already-known competition between gauge and chiral symmetries for anomalous
amplitudes. We elucidate this point by studying Dirac traces and how they
allow different results for the same integral. Differently, an
interpretation adopting one (specific) finite value for surface terms
implies that all trace manipulations provide a unique tensor. Although that
preserves the linearity of integration, it induces violating terms for even
and odd amplitudes. Our perspective on low-energy implications offers a
clear understanding of this subject.

We organized the work as follows. Section (\ref{ModlDef}) introduces the
model while presenting definitions and preliminary discussions. Section (\ref%
{IREG}) discusses the strategy to handle divergences, establishing the
required mathematical tools. Section (\ref{2Dim2Pt}) studies the role of
Dirac traces, surface terms, finite integrals, and low-energy theorems for
anomalies in a two-dimensional theory. We develop a similar discussion in
Section (\ref{4Dim3Pt}); however, four-dimensional calculations are more
complex and allow detailing some aspects. Then, Section (\ref{6Dim4Pt})
extends the analysis to six dimensions to indicate the generality of this
investigation. Section (\ref{finalremarks}) presents our final remarks and
prospects.

\section{Model and Notations}

\label{ModlDef}

Feynman rules employed in this investigation come from a model where
fermionic densities couple to bosonic fields of even and odd parity $\{\Phi
,\Pi ,V_{\mu },A_{\mu },H_{\mu \nu },W_{\mu \nu }\}$ through the general
interacting action%
\begin{eqnarray}
\mathcal{S}_{I} &=&\int \mathrm{d}^{2n}x\left[ e_{S}j\left( x\right) \Phi
\left( x\right) +e_{P}j_{\ast }\left( x\right) \Pi \left( x\right)
+e_{V}j^{\mu }\left( x\right) V_{\mu }\left( x\right) \right.  \notag \\
&&\left. +e_{A}j_{\ast }^{\mu }\left( x\right) A_{\mu }\left( x\right)
+e_{T}j^{\mu \nu }\left( x\right) H_{\mu \nu }\left( x\right) +e_{\tilde{T}%
}j_{\ast }^{\mu \nu }\left( x\right) W_{\mu \nu }\left( x\right) \right] .
\label{Action}
\end{eqnarray}%
Even if some couplings do not directly concern the investigated perturbative
corrections, we will see that they emerge in substructures of these
amplitudes. That is the case of tensor and pseudotensor couplings in the
six-dimensional box, as already noted in reference \cite{Fonseca2014} for
the pseudotensor case.

Respectively, each term of this functional involves scalar, pseudoscalar,
vector, axial, tensor, and pseudotensor quantities. This information
reflects in indexes $\{S,P,V,A,T,\tilde{T}\}$ attributed to coupling
constants $e_{i}$, taken as the unit for convenience. The currents $%
\{j,j_{\ast },j_{\mu },j_{\ast \mu },j_{\mu \nu },j_{\ast \mu \nu }\}$ are
bilinears in the fermionic fields $j_{i}=\bar{\psi}\left( x\right) \Gamma
_{i}\psi \left( x\right) $. They deliver the vertices proportional to%
\begin{equation}
\Gamma _{i}\in (S,P,V,A,T,\tilde{T})=(1,\gamma _{\ast },\gamma _{\mu
},\gamma _{\ast }\gamma _{\mu },\gamma _{\lbrack \mu \nu ]},\gamma _{\ast
}\gamma _{\lbrack \mu \nu ]}),  \label{SetofVertexes}
\end{equation}%
where $\gamma _{\mu }$ are generators of the Clifford algebra satisfying $%
\{\gamma ^{\mu },\gamma ^{\nu }\}=2g^{\mu \nu }$. The chiral matrix, which
is the algebra's highest-weight element, satisfies $\{\gamma _{\ast },\gamma
^{\mu }\}=0$ and assumes the form 
\begin{equation}
\gamma _{\ast }=i^{n-1}\gamma _{0}\gamma _{1}\cdots \gamma _{2n-1}=\frac{%
i^{n-1}}{\left( 2n\right) !}\varepsilon _{\nu _{1}\cdots \nu _{2n}}\gamma
^{\nu _{1}\cdots \nu _{2n}}.
\end{equation}%
We often adopt a merging notation to products of matrices $\gamma ^{\nu
_{1}\cdots \nu _{2n}}=\gamma ^{\nu _{1}}\gamma ^{\nu _{2}}\cdots \gamma
^{\nu _{2n}}$, adapting to Lorentz indexes $\nu _{1}\nu _{2}\cdots \nu
_{s}=\nu _{12\cdots s}$ whenever convenient. The behavior under the
permutation of indexes is determined by the objects: $g_{\mu _{12}}=g_{\mu
_{21}}$ or $\varepsilon _{\mu _{12\cdots 2n}}=-\varepsilon _{\mu _{21}\cdots
\mu _{2n}}$. For $2n$ dimensions, follow the normalization $\varepsilon
^{0123\cdots 2n-1}=1$.

The algebra elements are the antisymmetrized products of gamma matrices
given by%
\begin{equation}
\gamma _{\left[ \mu _{1}\cdots \mu _{r}\right] }=\frac{1}{r!}\sum_{\pi \in
S_{r}}\mathrm{sign}\left( \pi \right) \gamma _{\mu _{\pi \left( 1\right)
}\cdots \mu _{\pi \left( r\right) }},
\end{equation}%
which satisfy identities as seen in the appendix of the reference \cite%
{deWit1986}: 
\begin{equation}
\gamma _{\ast }\gamma _{\left[ \mu _{1}\cdots \mu _{r}\right] }=\frac{%
i^{n-1+r\left( r+1\right) }}{\left( 2n-r\right) !}\varepsilon _{\mu
_{1}\cdots \mu _{r}}^{\hspace{0.95cm}\nu _{r+1}\cdots \nu _{2n}}\gamma _{%
\left[ \nu _{r+1}\cdots \nu _{2n}\right] }.  \label{Chiral-Id}
\end{equation}%
These identities are used when taking traces involving the chiral matrix.
And the notation of antisymmetrization for products of tensors follows 
\begin{equation}
A_{[\alpha _{1}\cdots \alpha _{r}}B_{\alpha _{r+1}\cdots \alpha _{s}]}=\frac{%
1}{s!}\sum_{\pi \in S_{s}}\mathrm{sign}(\pi )A_{\alpha _{\pi \left( 1\right)
}\cdots \alpha _{\pi \left( r\right) }}B_{\alpha _{\pi \left( r+1\right)
}\cdots \alpha _{\pi \left( s\right) }},
\end{equation}%
whose factor of normalization does not interfere with used identities.

Spinorial Feynman propagators come from the standard kinetic term of Dirac
fermions%
\begin{equation}
S_{F}\left( K_{i}\right) =S_{F}\left( i\right) =\frac{1}{(\slashed{K}%
_{i}-m+i0^{+})}=\frac{(\slashed{K}_{i}+m)}{D_{i}},  \label{Prop}
\end{equation}%
where $D_{i}=K_{i}^{2}-m^{2}$ with $K_{i}=k+k_{i}$. We use the numerical
index $i$ to represent all parameters of the corresponding line in the
simplified notation $S_{F}\left( i\right) $. The variable $k$ is the
unrestricted loop momentum while $k_{i}$ are routings that keep track of the
flux of external momenta through the graph. These routings are arbitrary
quantities \cite{Sterman1993}\footnote{%
Consult Section (4.1) for a comment on the arbitrariness of these routings.}%
. They cannot be reduced by shifts as functions of the kinematical data in
divergent integrals, cases in which our approach uses them to codify
conditions of the satisfaction of symmetries or lack thereof. Nonetheless,
using momenta conservation in the vertices of the diagram in Figure \ref%
{diag1} relates their differences with external momenta through the
definition%
\begin{equation}
p_{ij}=K_{i}-K_{j}=k_{i}-k_{j}.  \label{pij}
\end{equation}%
\begin{figure}[tbph]
\begin{equation*}
T^{\Gamma _{1}\Gamma _{2}\cdots \Gamma _{n_{1}}}=%
\begin{array}{c}
\includegraphics[scale=0.8]{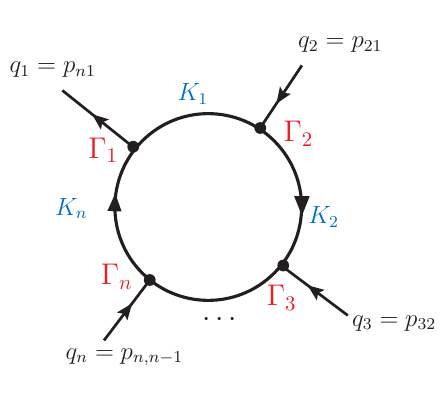}%
\end{array}%
\end{equation*}%
\caption{General diagram for the one-loop amplitudes of this work. }
\label{diag1}
\end{figure}

Processes of interest have exclusively bosonic external legs; therefore,
next-to-leading order corrections correspond to pure fermion loops. Setting
aside the minus signs for closed loops, Feynman rules allow expressing their
integrands as 
\begin{equation}
t^{\Gamma _{1}\Gamma _{2}\cdots \Gamma _{s}}\left( k_{1},\ldots
,k_{s};k\right) =\text{\textrm{tr}}[\Gamma _{1}S_{F}(1)\Gamma
_{2}S_{F}(2)\cdots \Gamma _{s}S_{F}(s)].  \label{t}
\end{equation}%
They are well-defined functions dependent on both external and internal
momenta. The internal ones are arbitrary because they are not constrained by
momentum conservation. Hence, we express them through sums of routings
following the general structure:%
\begin{equation}
P_{i_{1}i_{2}\cdots i_{r}}=k_{i_{1}}+k_{i_{2}}+\cdots +k_{i_{r}}.
\label{Pij}
\end{equation}

The total amplitude comes from the last Feynman rule, the integration over
the loop momenta:%
\begin{equation}
T^{\Gamma _{1}\Gamma _{2}\cdots \Gamma _{s}}\left( 1,\cdots ,s\right) =\int 
\frac{\mathrm{d}^{2n}k}{(2\pi )^{2n}}t^{\Gamma _{1}\Gamma _{2}\cdots \Gamma
_{s}}\left( 1,\cdots ,s\right) .  \label{T}
\end{equation}%
Distinguishing these two stages enables a preliminary discussion of
integrands without worrying about divergent structures arising posteriorly.
When replacing vertex operators $\Gamma _{i}$ by (\ref{SetofVertexes}), the
notation accompanies Lorentz indexes in order with the operators.

In the sequence, we derive identities involving integrands of amplitudes
displaying \textit{vector} and \textit{axial} vertices. If satisfied after
integration, they become proper relations among Green functions (RAGFs).
Their study has a crucial role in investigations using Implicit
Regularization (IReg) \cite{Battistel2002a, Battistel2012, Battistel2014},
which also occurs in this work. They embody the integration linearity even
before Ward Identities (WIs) are asked to play a role in perturbation
amplitudes.

Let us take the $r$-point amplitude $AV^{r-1}$ to introduce these relations%
\begin{equation}
t_{\mu _{1}\mu _{2}\cdots \mu _{r}}^{AV\cdots V}=\text{\textrm{tr}}[\gamma
_{\ast }\gamma _{\mu _{1}}S_{F}\left( 1\right) \gamma _{\mu _{2}}S_{F}\left(
2\right) \cdots \gamma _{\mu _{r}}S_{F}\left( r\right) ],  \label{tt}
\end{equation}%
starting with an example of vector contraction. The fundamental idea is to
recognize $S_{F}^{-1}\left( i\right) =\slashed{K}_{i}-m$ after using
Equation (\ref{pij})%
\begin{equation}
\slashed{p}_{ab}=\slashed{K}_{a}-\slashed{K}_{b}=S_{F}^{-1}\left( a\right)
-S_{F}^{-1}\left( b\right) ,
\end{equation}%
generating a standard manipulation to remove one propagator of the
expression. Observe how this works for the contraction with $p_{21}^{\mu
_{2}}$, producing $S_{F}\left( 1\right) \slashed{p}_{21}S_{F}\left( 2\right)
=S_{F}\left( 1\right) -S_{F}\left( 2\right) $. That leads to one \textit{%
vector }relation%
\begin{equation}
p_{21}^{\mu _{2}}t_{\mu _{1}\mu _{2}\cdots \mu _{r}}^{AV\cdots V}\equiv
t_{\mu _{1}\hat{\mu}_{2}\cdots \mu _{r}}^{AV\cdots V}(1,\hat{2},\cdots
,r)-t_{\mu _{1}\hat{\mu}_{2}\cdots \mu _{r}}^{AV\cdots V}(\hat{1},2,\cdots
,r),  \label{RAGF1}
\end{equation}%
where "hats" mean the omission of the propagator corresponding to that
routing and vertices to the Lorentz indexes. The RHS contains lower-point
functions, possibly more singular under integration. This procedure works
for axial contractions $p_{r1}^{\mu _{1}}$, but an additional contribution
emerges from permuting the chiral matrix $S_{F}\gamma _{\ast
}S_{F}^{-1}=-\left( 1+2mS_{F}\right) \gamma _{\ast }$. Following this
strategy, we obtain the \textit{axial }relation as%
\begin{equation}
p_{r1}^{\mu _{1}}t_{\mu _{12}\cdots \mu _{r}}^{AV\cdots V}\equiv t_{\mu _{r}%
\hat{\mu}_{1}\mu _{2}\cdots \mu _{r-1}}^{AV\cdots V}(1,2,\cdots ,\hat{r})-t_{%
\hat{\mu}_{1}\mu _{2}\cdots \mu _{r}}^{AV\cdots V}(\hat{1},2,\cdots
,r)-2mt_{\mu _{2}\cdots \mu _{r}}^{PV\cdots V}.  \label{RAGF2}
\end{equation}

As mentioned, integration turns true identities derived above into RAGFs:%
\begin{eqnarray}
p_{r1}^{\mu _{1}}T_{\mu _{12}\cdots \mu _{r}}^{AV\cdots V} &=&T_{\mu _{r}%
\hat{\mu}_{1}\cdots \mu _{r-1}}^{AV\cdots V}(1,2,\cdots ,\hat{r})-T_{\hat{\mu%
}_{1}\cdots \mu _{r}}^{AV\cdots V}(\hat{1},2,\cdots ,r)-2mT_{\mu _{2}\cdots
\mu _{r}}^{PV\cdots V}, \\
p_{21}^{\mu _{2}}T_{\mu _{12}\cdots \mu _{r}}^{AV\cdots V} &=&T_{\mu _{1}%
\hat{\mu}_{2}\cdots \mu _{r}}^{AV\cdots V}(1,\hat{2},\cdots ,r)-T_{\mu _{1}%
\hat{\mu}_{2}\cdots \mu _{r}}^{AV\cdots V}(\hat{1},2,\cdots ,r).
\end{eqnarray}%
Although they carry assumptions of linearity of integration in perturbative
computations, this property is not guaranteed for divergent amplitudes. They
are structural properties, not linked a priori to the particularities of the
model and its symmetries. At the same time, after summing up contributions
from crossed diagrams (if applicable and indicated by the arrow notation
below), RAGFs should coincide with symmetry implications through Ward
identities (WIs). These constraints arise from the joint application of the
algebra of quantized currents and equations of motion to these currents: $%
\partial _{\mu }j^{\mu }=0$ and $\partial _{\mu }j_{\ast }^{\mu
}=-2imj_{\ast }$. Their expressions in the position space for \textit{axial}
and one of the \textit{vector} WIs are%
\begin{eqnarray}
\partial _{\mu _{1}}^{x_{1}}\left\langle j_{\ast }^{\mu _{1}}\left(
x_{1}\right) j^{\mu _{2}}\left( x_{2}\right) \cdots j^{\mu _{r}}\left(
x_{r}\right) \right\rangle &=&-2im\left\langle j_{\ast }\left( x_{1}\right)
j^{\mu _{2}}\left( x_{2}\right) \cdots j^{\mu _{r}}\left( x_{r}\right)
\right\rangle ,  \label{AWI} \\
\partial _{\mu _{2}}^{x_{2}}\left\langle j_{\ast }^{\mu _{1}}\left(
x_{1}\right) j^{\mu _{2}}\left( x_{2}\right) \cdots j^{\mu _{r}}\left(
x_{r}\right) \right\rangle &=&0,  \label{VWI}
\end{eqnarray}%
where $\left\langle \cdots \right\rangle =\left\langle 0\left\vert T\left[
\cdots \right] \right\vert 0\right\rangle $ is an abbreviation for the time
ordering of the currents. We cast analogous equations using the notation for
perturbative amplitudes:%
\begin{eqnarray}
q_{1}^{\mu _{1}}T_{\mu _{12}\cdots \mu _{r}}^{A\rightarrow V\cdots V}
&=&-2mT_{\mu _{2}\cdots \mu _{r}}^{P\rightarrow V\cdots V},  \notag \\
q_{2}^{\mu _{2}}T_{\mu _{12}\cdots \mu _{r}}^{A\rightarrow V\cdots V}
&=&0,...,q_{r}^{\mu _{r}}T_{\mu _{12}\cdots \mu _{r}}^{A\rightarrow V\cdots
V}=0.
\end{eqnarray}%
Again, we use the first vector contraction to illustrate the series of
relations of this type. Comparing these definitions with Figure \ref{diag1},
one identifies the external momenta $q_{1}=p_{r1}$, $q_{2}=p_{21}$.

Breaking these symmetry implications characterizes \textit{anomalous
amplitudes}. Since the connection involving RAGFs and WIs is
straightforward, violations of the first imply violations of the second.
This way, maintaining all WIs depends on satisfying all RAGFs while having
translational invariance in the momentum space. We show how this requirement
is impossible for a class of amplitudes referred as \textit{axial amplitudes}%
:

\begin{itemize}
\item Initial discussion uses bubbles in $2D$: $AV$ and $VA$;

\item Main argumentation uses triangles in $4D$: $AVV$, $VAV$, $VVA$, and $%
AAA$;

\item Generalization uses one box in $6D$: $AVVV$;
\end{itemize}

As a feature of $2n$ dimensions, we will show that momenta contractions over
these amplitudes lead to lower-order ones with the general form%
\begin{equation}
T_{\mu _{1\cdots n}}^{AV^{n-1}}\left( 1,\cdots ,n\right) =\left(
2^{n}i^{n-1}/n\right) \varepsilon _{\mu _{1\cdots n}\nu _{1\cdots n}}\left(
p_{21}^{\nu _{2}}\cdots p_{n,1}^{\nu _{n}}\right) \left( P_{12\cdots
n}\right) _{\nu _{n+1}}\Delta _{n+1}^{\nu _{1}\nu _{n+1}},
\end{equation}%
with objects $\Delta _{n+1}$ representing surface terms (\ref{delta2n}).
Meanwhile, we will also find purely finite integrals when performing axial
contractions. That allows discussing a crucial point of this investigation
by exploring the low-energy behavior of anomalous amplitudes and offering an
interpretation of anomalies through their connection with finite amplitudes.

Considering these purposes, we must take Dirac traces to compute all
mentioned amplitudes. When integrated, they become linear combinations of
bare Feynman integrals following the definition\footnote{%
We also simplify the arguments of these functions when clear $f\left(
k_{1},k_{2},\cdots \right) =f\left( 1,2,\cdots \right) $. Changing the
reference routing $k_{j}$ to another $k_{i}$ is a matter of recognizing the $%
p_{ij}$ definition (\ref{pij}) and writing $K_{i}=K_{j}+p_{ij}$.}:%
\begin{equation}
\bar{J}_{n_{2}}^{\mu _{1}\mu _{2}\cdots \mu _{n_{1}}}\left( 1,2,\cdots
,n_{2}\right) =\int \frac{\mathrm{d}^{2n}k}{\left( 2\pi \right) ^{2n}}\frac{%
K_{i}^{\mu _{1}}\cdots K_{i}^{\mu _{n_{1}}}}{D_{1}D_{2}\cdots D_{n_{2}}}.
\label{JInt}
\end{equation}%
These integrals have power counting $\omega =2n+n_{1}-2n_{2}$, where $n_{1}$
is the tensor rank, and $n_{2}$ is the number of denominators. One set of
five integrals arises within each amplitude, whose evaluation is the subject
of Subsection (\ref{FinFcts}). Before that, let us develop a strategy to
deal with divergent quantities emerging with this operation.

\section{Strategy}

\label{IREG}Before presenting the strategy to solve amplitudes, let us
digress into\ the issue of divergent integrals in QFT. It is well-known that
products of propagators (that are not regular distributions) are generally
ill-defined. A good example is the following equation%
\begin{equation}
\int \frac{\mathrm{d}^{4}k}{\left( 2\pi \right) ^{4}}\mathrm{tr}[S_{F}\left(
k\right) S_{F}\left( k-p\right) ]=\int \mathrm{d}^{4}x\mathrm{tr}[\hat{S}%
_{F}\left( x\right) \hat{S}_{F}\left( -x\right) ]\mathrm{e}^{ip\cdot x}.
\end{equation}%
While the LHS displays a divergent convolution of two Feynman propagators in
momentum space, the RHS presents the Fourier transform of a product of
propagators in position space. Both sides do not define distributions
because when the point-wise product of distributions does not exist, the
convolution product of their Fourier transform does not also.

These short-distance UV singularities manifest through divergences in
loop-momentum integrals. Their origins trace back to multiplications of
distributions by a discontinuous step function in the chronological ordering
of operators in the interaction picture. That leads, through the Wick
theorem, to Feynman rules; see \cite{Scharf2010, Scharf2014}, originally in
Epstein and Glaser \cite{EpsteinGlaser1973}. Although the undefined Feynman
diagrams can be circumvented by carefully studying the splitting of
distributions with causal support in the setting of causal perturbation
theory \cite{Aste1997, Aste2003} (where no divergent integral appears at
all), let us work with Feynman rules in the context of regularizations.

We use the procedure known as Implicit Regularization (IReg) to handle
divergences. Its development dates back to the late 1990s in the Ph.D.
thesis of O. A. Battistel \cite{PhdBattistel1999}, having its first
investigations in references \cite{Battistel1997, BattistelNemes1999}. Its
goal is to keep the connection at all times with the expression of "bare"
Feynman rules while removing physical parameters (i.e., routings and masses)
from divergent integrals and putting them in strictly finite integrals. The
divergent ones do not receive any modification besides an organization
through surface terms and irreducible scalar integrals.

This objective is realized by noticing that all Feynman integrals depend on
propagator-like structures $D_{i}=[\left( k+k_{i}\right) ^{2}-m^{2}]$
defined in Eq. (\ref{Prop}). Thus, by introducing a parameter $\lambda ^{2}$%
, constructing an identity to separate quantities depending on physical
parameters is possible%
\begin{equation}
\frac{1}{D_{i}}=\frac{1}{D_{\lambda }+A_{i}}=\frac{1}{D_{\lambda }}\frac{1}{%
[1-(-A_{i}/D_{\lambda })]},  \label{DecompDi}
\end{equation}%
where $D_{\lambda }=(k^{2}-\lambda ^{2})$ and $A_{i}=2k\cdot
k_{i}+(k_{i}^{2}+\lambda ^{2}-m^{2})$. Now, we use the sum of the geometric
progression of order $N$ and ratio $(-A_{i}/D_{\lambda })$ to write%
\begin{equation}
\frac{1}{[1-(-A_{i}/D_{\lambda })]}=\sum_{r=0}^{N}(-A_{i}/D_{\lambda
})^{r}+(-A_{i}/D_{\lambda })^{N+1}\frac{1}{[1-(-A_{i}/D_{\lambda })]}.
\label{IdentGeom}
\end{equation}%
Immediately, one determines the asymptotic behavior at infinity of the
powers $(-A_{i}/D_{\lambda })^{r}$ as $\left\Vert k\right\Vert ^{-r}$.
Observe that terms in the summation sign depend on routings only through a
polynomial in the numerator.

Putting the last two equations together leads to the following identity:%
\begin{equation}
\frac{1}{D_{i}}=\sum_{r=0}^{N}\left( -1\right) ^{r}\frac{A_{i}^{r}}{%
D_{\lambda }^{r+1}}+\left( -1\right) ^{N+1}\frac{A_{i}^{N+1}}{D_{\lambda
}^{N+1}D_{i}}\text{, with }N\in \mathbb{N}\text{.}  \label{id}
\end{equation}%
We can choose $N$ as equal to or greater than the power counting. Hence, at
least the last term of this expansion leads to a finite contribution
dependent on external momenta when treating a product of propagators.
Applying the corresponding derivative shows this identity does not depend on
the parameter $\lambda ^{2}$. Meanwhile, $\lambda ^{2}$ connects divergent
and finite parts of integrals implying specific behavior to divergent scalar
integrals, and this behavior is straightforwardly satisfied. Thus, without
loss of generality, we adopt the propagator mass as the scale ($\lambda
^{2}=m^{2}$).

In the sequence, we cast elements associated with the systematization
proposed by IReg. The first subsection organizes divergences without
modifications. Then, finite functions necessary to express perturbative
amplitudes are introduced. Lastly, we define integrals pertinent to this
work and discuss some examples.

\subsection{Divergent Terms\label{DivTerms}}

After applying the separation identity (\ref{id}), divergent terms follow
the structure of the summation part. They appear as a set of pure
integration-momentum integrals through the following tensor structures:%
\begin{equation}
\int \frac{\mathrm{d}^{2n}k}{\left( 2\pi \right) ^{2n}}\frac{1}{D_{\lambda
}^{a}}\text{, }\int \frac{\mathrm{d}^{2n}k}{\left( 2\pi \right) ^{2n}}\frac{%
k_{\mu _{1}}k_{\mu _{2}}}{D_{\lambda }^{a+1}}\text{,}\cdots \int \frac{%
\mathrm{d}^{2n}k}{\left( 2\pi \right) ^{2n}}\frac{k_{\mu _{1}}k_{\mu
_{2}}\cdots k_{\mu _{2b-1}}k_{\mu _{2b}}}{D_{\lambda }^{a+b}}\text{, with }%
n\geq a\text{.}
\end{equation}%
As we cast these objects with the same power counting, combining them into
surface terms is direct%
\begin{equation}
-\frac{\partial }{\partial k_{\mu _{1}}}\frac{k_{\mu _{2}}\cdots k_{\mu
_{2n}}}{D_{\lambda }^{a}}=2a\frac{k_{\mu _{1}}k_{\mu _{2}}\cdots k_{\mu
_{2n}}}{D_{\lambda }^{a+1}}-g_{\mu _{1}\mu _{2}}\frac{k_{\mu _{3}}\cdots
k_{\mu _{2n}}}{D_{\lambda }^{a}}-\text{permutations.}
\end{equation}

Observe that the equation above exhibits lower-order surface terms inside
higher-order ones. That produces a chain of associations, leading to scalar
integrals that encode the divergent content of the original expression.
These terms carry information about shifting the integration variable. We
are trading the freedom of the operation of translation in the momentum
space for the arbitrary choice of routings in these perturbative corrections.%
\textbf{\ }They are always present for linear\ or higher divergent integrals
and logarithmic-divergent tensor ones. Although their coefficients depend on
ambiguous momenta $P_{ij}=k_{i}+k_{j}$ in the first case, only external
momenta $p_{ij}=k_{i}-k_{j}$ may appear in the second. For this
investigation, combinations arising in $2n$-dimensional calculations are%
\begin{equation}
\Delta _{\left( n+1\right) \mu \nu }^{\left( 2n\right) }(\lambda ^{2})=\int 
\frac{\mathrm{d}^{2n}k}{\left( 2\pi \right) ^{2n}}\left( \frac{2nk_{\mu
}k_{\nu }}{D_{\lambda }^{n+1}}-g_{\mu \nu }\frac{1}{D_{\lambda }^{n}}\right)
=-\int \frac{\mathrm{d}^{2n}k}{\left( 2\pi \right) ^{2n}}\frac{\partial }{%
\partial k_{\mu }}\frac{k_{\nu }}{D_{\lambda }^{n}},  \label{delta2n}
\end{equation}%
with the corresponding irreducible scalars coming from the definition%
\begin{equation}
I_{\log }^{\left( 2n\right) }\left( \lambda ^{2}\right) =\int \frac{\mathrm{d%
}^{2n}k}{\left( 2\pi \right) ^{2n}}\frac{1}{D_{\lambda }^{n}}.
\end{equation}

The separation highlights diverging structures and organizes them without
performing any analytic operation. Moreover, it makes evident that the
divergent content is a local polynomial in the ambiguous and physical
momenta obtained without expansions or limits.

\subsection{Finite Functions\label{FinFcts}}

After separating the finite part, solving the corresponding integrals
through techniques of perturbative calculations is necessary. For two-point
amplitudes, we project results into the following families of functions%
\begin{eqnarray}
Z_{n_{1}}^{\left( -1\right) } &=&\int_{0}^{1}\mathrm{d}x_{1}\frac{%
x_{1}^{n_{1}}}{Q}, \\
Z_{n_{1}}^{\left( 0\right) } &=&\int_{0}^{1}\mathrm{d}x_{1}\text{ }%
x_{1}^{n_{1}}\log \frac{Q}{-m^{2}},
\end{eqnarray}%
where powers are $n_{i}\in \mathbb{N}$ and $Q$\footnote{%
These polynomials can be written in terms of Symanzik polynomials
constructed using the spanning trees and two-forests of the graph.} is a
polynomial on Feynman parameters $x_{i}$ 
\begin{equation}
Q=p^{2}x\left( 1-x\right) -m^{2}.
\end{equation}%
Given our interest in investigating the low-energy behavior of axial
amplitudes, let us observe the value of these functions when bilinears on
the momentum are null (i.e., $p^{2}=0$):%
\begin{equation}
Z_{n_{1}}^{\left( -1\right) }\left( 0\right) =-\frac{1}{m^{2}\left(
n_{1}+1\right) },\quad Z_{n_{1}}^{\left( 0\right) }\left( 0\right) =0.
\label{LetZ2D}
\end{equation}

Similarly, basic functions associated with three-point amplitudes arise as%
\begin{eqnarray}
Z_{n_{1}n_{2}}^{\left( -1\right) } &=&\int_{0}^{1}\mathrm{d}%
x_{1}\int_{0}^{1-x_{1}}\mathrm{d}x_{2}\frac{x_{1}^{n_{1}}x_{2}^{n_{2}}}{Q},
\label{Znm(-1)} \\
Z_{n_{1}n_{2}}^{\left( 0\right) } &=&\int_{0}^{1}\mathrm{d}%
x_{1}\int_{0}^{1-x_{1}}\mathrm{d}x_{2}\text{ }x_{1}^{n_{1}}x_{2}^{n_{2}}\log 
\frac{Q}{-m^{2}},
\end{eqnarray}%
with the polynomial $Q$ assuming the form 
\begin{equation}
Q=p^{2}x_{1}\left( 1-x_{1}\right) +q^{2}x_{2}\left( 1-x_{2}\right) -2\left(
p\cdot q\right) x_{1}x_{2}-m^{2}.
\end{equation}%
At the point where all momenta bilinears are zero, they satisfy%
\begin{equation}
Z_{n_{1}n_{2}}^{\left( -1\right) }\left( 0\right) =-\frac{n_{1}!n_{2}!}{m^{2}%
\left[ \left( n_{1}+n_{2}+2\right) !\right] },\quad Z_{n_{1}n_{2}}^{\left(
0\right) }\left( 0\right) =0.  \label{LetZ4D}
\end{equation}%
Lastly, four-point amplitudes lead to the following basic functions%
\begin{eqnarray}
Z_{n_{1}n_{2}n_{3}}^{\left( -1\right) } &=&\int_{0}^{1}\mathrm{d}%
x_{1}\int_{0}^{1-x_{1}}\mathrm{d}x_{2}\int_{0}^{1-x_{1}-x_{2}}\mathrm{d}x_{3}%
\frac{x_{1}^{n_{1}}x_{2}^{n_{2}}x_{3}^{n_{3}}}{Q},  \label{Zmnp(-1)} \\
Z_{n_{1}n_{2}n_{3}}^{\left( 0\right) } &=&\int_{0}^{1}\mathrm{d}%
x_{1}\int_{0}^{1-x_{1}}\mathrm{d}x_{2}\int_{0}^{1-x_{1}-x_{2}}\mathrm{d}x_{3}%
\text{ }x_{1}^{n_{1}}x_{2}^{n_{2}}x_{3}^{n_{3}}\log \frac{Q}{-m^{2}},
\end{eqnarray}%
whose corresponding polynomial is given by%
\begin{eqnarray}
Q &=&p^{2}x_{1}\left( 1-x_{1}\right) +q^{2}x_{2}\left( 1-x_{2}\right)
+r^{2}x_{3}\left( 1-x_{3}\right)  \notag \\
&&-2\left( p\cdot q\right) x_{1}x_{2}-2\left( p\cdot r\right)
x_{1}x_{3}-2\left( q\cdot r\right) x_{2}x_{3}-m^{2}.
\end{eqnarray}%
Once more, their values to vanishing bilinears are%
\begin{equation}
Z_{n_{1}n_{2}n_{3}}^{\left( -1\right) }\left( 0\right) =-\frac{1}{m^{2}}%
\frac{n_{1}!n_{2}!n_{3}!}{\left( n_{1}+n_{2}+n_{3}+3\right) !};\quad
Z_{n_{1}n_{2}n_{3}}^{\left( 0\right) }\left( 0\right) =0.  \label{LetZ6D}
\end{equation}

By writing parameters in terms of derivatives of polynomials and using
partial integration, relations among these finite functions appear. Such
operations relate to momenta contractions or traces seen throughout our
calculations and imply reductions of the sum of parameter powers $\Sigma
n_{i}$. We cast the employed properties together with the solutions achieved
for Feynman integrals. They were approached in references \cite%
{Battistel2006, Dallabona2012, SunYi2012}; however, we introduce a new
perspective that manipulates groups of functions instead of individual cases.

\subsection{Feynman Integrals\label{BasisFI}}

At the end of Section (\ref{ModlDef}), we introduced a set of $\left(
n+1\right) $-point amplitudes in $2n$ dimensions. Given their character as
odd tensors, we refer to them as axial amplitudes in this work. In the same
context, Eq. (\ref{JInt}) presented a general definition for Feynman
integrals that appear after taking Dirac traces. We describe in a nutshell
those that arise within the mentioned amplitudes:%
\begin{eqnarray}
(\bar{J}_{n};\bar{J}_{n}^{\mu }) &=&\int \frac{\mathrm{d}^{2n}k}{\left( 2\pi
\right) ^{2n}}\frac{(1;\ K_{1}^{\mu })}{D_{12\cdots n}}, \\
(\bar{J}_{n+1};\bar{J}_{n+1}^{\mu };\bar{J}_{n+1}^{\mu _{1}\mu _{2}})
&=&\int \frac{\mathrm{d}^{2n}k}{\left( 2\pi \right) ^{2n}}\frac{(1;\
K_{1}^{\mu };\ K_{1}^{\mu _{1}}K_{1}^{\mu _{2}})}{D_{12\cdots n+1}},
\end{eqnarray}%
with the conventions $D_{12\cdots i}=D_{1}D_{2}\cdots D_{i}$ and $%
K_{i}=k+k_{i}$. We adopted the overbar notation to emphasize the presence of
divergences since some integrals exhibit linear power counting $\omega (\bar{%
J}_{n}^{\mu })=1$ or logarithmic one $\omega \left( \bar{J}_{n}\right)
=\omega (\bar{J}_{n+1}^{\mu _{1}\mu _{2}})=0$. For instance, the presence of
the overbar distinguishes the full integral $\bar{J}_{n}$ from its finite
contributions labeled as $J_{n}$. That also means they coincide for strictly
finite integrals, namely $\bar{J}_{n+1}^{\mu }=J_{n+1}^{\mu }$ and $\bar{J}%
_{n+1}=J_{n+1}$.

We compute the cases with linear power counting in their respective
dimension to illustrate some features of our treatment, which requires using
the $N=1$ version of identity (\ref{id}):%
\begin{equation}
\frac{1}{D_{i}}=\frac{1}{D_{\lambda }}-\frac{A_{i}}{D_{\lambda }^{2}}+\frac{%
A_{i}^{2}}{D_{\lambda }^{2}D_{i}}.
\end{equation}%
Let us start with the four-dimensional integral%
\begin{equation}
\bar{J}_{2}^{\left( 4\right) \mu }=\int \frac{\mathrm{d}^{4}k}{\left( 2\pi
\right) ^{4}}\frac{K_{1}^{\mu }}{D_{12}},
\end{equation}%
whose separation allows rewriting the integrand%
\begin{eqnarray}
\frac{K_{1}^{\mu }}{D_{12}} &=&\left[ \frac{1}{D_{\lambda }^{2}}-\frac{%
\left( A_{1}+A_{2}\right) }{D_{\lambda }^{3}}\right] K_{1}^{\mu }  \notag \\
&&+\left[ \frac{A_{1}A_{2}}{D_{\lambda }^{4}}+\frac{A_{1}^{2}}{D_{\lambda
}^{3}D_{1}}+\frac{A_{2}^{2}}{D_{\lambda }^{3}D_{2}}-\frac{A_{1}A_{2}^{2}}{%
D_{\lambda }^{4}D_{2}}-\frac{A_{2}A_{1}^{2}}{D_{\lambda }^{4}D_{1}}+\frac{%
A_{1}^{2}A_{2}^{2}}{D_{\lambda }^{4}D_{12}}\right] K_{1}^{\mu }.
\end{eqnarray}%
After applying the integration sign, we gather purely divergent integrals
(first row) and organize them through surface terms and irreducible scalars;
see Subsection (\ref{DivTerms}). Then, following the notation from Eq. (\ref%
{Pij}), we express the sum of labels appearing in the coefficient as $%
P_{21}=k_{2}+k_{1}$: 
\begin{equation}
\bar{J}_{2}^{\left( 4\right) \mu }=J_{2}^{\left( 4\right) \mu }\left(
p_{21}\right) -\frac{1}{2}[P_{21}^{\nu }\Delta _{3\nu }^{\left( 4\right) \mu
}+p_{21}^{\mu }I_{\log }^{\left( 4\right) }].  \label{J2bar4D}
\end{equation}%
The remaining terms correspond to finite integrals denoted by $J$ without
overbar. Their integration occurs without restrictions and yields%
\begin{equation}
J_{2}^{\left( 4\right) \mu }\left( p_{21}\right) =i\left( 4\pi \right)
^{-2}p_{21}^{\mu }Z_{1}^{\left( 0\right) }\left( p_{21}\right) .
\end{equation}

Following the same strategy, the six-dimensional integral assumes the
organization%
\begin{equation}
\bar{J}_{3}^{\left( 6\right) \mu }=\int \frac{\mathrm{d}^{6}k}{\left( 2\pi
\right) ^{6}}\frac{K_{1}^{\mu }}{D_{123}}=J_{3}^{\left( 6\right) \mu }-\frac{%
1}{3}[P_{123}^{\nu }\Delta _{4\nu }^{\left( 6\right) \mu }+(p_{21}^{\mu
}+p_{31}^{\mu })I_{\log }^{\left( 6\right) }],
\end{equation}%
with $P_{123}=k_{1}+k_{2}+k_{3}$ and the finite contributions resulting in%
\begin{equation}
J_{3}^{\left( 6\right) \mu }\left( p_{21},p_{31}\right) =i\left( 4\pi
\right) ^{-3}[p_{21}^{\mu }Z_{10}^{\left( 0\right) }\left(
p_{21},p_{31}\right) +p_{31}^{\mu }Z_{01}^{\left( 0\right) }\left(
p_{21},p_{31}\right) ].
\end{equation}%
Lastly, we observe that the two-dimensional integral corresponds to a pure
surface term: 
\begin{equation}
\bar{J}_{1}^{\left( 2\right) \mu }\left( k_{1}\right) =\int \frac{\mathrm{d}%
^{2}k}{\left( 2\pi \right) ^{2}}\frac{K_{1}^{\mu }}{D_{1}}=-k_{1}^{\nu
}\Delta _{2\nu }^{\left( 2\right) \mu }.
\end{equation}

As anticipated, these integrals contain surface terms proportional to
ambiguous combinations of labels since they exhibit linear power counting.
The same situation manifests in logarithmically divergent integrals, albeit
without ambiguities. For all explicit results, see Appendices \ref{AppInt2D}%
, \ref{AppInt4D}, and \ref{AppInt6D}. In cases where the space-time
dimension is transparent, we drop the superindex indicating this feature.

\section{Two-Dimensional Amplitudes\label{2Dim2Pt}}

In this section, we compute axial amplitudes of two Lorentz indexes ($AV$\
and $VA$) to establish the connection between the linearity of integration
and symmetries, which materializes through relations among Green functions
(RAGFs) and Ward identities (WIs). We also initiate discussions about
low-energy implications and uniqueness, which will be fundamental topics in
the four-dimensional analysis.

\begin{figure}[h]
\includegraphics[scale=0.6]{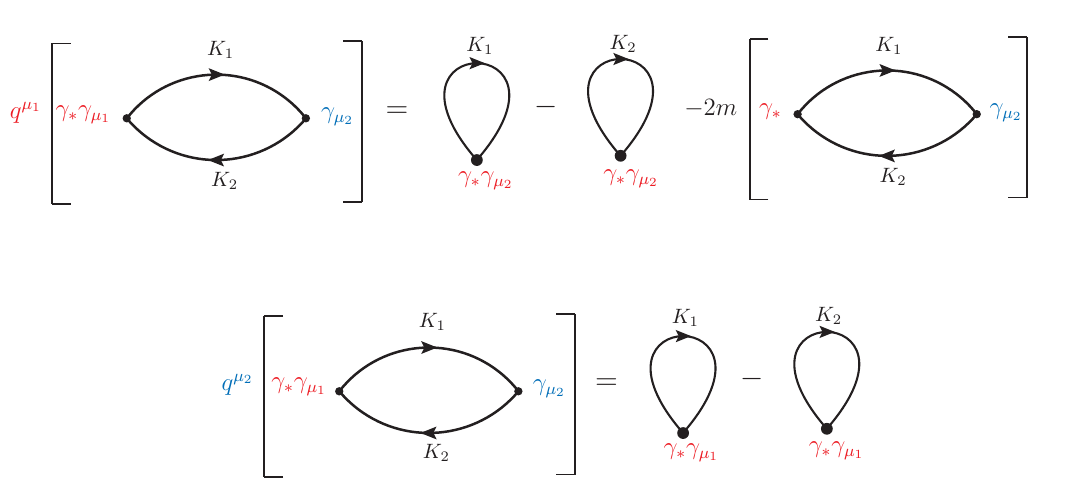}
\caption{RAGFs to $T_{\protect\mu _{12}}^{AV}$}
\label{figAV}
\end{figure}

The mentioned connection manifests in contractions with the external
momentum $q=p_{21}=k_{2}-k_{1}$. After introducing the model (\ref{ModlDef}%
), we derived identities involving integrands of amplitudes (\ref{RAGF1})-(%
\ref{RAGF2}). For the cases in analysis, the integration should produce
RAGFs for the vector vertex%
\begin{eqnarray}
q^{\mu _{2}}T_{\mu _{12}}^{AV} &=&T_{\mu _{1}}^{A}\left( 1\right) -T_{\mu
_{1}}^{A}\left( 2\right) ,  \label{p1AV} \\
q^{\mu _{1}}T_{\mu _{12}}^{VA} &=&T_{\mu _{2}}^{A}\left( 1\right) -T_{\mu
_{2}}^{A}\left( 2\right) ,  \label{p1VA}
\end{eqnarray}%
and for the axial vertex%
\begin{eqnarray}
q^{\mu _{1}}T_{\mu _{12}}^{AV} &=&T_{\mu _{2}}^{A}\left( 1\right) -T_{\mu
_{2}}^{A}\left( 2\right) -2mT_{\mu _{2}}^{PV},  \label{p2AV} \\
q^{\mu _{2}}T_{\mu _{12}}^{VA} &=&T_{\mu _{1}}^{A}\left( 1\right) -T_{\mu
_{1}}^{A}\left( 2\right) +2mT_{\mu _{1}}^{VP}.  \label{p2VA}
\end{eqnarray}%
Their satisfaction is necessary to maintain the linearity of integration.
Figure \ref{figAV} uses the $AV$ amplitude to\textbf{\ }illustrate these
relations. Meanwhile, WIs imply vanishing the one-point functions above as
required by the formal current-conservation equations (\ref{VWI})-(\ref{AWI}%
). Part of our objective consists of verifying these expectations
explicitly, even if they are not entirely contemplated since we deal with
anomalous amplitudes. 

On the other hand, if symmetry constraints were valid, the general structure
of these amplitudes as odd tensors implies kinematic properties to
invariants. Let us take the $AV$ structure as an example%
\begin{equation}
T_{\mu _{12}}^{AV}\rightarrow F_{\mu _{12}}=\varepsilon _{\mu
_{12}}F_{1}+\varepsilon _{\mu _{1}\nu }q^{\nu }q_{\mu _{2}}F_{2}+\varepsilon
_{\mu _{2}\nu }q^{\nu }q_{\mu _{1}}F_{3},  \label{AVForm}
\end{equation}%
where $F_{i}$ are scalar invariants. Since two-point amplitudes exhibit
logarithmic power counting in a two-dimensional setting, we only considered
dependence on the external momentum. Then, performing momenta contractions
yields%
\begin{eqnarray}
q^{\mu _{2}}T_{\mu _{12}}^{AV} &=&\varepsilon _{\mu _{1}\nu }q^{\nu
}(q^{2}F_{2}+F_{1}),  \label{q2ContAV} \\
q^{\mu _{1}}T_{\mu _{12}}^{AV} &=&\varepsilon _{\mu _{2}\nu }q^{\nu
}(q^{2}F_{3}-F_{1}).  \label{q1ContAV}
\end{eqnarray}%
The vector conservation in the first equation implies $F_{1}=-q^{2}F_{2}$,
whose replacement in the second equation produces 
\begin{equation}
q^{\mu _{1}}T_{\mu _{12}}^{AV}=\varepsilon _{\mu _{2}\nu }q^{\nu
}q^{2}(F_{3}+F_{2}).
\end{equation}%
Hence, if the invariants do not have poles in $q^{2}=0$, we have a
low-energy implication for the axial contraction. This falls on the $PV$\
amplitude if the axial WI is satisfied 
\begin{equation}
q^{\mu _{1}}T_{\mu _{12}}^{AV}|_{q^{2}=0}=0=\left. -2mT_{\mu
_{2}}^{PV}\right\vert _{q^{2}=0}=:\varepsilon _{\mu _{2}\nu }q^{\nu }\Omega
^{PV}(q^{2}=0),  \label{lowAV}
\end{equation}%
with $\Omega ^{PV}$ being the form factor associated with $PV$. As the
deduction of this last behavior requires the validity of both WIs, it has
the same status as a symmetry property.

The reciprocal form of this statement appears by exchanging the order of the
arguments. If the axial WI is selected first, it implies $%
F_{1}=q^{2}F_{3}-\Omega ^{PV}$ in (\ref{q1ContAV}). Its replacement in the
vector contraction (\ref{q2ContAV}) gives the low-energy implication for the
contraction with the index of the vector current%
\begin{equation}
\left. q^{\mu _{2}}T_{\mu _{12}}^{AV}\right\vert _{q^{2}=0}=-\varepsilon
_{\mu _{1}\nu }q^{\nu }\Omega ^{PV}(q^{2}=0).  \label{lowAV2}
\end{equation}

With this scenario in hands, our objective is the analysis in the light of
explicit integration (\ref{T}). Consulting the definition (\ref{t}), we
write the general integrand of two-point amplitudes%
\begin{eqnarray}
t^{\Gamma _{1}\Gamma _{2}} &=&K_{12}^{\nu _{12}}\text{\textrm{tr}}(\Gamma
_{1}\gamma _{\nu _{1}}\Gamma _{2}\gamma _{\nu _{2}})/D_{12}+m^{2}\text{%
\textrm{tr}}(\Gamma _{1}\Gamma _{2})/D_{12}  \notag \\
&&+mK_{1}^{\nu }\text{\textrm{tr}}(\Gamma _{1}\gamma _{\nu }\Gamma
_{2})/D_{12}+mK_{2}^{\nu }\text{\textrm{tr}}(\Gamma _{1}\Gamma _{2}\gamma
_{\nu })/D_{12};  \label{2ptexp}
\end{eqnarray}%
thus, specific versions emerge after choosing vertices and keeping the
nonzero traces: 
\begin{eqnarray}
t_{\mu _{12}}^{AV} &=&K_{12}^{\nu _{12}}\mathrm{tr}(\gamma _{\ast }\gamma
_{\mu _{1}\nu _{1}\mu _{2}\nu _{2}})/D_{12}+m^{2}\mathrm{tr}(\gamma _{\ast
}\gamma _{\mu _{12}})/D_{12}, \\
t_{\mu _{12}}^{VA} &=&K_{12}^{\nu _{12}}\mathrm{tr}(\gamma _{\ast }\gamma
_{\mu _{1}\nu _{1}\mu _{2}\nu _{2}})/D_{12}-m^{2}\mathrm{tr}(\gamma _{\ast
}\gamma _{\mu _{12}})/D_{12}.
\end{eqnarray}

The next step consists of taking Dirac traces, with the lower-rank one
resulting in $\mathrm{tr}(\gamma _{\ast }\gamma _{\mu _{12}})=-2\varepsilon
_{\mu _{12}}$. The trace of four gamma matrices is a linear combination of
the metric and the Levi-Civita tensor, so various expressions emerge through
substitutions involving the following versions of identity (\ref{Chiral-Id})
restricted to two dimensions:%
\begin{equation*}
2\gamma _{\ast }=\varepsilon _{\nu _{12}}\gamma ^{\nu _{12}};\text{\quad }%
\gamma _{\ast }\gamma _{\mu }=-\varepsilon _{\mu \nu }\gamma ^{\nu };\text{%
\quad }\gamma _{\ast }\gamma _{\left[ \mu \nu \right] }=-\varepsilon _{\mu
\nu }.
\end{equation*}%
They lead to expressions that are not automatically equal after integration.
To unfold this rationale, let us apply the definition of the chiral matrix
in the form $2\gamma _{\ast }=\varepsilon _{\nu _{12}}\gamma ^{\nu _{12}}$
(first identity)\ to write%
\begin{equation}
\mathrm{tr}(\gamma _{\ast }\gamma _{abcd})=2(-g_{ab}\varepsilon
_{cd}+g_{ac}\varepsilon _{bd}-g_{ad}\varepsilon _{bc}-g_{bc}\varepsilon
_{ad}+g_{bd}\varepsilon _{ac}-g_{cd}\varepsilon _{ab}).
\end{equation}%
Here, we explore two sorting of indexes $\gamma _{\mu _{1}\nu _{1}\mu
_{2}\nu _{2}}$ and $\gamma _{\mu _{2}\nu _{2}\mu _{1}\nu _{1}}$,
corresponding to replacing the chiral matrix definition around the first and
second vertices. Albeit equivalent, these traces differ through signs of
some terms. We perform the contractions with $K_{12}^{\nu _{12}}=K_{1}^{\nu
_{1}}K_{2}^{\nu _{2}}$ to study them:%
\begin{eqnarray}
K_{12}^{\nu _{12}}\mathrm{tr}(\gamma _{\ast }\gamma _{\mu _{1}}\gamma _{\nu
_{1}}\gamma _{\mu _{2}}\gamma _{\nu _{2}}) &=&-2\varepsilon _{\mu _{1}\nu
_{1}}\left( K_{1\mu _{2}}K_{2}^{\nu _{1}}+K_{2\mu _{2}}K_{1}^{\nu
_{1}}\right) -2\varepsilon _{\mu _{2}\nu _{1}}\left( K_{1\mu _{1}}K_{2}^{\nu
_{1}}-K_{2\mu _{1}}K_{1}^{\nu _{1}}\right)  \notag \\
&&+2\varepsilon _{\mu _{1}\mu _{2}}\left( K_{1}\cdot K_{2}\right) +2g_{\mu
_{1}\mu _{2}}\varepsilon _{\nu _{1}\nu _{2}}K_{12}^{\nu _{12}}, \\
K_{12}^{\nu _{12}}\mathrm{tr}(\gamma _{\ast }\gamma _{\mu _{2}}\gamma _{\nu
_{2}}\gamma _{\mu _{1}}\gamma _{\nu _{1}}) &=&+2\varepsilon _{\mu _{1}\nu
_{1}}\left( K_{1\mu _{2}}K_{2}^{\nu _{1}}-K_{2\mu _{2}}K_{1}^{\nu
_{1}}\right) -2\varepsilon _{\mu _{2}\nu _{1}}\left( K_{1\mu _{1}}K_{2}^{\nu
_{1}}+K_{2\mu _{1}}K_{1}^{\nu _{1}}\right)  \notag \\
&&-2\varepsilon _{\mu _{1}\mu _{2}}\left( K_{1}\cdot K_{2}\right) -2g_{\mu
_{1}\mu _{2}}\varepsilon _{\nu _{1}\nu _{2}}K_{12}^{\nu _{12}}.
\end{eqnarray}

It is often possible to examine the tensor structure of one amplitude to
find less complex ones inside it. Despite an $\varepsilon _{\mu _{1}\mu
_{2}} $ factor, using the general form (\ref{2ptexp}) leads to scalar
two-point subamplitudes below when combining the bilinears above with
squared mass terms:%
\begin{equation}
t^{PP}=t^{SS}-4m^{2}\frac{1}{D_{12}}=q^{2}\frac{1}{D_{12}}-\frac{1}{D_{1}}-%
\frac{1}{D_{2}}.  \label{SS}
\end{equation}%
The following reduction was used to simplify their integrands%
\begin{equation}
2(K_{i}\cdot K_{j}-m^{2})=D_{i}+D_{j}-p_{ij}^{2}.  \label{Sij}
\end{equation}%
All other contributions receive an organization in terms of the same object,
a standard tensor present similarly in all explored dimensions:%
\begin{equation}
t_{\mu }^{\left( \pm \right) \nu }=\left( K_{1\mu }K_{2}^{\nu }\pm K_{2\mu
}K_{1}^{\nu }\right) /D_{12}.  \label{t(s)}
\end{equation}%
Nevertheless, anticipating a connection with higher dimensions, we opt to
write the last term as a pseudoscalar function $t^{SP}=-t^{PS}=\varepsilon
^{\nu _{12}}t_{\nu _{12}}^{\left( -\right) }$. Therefore, given both
versions for the four-matrix trace, we have the corresponding versions for
the $AV$ amplitude%
\begin{eqnarray}
(t_{\mu _{12}}^{AV})_{1} &=&-2\varepsilon _{\mu _{1}\nu }t_{\mu
_{2}}^{\left( +\right) \nu }-\varepsilon _{\mu _{12}}t^{PP}-2\varepsilon
_{\mu _{2}\nu }t_{\mu _{1}}^{\left( -\right) \nu }+g_{\mu _{12}}t^{SP}, \\
(t_{\mu _{12}}^{AV})_{2} &=&-2\varepsilon _{\mu _{2}\nu }t_{\mu
_{1}}^{\left( +\right) \nu }-\varepsilon _{\mu _{12}}t^{SS}+2\varepsilon
_{\mu _{1}\nu }t_{\mu _{2}}^{\left( -\right) \nu }-g_{\mu _{12}}t^{SP}.
\end{eqnarray}

As mentioned at the beginning of the section, integrated amplitudes depend
exclusively on the external momentum $q$. That precludes the construction of
2nd-order antisymmetric tensors, which cancels out terms like $t^{\left(
-\right) }$ and $SP$. Further examination of the general form (\ref{2ptexp})
allows the identification of even amplitudes%
\begin{equation}
t_{\mu _{12}}^{VV}=2t_{\mu _{12}}^{\left( +\right) }+g_{\mu _{12}}t^{PP}%
\text{\quad and\quad }t_{\mu _{12}}^{AA}=2t_{\mu _{12}}^{\left( +\right)
}-g_{\mu _{12}}t^{SS}.  \label{even}
\end{equation}%
Hence, the integration provides relations among odd and even amplitudes%
\begin{eqnarray}
(T_{\mu _{12}}^{AV})_{1} &=&-\varepsilon _{\mu _{1}}^{\quad \nu }T_{\nu \mu
_{2}}^{VV};\text{\qquad }(T_{\mu _{12}}^{AV})_{2}=-\varepsilon _{\mu
_{2}}^{\quad \nu }T_{\mu _{1}\nu }^{AA};  \label{oddav} \\
(T_{\mu _{12}}^{VA})_{1} &=&-\varepsilon _{\mu _{1}}^{\quad \nu }T_{\nu \mu
_{2}}^{AA};\text{\qquad }(T_{\mu _{12}}^{VA})_{2}=-\varepsilon _{\mu
_{2}}^{\quad \nu }T_{\mu _{1}\nu }^{VV}.  \label{oddva}
\end{eqnarray}%
Although we did not detail, following the same steps produced both $VA$
versions. These associations are directly achieved at the integrand level
using the second identity for Dirac matrices $\gamma _{\ast }\gamma _{\mu
}=-\varepsilon _{\mu }^{\quad \nu }\gamma _{\nu }$ in the adequate position.
Even so, we need a clear distinction among versions since their comparison
is not automatic for integrated amplitudes due to their diverging character.

Lastly, we use the third identity in the form $\gamma _{\ast }\gamma _{\mu
\nu }=-\varepsilon _{\mu \nu }+g_{\mu \nu }\gamma _{\ast }$ to introduce the
third version for the discussed amplitudes. Disregarding terms on the
antisymmetric tensor $t^{\left( -\right) }$, the integrated amplitude links
to previous versions as follows:%
\begin{eqnarray}
(T_{\mu _{12}}^{AV})_{3} &=&-\frac{1}{2}[\varepsilon _{\mu _{1}}^{\hspace{7pt%
}\nu }T_{\nu \mu _{2}}^{VV}+\varepsilon _{\mu _{2}}^{\hspace{7pt}\nu }T_{\mu
_{1}\nu }^{AA}]=\frac{1}{2}[(T_{\mu _{12}}^{AV})_{1}+(T_{\mu
_{12}}^{AV})_{2}],  \label{AV3} \\
(T_{\mu _{12}}^{VA})_{3} &=&-\frac{1}{2}[\varepsilon _{\mu _{1}}^{\hspace{7pt%
}\nu }T_{\nu \mu _{2}}^{AA}+\varepsilon _{\mu _{2}}^{\hspace{7pt}\nu }T_{\mu
_{1}\nu }^{VV}]=\frac{1}{2}[(T_{\mu _{12}}^{VA})_{1}+(T_{\mu
_{12}}^{VA})_{2}].  \label{VA3}
\end{eqnarray}%
This particular aspect receives further attention in the four-dimensional
setting, having the sole purpose of illustrating how any amplitude version
follows from versions one and two here. The investigation from reference 
\cite{Battistel2004} uses the third version in Eq. (85).

Obtaining explicit results occurs by replacing the results from Appendix \ref%
{AppInt2D} inside integrated expressions of structures derived above. Scalar
two-point functions assume the forms 
\begin{equation}
T^{PP}=T^{SS}-4m^{2}J_{2}=q^{2}J_{2}-2I_{\log },
\end{equation}%
and the symmetric sign tensor is 
\begin{equation}
T_{\mu _{12}}^{\left( +\right) }=2(\bar{J}_{2\mu _{12}}+q_{\mu _{1}}J_{2\mu
_{2}})=\Delta _{2\mu _{12}}+g_{\mu _{12}}I_{\log }+2\theta _{\mu
_{12}}\left( m^{2}J_{2}+i/4\pi \right) -\frac{1}{2}g_{\mu _{12}}q^{2}J_{2},
\end{equation}%
where $\theta _{\mu \nu }\left( q\right) =\left( g_{\mu \nu }q^{2}-q_{\mu
}q_{\nu }\right) /q^{2}$ is the transversal projector. We combine these
pieces into odd tensors\footnote{%
It is possible to obtain $VA$ versions by redefining indexes through $\mu
_{1}\longleftrightarrow \mu _{2}$.}%
\begin{eqnarray}
(T_{\mu _{12}}^{AV})_{1} &=&-\varepsilon _{\mu _{1}}^{\quad \nu }[2\Delta
_{2\mu _{2}\nu }+4\theta _{\mu _{2}\nu }\left( m^{2}J_{2}+i/4\pi \right) ],
\label{AV1} \\
(T_{\mu _{12}}^{AV})_{2} &=&-\varepsilon _{\mu _{2}}^{\quad \nu }[2\Delta
_{2\mu _{1}\nu }+4\theta _{\mu _{1}\nu }\left( m^{2}J_{2}+i/4\pi \right)
-g_{\mu _{1}\nu }\left( 4m^{2}J_{2}\right) ],  \label{AV2}
\end{eqnarray}%
with the objects between squared brackets being even tensors.

We also use this opportunity to introduce amplitudes emerging through
momenta contractions. They follow a strong pattern acknowledged in all RAGFs
seen in this investigation. Whereas additional functions arising in axial
relations are finite%
\begin{equation}
T_{\mu }^{PV}=-T_{\mu }^{VP}=\varepsilon _{\mu \nu }q^{\nu }\left[
-2mJ_{2}\left( q\right) \right] ,  \label{PV}
\end{equation}%
other functions are pure surface terms proportional to the arbitrary
routings $k_{i}$\ as follows%
\begin{equation}
T_{\mu }^{A}\left( i\right) =-\varepsilon _{\mu }^{\quad \nu _{1}}T_{\nu
_{1}}^{V}\left( i\right) =2\varepsilon _{\mu }^{\quad \nu _{1}}k_{i}^{\nu
_{2}}\Delta _{2\nu _{12}}.  \label{A}
\end{equation}%
The last structure is consistent with the linear power counting of one-point
amplitudes in a two-dimensional setting.

Even though integrands of amplitudes are equivalent, the same does not apply
to their integrated form. In the case of even and odd tensor amplitudes,
expressions depend on the prescription adopted to evaluate divergences
because they contain surface terms $\Delta _{2}$. Additionally, odd
amplitudes depend on the trace version since using the definition of the
chiral matrix around the first or the second vertices brings implications
for the index arrangement in finite and divergent parts. This perspective
produced identities originally, but now the connection is not automatic.
That becomes clear when we subtract one $AV$ version from the other%
\begin{eqnarray}
(T_{\mu _{12}}^{AV})_{1}-(T_{\mu _{12}}^{AV})_{2} &=&-2(\varepsilon _{\mu
_{1}\nu }\Delta _{2\mu _{2}}^{\nu }-\varepsilon _{\mu _{2}\nu }\Delta _{2\mu
_{1}}^{\nu })+4\varepsilon _{\mu _{12}}m^{2}J_{2}  \notag \\
&&-4(\varepsilon _{\mu _{1}\nu }\theta _{\mu _{2}}^{\nu }-\varepsilon _{\mu
_{2}\nu }\theta _{\mu _{1}}^{\nu })\left( m^{2}J_{2}+i/4\pi \right) .
\end{eqnarray}%
We use Schouten identities\footnote{%
The antisymmetry of the Levi-Civita tensor establishes:%
\begin{eqnarray}
\varepsilon _{\lbrack \mu _{1}\nu }\Delta _{2\mu _{2}]}^{\nu }
&=&\varepsilon _{\mu _{1}\nu }\Delta _{2\mu _{2}}^{\nu }+\varepsilon _{\mu
_{2}\mu _{1}}\Delta _{2\nu }^{\nu }+\varepsilon _{\nu \mu _{2}}\Delta _{2\mu
_{1}}^{\nu }=0,  \label{SchDiv} \\
\varepsilon _{\lbrack \mu _{1}\nu }\theta _{\mu _{2}]}^{\nu } &=&\varepsilon
_{\mu _{1}\nu }\theta _{\mu _{2}}^{\nu }+\varepsilon _{\mu _{2}\mu
_{1}}\theta _{\nu }^{\nu }+\varepsilon _{\nu \mu _{2}}\theta _{\mu
_{1}}^{\nu }=0.  \label{SchTeta}
\end{eqnarray}%
} in two dimensions to rearrange indexes in the transversal projector and
surface terms; therefore, the difference reduces to 
\begin{equation}
(T_{\mu _{12}}^{AV})_{1}-(T_{\mu _{12}}^{AV})_{2}=-\varepsilon _{\mu
_{12}}[2\Delta _{2\alpha }^{\alpha }+i/\pi ].  \label{Uni-2D}
\end{equation}

The linearity of integration requires this difference to vanish identically,
which would constrain the value of the object $\Delta _{2\alpha }^{\alpha }$%
. That represents a link between linearity and the uniqueness of
perturbative solutions. We consider these concepts while investigating the
original expectation in the subsections.

\subsection{Relations Among Green Functions (RAGFs)}

This subsection aims to perform momenta contractions with odd amplitudes to
test the validity of RAGFs. Firstly, let us comment on even amplitudes
because they appear inside odd ones in Equations (\ref{AV1})-(\ref{AV2}).
They also follow relations, whose proof only requires algebraic operations: 
\begin{eqnarray}
q^{\mu _{1}}T_{\mu _{12}}^{VV} &=&2q^{\nu }\Delta _{2\mu _{2}\nu }=[T_{\mu
_{2}}^{V}\left( 1\right) -T_{\mu _{2}}^{V}\left( 2\right) ],  \label{pVV} \\
q^{\mu _{1}}T_{\mu _{12}}^{AA}+2mT_{\mu _{2}}^{PA} &=&2q^{\nu }\Delta _{2\mu
_{2}\nu }=[T_{\mu _{2}}^{V}\left( 1\right) -T_{\mu _{2}}^{V}\left( 2\right)
].
\end{eqnarray}%
Furthermore, they are automatic because they apply identically; observe the
vector one-point amplitudes (\ref{A}).

Such a feature differs from odd amplitudes although they contain the same
elements, i.e., finite contributions and the surface term $\Delta _{2}$. Let
us perform the corresponding contractions to test relations (\ref{p1AV})-(%
\ref{p2VA}). Starting with the first $AV$ version (\ref{AV1}), its vector
contraction yields%
\begin{equation}
q^{\mu _{2}}(T_{\mu _{12}}^{AV})_{1}=-2\varepsilon _{\mu _{1}\nu _{1}}q^{\nu
_{2}}\Delta _{2\nu _{2}}^{\nu _{1}}=[T_{\mu _{1}}^{A}\left( 1\right) -T_{\mu
_{1}}^{A}\left( 2\right) ].  \label{qAV1}
\end{equation}%
Analogously to the case of even amplitudes, finite terms vanish because $%
q^{\mu _{2}}\theta _{\mu _{2}}^{\nu }=0$ while it is straightforward to
identify the axial amplitude (\ref{A}).

In another way, the axial contraction exhibits an inadequate tensor
arranging since the momentum couples to the Levi-Civita symbol: 
\begin{equation}
q^{\mu _{1}}(T_{\mu _{12}}^{AV})_{1}=-q^{\mu _{1}}\varepsilon _{\mu
_{1}}^{\quad \nu }[2\Delta _{2\mu _{2}\nu }+4\theta _{\mu _{2}\nu }\left(
m^{2}J_{2}+i/4\pi \right) ].
\end{equation}%
This circumstance demands index permutations through Schouten identities (%
\ref{SchDiv})-(\ref{SchTeta}) for the surface term and the projector. Then,
reminding the trace $\theta _{\nu }^{\nu }=1$, we identify the $PV$
amplitude (\ref{PV}) and the axials 
\begin{equation}
q^{\mu _{1}}(T_{\mu _{12}}^{AV})_{1}=[T_{\mu _{2}}^{A}\left( 1\right)
-T_{\mu _{2}}^{A}\left( 2\right) ]-2mT_{\mu _{2}}^{PV}+\varepsilon _{\mu
_{2}\nu }q^{\nu }[2\Delta _{2\alpha }^{\alpha }+i/\pi ].  \label{pAV1}
\end{equation}%
The last term prevents the automatic satisfaction of this relation, which
depends on the value assumed by the surface term.

We observed the same situation for the second $AV$ version (\ref{AV2});
however, the additional term appears on its vector contraction\footnote{%
Since the third version is a combination of the others (\ref{AV3}), both
vertices have additional terms.}%
\begin{eqnarray}
q^{\mu _{2}}(T_{\mu _{12}}^{AV})_{2} &=&[T_{\mu _{1}}^{A}\left( 1\right)
-T_{\mu _{1}}^{A}\left( 2\right) ]+\varepsilon _{\mu _{1}\nu }q^{\nu
}[2\Delta _{2\alpha }^{\alpha }+i/\pi ]  \label{pAV2} \\
q^{\mu _{1}}(T_{\mu _{12}}^{AV})_{2} &=&[T_{\mu _{2}}^{A}\left( 1\right)
-T_{\mu _{2}}^{A}\left( 2\right) ]-2mT_{\mu _{2}}^{PV}.
\end{eqnarray}%
This pattern repeats for the $VA$ amplitude regardless of the vertex
arrangement. Additional terms arise for the $\mu _{1}$-contraction (vector)
of the first version and the $\mu _{2}$-contraction (axial) of the second
version:%
\begin{eqnarray}
q^{\mu _{1}}(T_{\mu _{12}}^{VA})_{1} &=&[T_{\mu _{2}}^{A}\left( 1\right)
-T_{\mu _{2}}^{A}\left( 2\right) ]+\varepsilon _{\mu _{2}\nu }q^{\nu
}[2\Delta _{2\alpha }^{\alpha }+i/\pi ] \\
q^{\mu _{2}}(T_{\mu _{12}}^{VA})_{2} &=&[T_{\mu _{1}}^{A}\left( 1\right)
-T_{\mu _{1}}^{A}\left( 2\right) ]+2mT_{\mu _{1}}^{VP}+\varepsilon _{\mu
_{1}\nu }q^{\nu }[2\Delta _{2\alpha }^{\alpha }+i/\pi ].
\end{eqnarray}

The RAGFs, deduced as identities for integrands, represent the linearity of
integration within this context. Even amplitudes automatically satisfy their
relations as there is no dependence on the surface term value. On the other
hand, odd amplitudes exhibit a potentially-violating term, so linearity
would require the condition%
\begin{equation}
\Delta _{2\alpha }^{\alpha }=-i\left( 2\pi \right) ^{-1}.  \label{finite1}
\end{equation}%
This contribution emerges for the contraction with the vertex that defines
the amplitude version (the position of use of the chiral matrix definition).
Choosing this finite value for the surface term ensures that all versions
are equal (\ref{Uni-2D}), elucidating the relation between linearity and
uniqueness. Any formula to the Dirac traces leads to one unique answer that
respects the linearity of integration.

Nevertheless, this condition sets nonzero values for the one-point functions
(\ref{A}), affecting symmetry implications through WIs. That occurs for all
relations in this subsection since amplitudes depend on the surface term.
This subject receives attention in the sequence.

\subsection{Ward Identities (WIs) and Low-Energy Implications}

We discussed the divergence of axial and vector currents (\ref{AWI})-(\ref%
{VWI}), indicating implications through WIs for perturbative amplitudes. The
adopted strategy translates these implications into restrictions over RAGFs,
linking linearity and symmetries. This subsection analyses such a connection
focusing on anomalous amplitudes ($AV$ and $VA$), known for the
impossibility of satisfying all WIs simultaneously.

Adopting a prescription that eliminates surface terms reduces all RAGFs for
even amplitudes ($VV$ and $AA$) to the corresponding WIs. Regarding odd
amplitudes, this condition satisfies those WIs corresponding to automatic
relations while violating others. Observe the first version of the $AV$ to
elucidate this statement. Identifying the vector relation was automatic;
however, the axial relation has an additional term. Hence, the zero value
for the surface term satisfies the vector WI while violating the axial WI
through one anomalous contribution. We see the opposite for the second
version of the amplitude, which violates the vector WI. Both identities are
violated for the third version since it is a composition of the first two.
Table \ref{tab2d} shows the mentioned results for the $AV$ and some examples
of even amplitudes. 
\begin{table}[h]
\caption{Ward identities using the zero value for the surface term.}
\label{tab2d}\renewcommand{\baselinestretch}{1.1}{\normalsize \centering{\ }%
\ }$%
\begin{tabular}{|l|l|}
\hline
$q^{\nu }(T_{\nu \mu }^{AV})_{1}=-2mT_{\mu }^{PV}+\left( i/\pi \right)
\varepsilon _{\mu \nu }q^{\nu }$ & $q^{\nu }(T_{\mu \nu }^{AV})_{1}=0$ \\ 
\hline
$q^{\nu }(T_{\nu \mu }^{AV})_{2}=-2mT_{\mu }^{PV}$ & $q^{\nu }(T_{\mu \nu
}^{AV})_{2}=\left( i/\pi \right) \varepsilon _{\mu \nu }q^{\nu }$ \\ \hline
$q^{\nu }(T_{\nu \mu }^{AV})_{3}=-2mT_{\mu }^{PV}+\left( i/2\pi \right)
\varepsilon _{\mu \nu }q^{\nu }$ & $q^{\nu }(T_{\mu \nu }^{AV})_{3}=\left(
i/2\pi \right) \varepsilon _{\mu \nu }q^{\nu }$ \\ \hline
$q^{\nu }T_{\nu \mu }^{AA}=-2mT_{\mu }^{PA}$ & $q^{\nu }T_{\mu \nu }^{VV}=0$
\\ \hline
\end{tabular}%
$%
\end{table}

This argumentation applies to the $VA$ without changes regarding vertex
arrangement. Under this perspective, selecting an amplitude version would
set the vertex (or vertices) with one anomalous contribution. Furthermore,
this perspective breaks the linearity of integration in anomalous amplitudes
for violating non-automatic RAGFs.

On the other hand, choosing the value that preserves linearity (\ref{finite1}%
) collapses different amplitude versions into one unique form\footnote{%
The third $AV$version is independent of the surface term value.
Parametrizing $\Delta _{2\mu \nu }=ag_{\mu \nu }$ in its equation, we get an
expression independent of the coefficient $a$ and equal to the unique form.}
(\ref{Uni-2D}). Nevertheless, that violates all WIs for odd and even
amplitudes as they depend on the surface term value; see Table \ref{tabuniq}%
. 
\begin{table}[h]
\caption{Ward identities using the non-zero value for the surface term.}
\label{tabuniq}\renewcommand{\baselinestretch}{1.1}{\normalsize \centering{\ 
}\ }$%
\begin{tabular}{|l|l|}
\hline
$q^{\nu }T_{\nu \mu }^{AV}=-2mT_{\mu }^{PV}+\left( i/2\pi \right)
\varepsilon _{\mu \nu }q^{\nu }$ & $q^{\nu }T_{\mu \nu }^{AV}=\left( i/2\pi
\right) \varepsilon _{\mu \nu }q^{\nu }$ \\ \hline
$q^{\nu }T_{\nu \mu }^{AA}=-2mT_{\mu }^{PA}-\left( i/2\pi \right) q_{\mu }$
& $q^{\nu }T_{\nu \mu }^{VV}=-\left( i/2\pi \right) q_{\mu }$ \\ \hline
\end{tabular}%
$%
\end{table}

Low-energy properties of finite functions are crucial to deepen this
analysis. Under the hypothesis that both WIs for the $AV$ amplitude apply,
we established the kinematical behavior at zero of $\Omega ^{PV}$ as being
zero (\ref{lowAV}). Nonetheless, employing the $PV$ expression (\ref{PV})
together with the limit (\ref{LetZ2D}) yields a non-zero outcome: 
\begin{equation}
\Omega ^{PV}\left( 0\right) =\left. 4m^{2}J_{2}\right\vert _{0}=\frac{i}{\pi 
}m^{2}Z_{0}^{\left( -1\right) }\left( 0\right) =-\frac{i}{\pi }.
\end{equation}%
That means the hypothesis is false. Hence, when satisfying the vector WI,
the axial WI violation is the value corresponding to the negative of $\Omega
^{PV}(0)$. The other expectation (\ref{lowAV2}) leads to the reciprocal;
thus, satisfying the axial WI implies violating the vector WI.

Let us extend these ideas using the general structure of a 2nd-order odd
tensor (\ref{AVForm}). In both $AV$ and $VA$ cases, momenta contractions
lead to a set of functions written in terms of form factors%
\begin{eqnarray}
q^{\mu _{1}}F_{\mu _{12}} &=&\varepsilon _{\mu _{2}\nu }q^{\nu }V_{1}\left(
q^{2}\right) =\varepsilon _{\mu _{2}\nu }q^{\nu }\left(
q^{2}F_{3}-F_{1}\right) \\
q^{\mu _{2}}F_{\mu _{12}} &=&\varepsilon _{\mu _{1}\nu }q^{\nu }V_{2}\left(
q^{2}\right) =\varepsilon _{\mu _{1}\nu }q^{\nu }\left(
q^{2}F_{2}+F_{1}\right) .
\end{eqnarray}%
If form factors are free of kinematic singularities observed in the explicit
forms of amplitudes, the implication at zero follows 
\begin{equation}
V_{1}\left( 0\right) +V_{2}\left( 0\right) =0.  \label{id2D}
\end{equation}%
Thereby, if one term vanishes, the other must do so. Otherwise, if one term
relates to a finite function ($PV$ or $VP$), an additional constant must
appear as compensation within the last equation. These statements are
inconsistent with the satisfaction of both WIs, which only occurs if
linearity of integration holds with null surface terms. Thus, the low-energy
behavior of these finite functions is the source of anomalous terms in
amplitudes and not their perturbative ambiguity.

Nevertheless, ambiguities relate to these low-energy implications. Under the
condition of linearity and considering surface terms in the general tensor,
this limit implies the constraint $2\Delta _{2\alpha }^{\alpha }=\Omega
^{PV}(0)$. Such an aspect will be fully explored in the following section
considering axial triangles in the physical dimension. Conclusions similar
to those drawn here anticipate the presence of anomalies and linearity
breaking in this new context.

\section{Four-Dimensional Amplitudes}

\label{4Dim3Pt}The analysis developed in the physical dimension focuses on 
\textit{axial} amplitudes that are rank-3 tensors, namely $AVV$, $VAV$, $VVA$%
, and $AAA$. Their mathematical structures follow the same features observed
in two dimensions since computing the highest-order trace yields products
between the Levi-Civita symbol and the metric tensor. After integration,
that generates expressions that differ in their dependence on surface terms
and finite parts. We want to verify these prospects by evaluating the
triangles' basic versions. Once these resources are clear, we study how
symmetries, linearity of integration, and uniqueness manifest.

From Eqs. (\ref{t}) and (\ref{T}), integrated three-point amplitudes are
denoted through capital letters $T^{\Gamma _{1}\Gamma _{2}\Gamma _{3}}$ and
exhibit the integrand%
\begin{equation}
t^{\Gamma _{1}\Gamma _{2}\Gamma _{3}}=\text{\textrm{tr}}\left[ \Gamma
_{1}S_{F}\left( 1\right) \Gamma _{2}S_{F}\left( 2\right) \Gamma
_{3}S_{F}\left( 3\right) \right] .  \label{t3}
\end{equation}%
Thus, after replacing vertex operators and disregarding vanishing traces,
3rd-order amplitudes assume the forms%
\begin{eqnarray}
t_{\mu _{123}}^{AVV} &=&[K_{123}^{\nu _{123}}\mathrm{tr}(\gamma _{\ast \mu
_{1}\nu _{1}\mu _{2}\nu _{2}\mu _{3}\nu _{3}})+m^{2}\mathrm{tr}(\gamma
_{\ast \mu _{1}\mu _{2}\mu _{3}\nu _{1}})(K_{1}^{\nu _{1}}-K_{2}^{\nu
_{1}}+K_{3}^{\nu _{1}})]/D_{123},  \label{AVVexp} \\
t_{\mu _{123}}^{VAV} &=&[K_{123}^{\nu _{123}}\mathrm{tr}(\gamma _{\ast \mu
_{1}\nu _{1}\mu _{2}\nu _{2}\mu _{3}\nu _{3}})+m^{2}\mathrm{tr}(\gamma
_{\ast \mu _{1}\mu _{2}\mu _{3}\nu _{1}})(K_{1}^{\nu _{1}}+K_{2}^{\nu
_{1}}-K_{3}^{\nu _{1}})]/D_{123},  \label{VAVexp} \\
t_{\mu _{123}}^{VVA} &=&[K_{123}^{\nu _{123}}\mathrm{tr}(\gamma _{\ast \mu
_{1}\nu _{1}\mu _{2}\nu _{2}\mu _{3}\nu _{3}})-m^{2}\mathrm{tr}(\gamma
_{\ast \mu _{1}\mu _{2}\mu _{3}\nu _{1}})(K_{1}^{\nu _{1}}-K_{2}^{\nu
_{1}}-K_{3}^{\nu _{1}})]/D_{123},  \label{VVAexp} \\
t_{\mu _{123}}^{AAA} &=&[K_{123}^{\nu _{123}}\mathrm{tr}(\gamma _{\ast \mu
_{1}\nu _{1}\mu _{2}\nu _{2}\mu _{3}\nu _{3}})-m^{2}\mathrm{tr}(\gamma
_{\ast \mu _{1}\mu _{2}\mu _{3}\nu _{1}})(K_{1}^{\nu _{1}}+K_{2}^{\nu
_{1}}+K_{3}^{\nu _{1}})]/D_{123},  \label{AAAexp}
\end{eqnarray}%
where we recall conventions $K_{123}^{\nu _{123}}=K_{1}^{\nu _{1}}K_{2}^{\nu
_{2}}K_{3}^{\nu _{3}}$ and $D_{123}=D_{1}D_{2}D_{3}$.

Although the trace involving four Dirac matrices plus the chiral one is
univocal, different expressions are attributed to the leading trace when
considering identities (\ref{Chiral-Id}). Appendix \ref{Tr6G4D} shows that
forms achieved through definition $\gamma _{\ast }=i\varepsilon _{\nu
_{1234}}\gamma ^{\nu _{1234}}/4!$ are enough to compound any other; thus,
our starting point is on their general structure 
\begin{eqnarray}
\left( 4i\right) ^{-1}\mathrm{tr}(\gamma _{\ast abcdef})
&=&+g_{ab}\varepsilon _{cdef}+g_{ad}\varepsilon _{bcef}+g_{af}\varepsilon
_{bcde}  \notag \\
&&+g_{bc}\varepsilon _{adef}+g_{cd}\varepsilon _{abef}+g_{cf}\varepsilon
_{abde}  \notag \\
&&+g_{be}\varepsilon _{acdf}+g_{de}\varepsilon _{abcf}+g_{ef}\varepsilon
_{abcd}  \notag \\
&&-g_{bd}\varepsilon _{acef}-g_{df}\varepsilon _{abce}-g_{bf}\varepsilon
_{acde}  \notag \\
&&-g_{ac}\varepsilon _{bdef}-g_{ce}\varepsilon _{abdf}-g_{ae}\varepsilon
_{bcdf}.  \label{Mold6}
\end{eqnarray}%
There are three basic versions, each corresponding to replacing the chiral
matrix near a specific vertex operator designated by a numeric label: 
\begin{equation}
\lbrack \mathrm{tr}(\gamma _{\ast \mu _{1}\nu _{1}\mu _{2}\nu _{2}\mu
_{3}\nu _{3}})]_{1}=[\mathrm{tr}(\gamma _{\ast \mu _{2}\nu _{2}\mu _{3}\nu
_{3}\mu _{1}\nu _{1}})]_{2}=[\mathrm{tr}(\gamma _{\ast \mu _{3}\nu _{3}\mu
_{1}\nu _{1}\mu _{2}\nu _{2}})]_{3}.  \label{identraces}
\end{equation}%
One obtains their explicit forms when setting the index configurations in
the general trace, which brings sign differences for the monomials. This
property is clear in contractions cast in the sequence. Integrating these
structures leads to three not (automatically) equivalent expressions for
each triangle. 
\begin{eqnarray}
\lbrack K_{123}^{\nu _{123}}\mathrm{tr}(\gamma _{\ast \mu _{1}\nu _{1}\mu
_{2}\nu _{2}\mu _{3}\nu _{3}})]_{1} &=&-4i\varepsilon _{\mu _{23}\nu
_{12}}[K_{1\mu _{1}}K_{23}^{\nu _{12}}-K_{2\mu _{1}}K_{13}^{\nu
_{12}}+K_{3\mu _{1}}K_{12}^{\nu _{12}}]  \notag \\
&&-4i\varepsilon _{\mu _{13}\nu _{12}}[K_{1\mu _{2}}K_{23}^{\nu
_{12}}+K_{2\mu _{2}}K_{13}^{\nu _{12}}-K_{3\mu _{2}}K_{12}^{\nu _{12}}] 
\notag \\
&&+4i\varepsilon _{\mu _{12}\nu _{12}}[K_{1\mu _{3}}K_{23}^{\nu
_{12}}-K_{2\mu _{3}}K_{13}^{\nu _{12}}-K_{3\mu _{3}}K_{12}^{\nu _{12}}] 
\notag \\
&&-4i\varepsilon _{\mu _{123}\nu }[K_{1}^{\nu }(K_{2}\cdot K_{3})-K_{2}^{\nu
}(K_{1}\cdot K_{3})+K_{3}^{\nu }(K_{1}\cdot K_{2})]  \notag \\
&&+4i[-g_{\mu _{12}}\varepsilon _{\mu _{3}\nu _{123}}-g_{\mu
_{23}}\varepsilon _{\mu _{1}\nu _{123}}+g_{\mu _{13}}\varepsilon _{\mu
_{2}\nu _{123}}]K_{123}^{\nu _{123}}  \label{tr1}
\end{eqnarray}%
\begin{eqnarray}
\lbrack K_{123}^{\nu _{123}}\mathrm{tr}(\gamma _{\ast \mu _{2}\nu _{2}\mu
_{3}\nu _{3}\mu _{1}\nu _{1}})]_{2} &=&+4i\varepsilon _{\mu _{13}\nu
_{12}}[K_{1\mu _{2}}K_{23}^{\nu _{12}}-K_{2\mu _{2}}K_{13}^{\nu
_{12}}+K_{3\mu _{2}}K_{12}^{\nu _{12}}]  \notag \\
&&-4i\varepsilon _{\mu _{12}\nu _{12}}[K_{1\mu _{3}}K_{23}^{\nu
_{23}}+K_{2\mu _{3}}K_{13}^{\nu _{13}}+K_{3\mu _{3}}K_{12}^{\nu _{12}}] 
\notag \\
&&-4i\varepsilon _{\mu _{23}\nu _{12}}[K_{1\mu _{1}}K_{23}^{\nu
_{23}}+K_{2\mu _{1}}K_{13}^{\nu _{13}}-K_{3\mu _{1}}K_{12}^{\nu _{12}}] 
\notag \\
&&-4i\varepsilon _{\mu _{123}\nu }[K_{1}^{\nu }(K_{2}\cdot K_{3})+K_{2}^{\nu
}(K_{1}\cdot K_{3})-K_{3}^{\nu }(K_{1}\cdot K_{2})]  \notag \\
&&+4i[g_{\mu _{12}}\varepsilon _{\mu _{3}\nu _{123}}-g_{\mu
_{13}}\varepsilon _{\mu _{2}\nu _{123}}-g_{\mu _{23}}\varepsilon _{\mu
_{1}\nu _{123}}]K_{123}^{\nu _{123}}  \label{tr2}
\end{eqnarray}%
\begin{eqnarray}
\lbrack K_{123}^{\nu _{123}}\mathrm{tr}(\gamma _{\ast \mu _{3}\nu _{3}\mu
_{1}\nu _{1}\mu _{2}\nu _{2}})]_{3} &=&-4i\varepsilon _{\mu _{12}\nu
_{12}}[K_{1\mu _{3}}K_{23}^{\nu _{12}}-K_{2\mu _{3}}K_{13}^{\nu
_{12}}+K_{3\mu _{3}}K_{12}^{\nu _{12}}]  \notag \\
&&-4i\varepsilon _{\mu _{23}\nu _{12}}[K_{1\mu _{1}}K_{23}^{\nu
_{12}}-K_{2\mu _{1}}K_{13}^{\nu _{12}}-K_{3\mu _{1}}K_{12}^{\nu _{12}}] 
\notag \\
&&-4i\varepsilon _{\mu _{13}\nu _{12}}[K_{1\mu _{2}}K_{23}^{\nu
_{12}}+K_{2\mu _{2}}K_{13}^{\nu _{12}}+K_{3\mu _{2}}K_{12}^{\nu _{12}}] 
\notag \\
&&+4i\varepsilon _{\mu _{123}\nu }[K_{1}^{\nu }(K_{2}\cdot K_{3})-K_{2}^{\nu
}(K_{1}\cdot K_{3})-K_{3}^{\nu }(K_{1}\cdot K_{2})]  \notag \\
&&+4i[-g_{\mu _{12}}\varepsilon _{\mu _{3}\nu _{123}}-g_{\mu
_{13}}\varepsilon _{\mu _{2}\nu _{123}}+g_{\mu _{23}}\varepsilon _{\mu
_{1}\nu _{123}}]K_{123}^{\nu _{123}}  \label{tr3}
\end{eqnarray}

Analogously to two-dimensional calculations, our next task consists of
organizing and integrating the amplitudes. As the three first rows of the
above equations are similar to the object (\ref{t(s)}), let us define
another standard tensor%
\begin{equation}
\varepsilon _{\mu _{ab}\nu _{12}}t_{\mu _{c}}^{\nu _{12}\left(
s_{1}s_{2}\right) }=\varepsilon _{\mu _{ab}\nu _{12}}\left( K_{1\mu
_{c}}K_{23}^{\nu _{12}}+s_{1}K_{2\mu _{c}}K_{13}^{\nu _{12}}+s_{2}K_{3\mu
_{c}}K_{12}^{\nu _{12}}\right) /D_{123}
\end{equation}%
where $s_{i}=\pm 1$. We rewrite this equation using $K_{i}=K_{j}+p_{ij}$ and 
$\varepsilon _{\mu _{ab}\nu _{12}}K_{ij}^{\nu _{12}}=\varepsilon _{\mu
_{ab}\nu _{12}}p_{ji}^{\nu _{2}}K_{i}^{\nu _{1}}$ to achieve structures
introduced in Subsection (\ref{BasisFI}): 
\begin{eqnarray}
\varepsilon _{\mu _{ab}\nu _{12}}t_{\mu _{c}}^{\nu _{12}\left(
s_{1}s_{2}\right) } &=&\varepsilon _{\mu _{ab}\nu
_{12}}[(1+s_{1})p_{31}^{\nu _{2}}-(1-s_{2})p_{21}^{\nu _{2}}]K_{1}^{\nu
_{1}}K_{1\mu _{c}}/D_{123}  \notag \\
&&+\varepsilon _{\mu _{ab}\nu _{12}}[p_{21}^{\nu _{1}}p_{32}^{\nu
_{2}}K_{1\mu _{c}}+(s_{1}p_{21\mu _{c}}p_{31}^{\nu _{2}}+s_{2}p_{31\mu
_{c}}p_{21}^{\nu _{2}})K_{1}^{\nu _{1}}]/D_{123}.
\end{eqnarray}%
Hence, final expressions arise by replacing vector and tensor Feynman
integrals from Appendix \ref{AppInt4D}. Although four sign configurations
are available, the integral achieved by taking $s_{1}=-s_{2}=-1$ cancels
out. That is straightforward for the first row, but a closer look at the
vector integral is necessary to analyze the second: 
\begin{equation}
\overline{J}_{3}^{\mu }=J_{3}^{\mu }=i\left( 4\pi \right) ^{-2}[-p_{21}^{\mu
}Z_{10}^{\left( -1\right) }\left( p_{21},p_{31}\right) -p_{31}^{\mu
}Z_{01}^{\left( -1\right) }\left( p_{21},p_{31}\right) ].  \label{4DJ3mu}
\end{equation}%
Since it is proportional to external momenta, it leads to symmetric tensors
that vanish when contracted with the Levi-Civita symbol. We cast all sign
configurations in the sequence: 
\begin{eqnarray}
2\varepsilon _{\mu _{ab}\nu _{12}}T_{\mu _{c}}^{\nu _{12}\left( -+\right) }
&=&2\varepsilon _{\mu _{ab}\nu _{12}}[p_{21}^{\nu _{1}}p_{32}^{\nu
_{2}}J_{3\mu _{c}}+(-p_{21\mu _{c}}p_{31}^{\nu _{2}}+p_{31\mu
_{c}}p_{21}^{\nu _{2}})J_{3}^{\nu _{1}}]\equiv 0,  \label{T-+} \\
2\varepsilon _{\mu _{ab}\nu _{12}}T_{\mu _{c}}^{\nu _{12}(+-)}
&=&4\varepsilon _{\mu _{ab}\nu _{12}}[p_{31}^{\nu _{2}}(J_{3\mu _{c}}^{\nu
_{1}}+p_{21\mu _{c}}J_{3}^{\nu _{1}})-p_{21}^{\nu _{2}}(J_{3\mu _{c}}^{\nu
_{1}}+p_{31\mu _{c}}J_{3}^{\nu _{1}})]  \notag \\
&&+(\varepsilon _{\mu _{ab}\nu _{12}}p_{32}^{\nu _{2}}\Delta _{3\mu
_{c}}^{\nu _{1}}+\varepsilon _{\mu _{abc}\nu }p_{32}^{\nu }I_{\log }),
\label{T+-} \\
2\varepsilon _{\mu _{ab}\nu _{12}}T_{\mu _{c}}^{\nu _{12}\left( --\right) }
&=&-4\varepsilon _{\mu _{ab}\nu _{12}}p_{21}^{\nu _{2}}(J_{3\mu _{c}}^{\nu
_{1}}+p_{31\mu _{c}}J_{3}^{\nu _{1}})-(\varepsilon _{\mu _{ab}\nu
_{12}}p_{21}^{\nu _{2}}\Delta _{3\mu _{c}}^{\nu _{1}}+\varepsilon _{\mu
_{abc}\nu }p_{21}^{\nu }I_{\log }),  \label{T--} \\
2\varepsilon _{\mu _{ab}\nu _{12}}T_{\mu _{c}}^{\nu _{12}\left( ++\right) }
&=&+4\varepsilon _{\mu _{ab}\nu _{12}}p_{31}^{\nu _{2}}(J_{3\mu _{c}}^{\nu
_{1}}+p_{21\mu _{c}}J_{3}^{\nu _{1}})+(\varepsilon _{\mu _{ab}\nu
_{12}}p_{31}^{\nu _{2}}\Delta _{3\mu _{c}}^{\nu _{1}}+\varepsilon _{\mu
_{abc}\nu }p_{31}^{\nu }I_{\log }).  \label{T++}
\end{eqnarray}

Different tensor contributions appear for each trace version from (\ref{tr1}%
)-(\ref{tr3}); therefore, one identifies the ensuing combinations after
disregarding the vanishing contribution: 
\begin{eqnarray}
C_{1\mu _{123}} &=&-\varepsilon _{\mu _{13}\nu _{12}}T_{\mu _{2}}^{\nu
_{12}\left( +-\right) }+\varepsilon _{\mu _{12}\nu _{12}}T_{\mu _{3}}^{\nu
_{12}\left( --\right) },  \label{C1} \\
C_{2\mu _{123}} &=&-\varepsilon _{\mu _{12}\nu _{12}}T_{\mu _{3}}^{\nu
_{12}\left( ++\right) }-\varepsilon _{\mu _{23}\nu _{12}}T_{\mu _{1}}^{\nu
_{12}\left( +-\right) }, \\
C_{3\mu _{123}} &=&-\varepsilon _{\mu _{23}\nu _{12}}T_{\mu _{1}}^{\nu
_{12}\left( --\right) }-\varepsilon _{\mu _{13}\nu _{12}}T_{\mu _{2}}^{\nu
_{12}\left( ++\right) }.  \label{C3}
\end{eqnarray}%
The sampling of indexes reflects the absence of the version-defining index $%
\mu _{i}$ within standard tensors from $C_{i}$, enabling the anticipation of
violations of either WIs or RAGFs. This specific index appears in vanishing
contributions $\varepsilon _{\mu _{ab}\nu _{12}}T_{\mu _{i}}^{\nu
_{12}\left( -,+\right) }$, present in expressions above before integration.

Let us return to the last row of Eqs. (\ref{tr1})-(\ref{tr3}), which
corresponds to 1st-order parity-odd triangles. The precise identifications
among the twelve possibilities occur when replacing the vertex
configurations in the general integrand (\ref{t3}); however, all of them are
proportional to the same structure:%
\begin{equation}
t_{\mu _{i}}^{ASS}=4i\varepsilon _{\mu _{i}\nu _{123}}K_{123}^{\nu _{123}}%
\frac{1}{D_{123}}=4i\varepsilon _{\mu _{i}\nu _{123}}p_{21}^{\nu
_{2}}p_{31}^{\nu _{3}}K_{1}^{\nu _{1}}\frac{1}{D_{123}}.
\end{equation}%
We already performed simplifications through symmetry properties already
acknowledged in the tensor sector. The integrated amplitude depends on the
finite vector $\bar{J}_{3}^{\nu _{1}}=J_{3}^{\nu _{1}}$, whose contraction
vanishes for being proportional to external momenta:%
\begin{equation}
T_{\mu _{i}}^{ASS}=4i\varepsilon _{\mu _{i}\nu _{123}}p_{21}^{\nu
_{2}}p_{31}^{\nu _{3}}J_{3}^{\nu _{1}}=0.  \label{ASS}
\end{equation}%
For this reason, we omit this class of amplitudes from the final triangles.

Lastly, we still have to organize terms proportional to $\varepsilon _{\mu
_{123}\nu }$ within traces (\ref{tr1})-(\ref{tr3}). Together with mass terms
from the remaining trace, these bilinears lead to twelve different
subamplitudes identified after comparing vertex arrangements in (\ref{t3}).
This result is general: besides the common tensors $C_{i}$, rank-1
parity-even subamplitudes appear inside each version of rank-3 \textit{axial}
amplitudes. Table \ref{tabversions} accounts for all of these possibilities,
while Appendix \ref{AppSub} presents explicit expressions. 
\begin{table}[tbph]
\caption{Even subamplitudes related to each version of 3rd-order axial
amplitudes.}
\label{tabversions}\renewcommand{\baselinestretch}{1.4}{\normalsize %
\centering{\ }}$%
\begin{tabular}{|c|c|c|c|c|}
\hline
$\text{Version/Type}$ & $AVV$ & $VAV$ & $VVA$ & $AAA$ \\ \hline
1 & $+VPP$ & $+ASP$ & $-APS$ & $-VSS$ \\ \hline
2 & $-SAP$ & $+PVP$ & $+PAS$ & $-SVS$ \\ \hline
3 & $+SPA$ & $-PSA$ & $+PPV$ & $-SSV$ \\ \hline
\end{tabular}%
${\normalsize \ }
\end{table}

Let us consider the first $AVV$ version to illustrate. After combining mass
terms from Eq. (\ref{AVVexp}) with bilinears from Eq. (\ref{tr1}), we find
the $VPP$ subamplitude 
\begin{equation}
\mathrm{sub}(t_{\mu _{123}}^{AVV})_{1}=i\varepsilon _{\mu _{123}\nu
}(t^{VPP})^{\nu }.
\end{equation}%
Integrating the corresponding structure yields the form%
\begin{equation}
(t^{VPP})^{\nu }=\text{\textrm{tr}}[\gamma ^{\nu }S_{F}\left( 1\right)
\gamma _{\ast }S_{F}\left( 2\right) \gamma _{\ast }S_{F}\left( 3\right)
]=4(-K_{1}^{\nu }S_{23}+K_{2}^{\nu }S_{13}-K_{3}^{\nu }S_{12})/D_{123},
\end{equation}%
where combinations $S_{ij}=K_{i}\cdot K_{j}-m^{2}$ come from definition (\ref%
{Sij}). After reducing the denominator, we perform the integration%
\begin{eqnarray}
(T^{VPP})^{\nu } &=&2[P_{31}^{\alpha }\Delta _{3\alpha }^{\nu }+(p_{21}^{\nu
}-p_{32}^{\nu })I_{\log }]-4\left( p_{21}\cdot p_{32}\right) J_{3}^{\nu } 
\notag \\
&&+2[(p_{31}^{\nu }p_{21}^{2}-p_{21}^{\nu }p_{31}^{2})J_{3}+p_{21}^{\nu
}J_{2}\left( p_{21}\right) -p_{32}^{\nu }J_{2}\left( p_{32}\right) ].
\label{TVPP}
\end{eqnarray}

We also use this opportunity to elucidate the final form of \textit{axial}
amplitudes. In general, the $i$-th version of the amplitude arises as a
combination between the common tensor $C_{i}$ and one specific vector
subamplitude. For instance, consulting Table \ref{tabversions}, one writes
the three basic versions of the $AVV$ triangle%
\begin{eqnarray}
(T_{\mu _{123}}^{AVV})_{1} &=&4iC_{1\mu _{123}}+i\varepsilon _{\mu _{123}\nu
}(T^{VPP})^{\nu },  \label{AVV1} \\
(T_{\mu _{123}}^{AVV})_{2} &=&4iC_{2\mu _{123}}-i\varepsilon _{\mu _{123}\nu
}(T^{SAP})^{\nu }, \\
(T_{\mu _{123}}^{AVV})_{3} &=&4iC_{3\mu _{123}}+i\varepsilon _{\mu _{123}\nu
}(T^{SPA})^{\nu }.
\end{eqnarray}%
It is straightforward to attribute an expression that comprises all vertices
configurations: 
\begin{equation}
(T_{\mu _{123}}^{\Gamma _{1}\Gamma _{2}\Gamma _{3}})_{i}=4iC_{i\mu
_{123}}\pm i\varepsilon _{\mu _{123}\nu }\left( \text{Corresponding
subamplitude}\right) ^{\nu }.
\end{equation}

To detail crucial points about these amplitudes, let us use the tools
developed in this section to build up the first $AVV$ version%
\begin{eqnarray}
(T_{\mu _{123}}^{AVV})_{1} &=&S_{1\mu _{123}}-8i\varepsilon _{\mu _{12}\nu
_{12}}p_{21}^{\nu _{2}}(J_{3\mu _{3}}^{\nu _{1}}+p_{31\mu _{3}}J_{3}^{\nu
_{1}})  \notag \\
&&-8i\varepsilon _{\mu _{13}\nu _{12}}[p_{31}^{\nu _{2}}(J_{3\mu _{2}}^{\nu
_{1}}+p_{21\mu _{2}}J_{3}^{\nu _{1}})-p_{21}^{\nu _{2}}(J_{3\mu _{2}}^{\nu
_{1}}+p_{31\mu _{2}}J_{3}^{\nu _{1}})]  \notag \\
&&-4i\varepsilon _{\mu _{123}\nu }(p_{21}\cdot p_{32})J_{3}^{\nu
}+2i\varepsilon _{\mu _{123}\nu }[(p_{31}^{\nu }p_{21}^{2}-p_{21}^{\nu
}p_{31}^{2})]J_{3}  \notag \\
&&+2i\varepsilon _{\mu _{123}\nu }[p_{21}^{\nu }J_{2}\left( p_{21}\right)
-p_{32}^{\nu }J_{2}\left( p_{32}\right) ].  \label{AVV1complete}
\end{eqnarray}%
The divergent part of the common tensor (\ref{C1}) comes from Eqs. (\ref{T+-}%
) and (\ref{T--}) as 
\begin{equation}
4iC_{1\mu _{123}}=-2i[\varepsilon _{\mu _{13}\nu _{12}}p_{32}^{\nu
_{2}}\Delta _{3\mu _{2}}^{\nu _{1}}+\varepsilon _{\mu _{12}\nu
_{12}}p_{21}^{\nu _{2}}\Delta _{3\mu _{3}}^{\nu _{1}}+\varepsilon _{\mu
_{123}\nu }(p_{21}^{\nu }-p_{32}^{\nu })I_{\log }].
\end{equation}%
When combined with the $VPP$ subamplitude, we acknowledge the exact
cancellation of the object $I_{\log }$ as it occured for all investigated
versions of all amplitudes. Thus, surface terms compound the whole structure
of divergences%
\begin{equation}
S_{1\mu _{123}}=2i(\varepsilon _{\mu _{13}\nu _{12}}p_{23}^{\nu _{2}}\Delta
_{3\mu _{2}}^{\nu _{1}}+\varepsilon _{\mu _{12}\nu _{12}}p_{12}^{\nu
_{2}}\Delta _{3\mu _{3}}^{\nu _{1}}+\varepsilon _{\mu _{123}\nu
_{1}}P_{31}^{\nu _{2}}\Delta _{3\nu _{2}}^{\nu _{1}}).  \label{ST1}
\end{equation}%
Moreover, contributions from vector subamplitudes exhibit arbitrary momenta $%
P_{ij}=k_{i}+k_{j}$ as coefficients. We stress that the divergent content is
shared: regardless of the vertex arrangement, the first version of each 
\textit{axial} amplitude contains the same structure (\ref{ST1}). For later
use, we define the other sets of surface terms:%
\begin{eqnarray}
S_{2\mu _{123}} &=&2i(\varepsilon _{\mu _{12}\nu _{12}}p_{13}^{\nu
_{2}}\Delta _{3\mu _{3}}^{\nu _{1}}+\varepsilon _{\mu _{23}\nu
_{12}}p_{23}^{\nu _{2}}\Delta _{3\mu _{1}}^{\nu _{1}}+\varepsilon _{\mu
_{123}\nu _{1}}P_{21}^{\nu _{2}}\Delta _{3\nu _{2}}^{\nu _{1}}),  \label{ST2}
\\
S_{3\mu _{123}} &=&2i(\varepsilon _{\mu _{13}\nu _{12}}p_{13}^{\nu
_{2}}\Delta _{3\mu _{2}}^{\nu _{1}}+\varepsilon _{\mu _{23}\nu
_{12}}p_{21}^{\nu _{2}}\Delta _{3\mu _{1}}^{\nu _{1}}+\varepsilon _{\mu
_{123}\nu _{1}}P_{32}^{\nu _{2}}\Delta _{3\nu _{2}}^{\nu _{1}}).  \label{ST3}
\end{eqnarray}%
That concludes the preliminary discussion on rank-3 triangles, so
investigating momenta contractions is attainable. That is the subject of the
following subsections.

\subsection{Relations Among Green Functions (RAGFs)\label{unique}}

The next step is to perform momenta contractions to find RAGFs following the
recipes in (\ref{RAGF1})-(\ref{RAGF2}). Although they are algebraic
identities at the integrand level, their satisfaction is not automatic after
integration. In parallel with the two-dimensional case, possibilities for
Dirac traces and values of surface terms bring implications for this
analysis. Given the differences\footnote{%
Although other configurations appear, it is easy to verify their redundancy
using the antisymmetric character of this amplitude (below). By exchanging
the position of matrices within the trace, one permutes free indexes to show
that $t_{\mu _{ij}}^{AV}\left( a,b\right) =-t_{\mu _{ji}}^{AV}\left(
a,b\right) $ and summed indices to achieve $t_{\mu _{ij}}^{AV}\left(
a,b\right) =-t_{\mu _{ij}}^{AV}\left( b,a\right) $.%
\begin{equation*}
t_{\mu _{ij}}^{AV}\left( a,b\right) =K_{a}^{\nu _{1}}K_{b}^{\nu _{2}}\mathrm{%
tr}(\gamma _{\ast }\gamma _{\mu _{i}\nu _{1}\mu _{j}\nu _{2}})/D_{ij}
\end{equation*}%
}%
\begin{eqnarray}
t_{1\left( -\right) \mu _{23}}^{AV} &=&t_{\mu _{23}}^{AV}\left( 2,1\right)
-t_{\mu _{23}}^{AV}\left( 2,3\right) , \\
t_{2\left( -\right) \mu _{13}}^{AV} &=&t_{\mu _{13}}^{AV}\left( 1,3\right)
-t_{\mu _{13}}^{AV}\left( 2,3\right) , \\
t_{3\left( -\right) \mu _{12}}^{AV} &=&t_{\mu _{12}}^{AV}\left( 1,2\right)
-t_{\mu _{12}}^{AV}\left( 1,3\right) ,
\end{eqnarray}%
we introduce the mentioned identities: 
\begin{eqnarray}
p_{31}^{\mu _{1}}t_{\mu _{123}}^{AVV} &=&t_{1\left( -\right) \mu
_{23}}^{AV}-2mt_{\mu _{23}}^{PVV}  \label{AVVragfs} \\
p_{21}^{\mu _{2}}t_{\mu _{123}}^{AVV} &=&t_{2\left( -\right) \mu _{13}}^{AV}
\\
p_{32}^{\mu _{3}}t_{\mu _{123}}^{AVV} &=&t_{3\left( -\right) \mu _{12}}^{AV}
\end{eqnarray}%
\begin{eqnarray}
p_{31}^{\mu _{1}}t_{\mu _{123}}^{VAV} &=&t_{1\left( -\right) \mu _{23}}^{AV}
\label{VAVragfs} \\
p_{21}^{\mu _{2}}t_{\mu _{123}}^{VAV} &=&t_{2\left( -\right) \mu
_{13}}^{AV}+2mt_{\mu _{13}}^{VPV} \\
p_{32}^{\mu _{3}}t_{\mu _{123}}^{VAV} &=&t_{3\left( -\right) \mu _{12}}^{AV}
\end{eqnarray}%
\begin{eqnarray}
p_{31}^{\mu _{1}}t_{\mu _{123}}^{VVA} &=&t_{1\left( -\right) \mu _{23}}^{AV}
\label{VVAragfs} \\
p_{21}^{\mu _{2}}t_{\mu _{123}}^{VVA} &=&t_{2\left( -\right) \mu _{13}}^{AV}
\\
p_{32}^{\mu _{3}}t_{\mu _{123}}^{VVA} &=&t_{3\left( -\right) \mu
_{12}}^{AV}+2mt_{\mu _{12}}^{VVP}
\end{eqnarray}%
\begin{eqnarray}
p_{31}^{\mu _{1}}t_{\mu _{123}}^{AAA} &=&t_{1\left( -\right) \mu
_{23}}^{AV}-2mt_{\mu _{23}}^{PAA}  \label{AAAragfs} \\
p_{21}^{\mu _{2}}t_{\mu _{123}}^{AAA} &=&t_{2\left( -\right) \mu
_{13}}^{AV}+2mt_{\mu _{13}}^{APA} \\
p_{32}^{\mu _{3}}t_{\mu _{123}}^{AAA} &=&t_{3\left( -\right) \mu
_{12}}^{AV}+2mt_{\mu _{12}}^{AAP}.
\end{eqnarray}

Let us introduce the structures emerging within the relations above. First,
axial contractions generate three-point functions that are finite tensors
depending on external momenta. This feature is transparent due to their
connection with finite Feynman integrals introduced in Appendix \ref%
{AppInt4D}, as highlighted by removing the overbar notation in $\bar{J}%
_{3}^{\nu _{1}}=J_{3}^{\nu _{1}}$ and $\bar{J}_{3}=J_{3}$. We have for
single-axial triangles%
\begin{eqnarray}
-2mT_{\mu _{23}}^{PVV} &=&\varepsilon _{\mu _{23}\nu _{12}}p_{21}^{\nu
_{1}}p_{32}^{\nu _{2}}(8im^{2}J_{3}),  \label{PVV} \\
2mT_{\mu _{13}}^{VPV} &=&\varepsilon _{\mu _{13}\nu _{12}}p_{21}^{\nu
_{1}}p_{32}^{\nu _{2}}(8im^{2}J_{3}), \\
2mT_{\mu _{12}}^{VVP} &=&\varepsilon _{\mu _{12}\nu _{12}}p_{21}^{\nu
_{1}}p_{32}^{\nu _{2}}(-8im^{2}J_{3}),
\end{eqnarray}%
while momenta contractions for the triple-axial triangle lead to%
\begin{eqnarray}
-2mT_{\mu _{23}}^{PAA} &=&\varepsilon _{\mu _{23}\nu _{12}}p_{31}^{\nu
_{2}}[8im^{2}(2J_{3}^{\nu _{1}}+p_{21}^{\nu _{1}}J_{3})], \\
2mT_{\mu _{13}}^{APA} &=&\varepsilon _{\mu _{13}\nu _{12}}p_{21}^{\nu
_{2}}[-8im^{2}(2J_{3}^{\nu _{1}}+p_{31}^{\nu _{1}}J_{3})], \\
2mT_{\mu _{12}}^{AAP} &=&\varepsilon _{\mu _{12}\nu _{12}}p_{32}^{\nu
_{2}}[8im^{2}(2J_{3}^{\nu _{1}}+p_{21}^{\nu _{1}}J_{3})].  \label{AAP}
\end{eqnarray}

In future subsections, we explore the connection of RAGFs with the
low-energy limits of these finite amplitudes. They depend on basic functions
observed within the scalar $J_{3}=i(4\pi )^{-2}Z_{00}^{\left( -1\right) }$
and the vector employed anteriorly (\ref{4DJ3mu}). Hence, one uses (\ref%
{LetZ4D}) to determine their kinematical behavior when all momenta bilinears
are zero: 
\begin{eqnarray}
\left. -2mT_{\mu _{23}}^{PVV}\right\vert _{0} &\rightarrow &\frac{1}{(2\pi
)^{2}};\quad \left. 2mT_{\mu _{13}}^{VPV}\right\vert _{0}\rightarrow \frac{1%
}{(2\pi )^{2}};\quad \left. 2mT_{\mu _{12}}^{VVP}\right\vert _{0}\rightarrow
-\frac{1}{(2\pi )^{2}};  \label{LEPVV} \\
\left. -2mT_{\mu _{23}}^{PAA}\right\vert _{0} &\rightarrow &\frac{1}{3(2\pi
)^{2}};\quad \left. 2mT_{\mu _{13}}^{APA}\right\vert _{0}\rightarrow \frac{1%
}{3(2\pi )^{2}};\quad \left. 2mT_{\mu _{12}}^{AAP}\right\vert
_{0}\rightarrow -\frac{1}{3(2\pi )^{2}}.  \label{LEPAA}
\end{eqnarray}%
Each term above is multiplied by the corresponding tensor $\varepsilon _{\mu
_{kl}\nu _{12}}p_{21}^{\nu _{1}}p_{32}^{\nu _{2}}$ with $k<l$.

Second, other structures appearing in RAGFs are $AV$ functions, which are
proportional to two-point vector integrals (\ref{J2bar4D}). As contributions
exclusively on the external momentum cancel out in the contraction, they are
pure surface terms proportional to arbitrary label combinations:%
\begin{equation}
T_{\mu _{ab}}^{AV}\left( i,j\right) =-4i\varepsilon _{\mu _{ab}\nu
_{12}}p_{ji}^{\nu _{2}}\bar{J}_{2}^{\nu _{1}}\left( i,j\right)
=2i\varepsilon _{\mu _{ab}\nu _{12}}p_{ji}^{\nu _{2}}P_{ji}^{\nu _{3}}\Delta
_{3\nu _{3}}^{\nu _{1}}.  \label{AV4D}
\end{equation}%
After replacing the adequate labels ($k_{i}$ and $k_{j}$), combinations seen
in the RAGFs above arise: 
\begin{eqnarray}
T_{1\left( -\right) \mu _{23}}^{AV} &=&2i\varepsilon _{\mu _{23}\nu
_{12}}\left( p_{12}^{\nu _{2}}P_{12}^{\nu _{3}}-p_{32}^{\nu _{2}}P_{32}^{\nu
_{3}}\right) \Delta _{3\nu _{3}}^{\nu _{1}},  \label{AV(-)1} \\
T_{2\left( -\right) \mu _{13}}^{AV} &=&2i\varepsilon _{\mu _{13}\nu
_{12}}\left( p_{31}^{\nu _{2}}P_{31}^{\nu _{3}}-p_{32}^{\nu _{2}}P_{32}^{\nu
_{3}}\right) \Delta _{3\nu _{3}}^{\nu _{1}},  \label{AV(-)2} \\
T_{3\left( -\right) \mu _{12}}^{AV} &=&2i\varepsilon _{\mu _{12}\nu
_{12}}\left( p_{21}^{\nu _{2}}P_{21}^{\nu _{3}}-p_{31}^{\nu _{2}}P_{31}^{\nu
_{3}}\right) \Delta _{3\nu _{3}}^{\nu _{1}}.  \label{AV(-)3}
\end{eqnarray}%
We stress that these forms do not depend on the specific \textit{axial}
amplitude. The numerical subindex in $T_{i\left( -\right) }^{AV}$ indicates
that this is the structure characteristic of the $i$-th contraction.

Next, we must contract external momenta with the integrated amplitudes to
verify RAGFs. Observe the first $AVV$ version (\ref{AVV1complete}) to
anticipate operations involving finite contributions. Some terms vanish due
to symmetry properties in the contraction; then, we manipulate the remaining
ones using tools developed in Appendix \ref{AppInt4D}. The procedure reduces 
$J$-tensors to identify finite 2nd-order amplitudes or achieve
cancellations. These reductions are well-defined relations involving finite
tensors 
\begin{eqnarray}
2p_{21}^{\alpha }J_{3\alpha }^{\nu } &=&-p_{21}^{2}J_{3}^{\nu }+J_{2}^{\nu
}\left( p_{31}\right) +J_{2}^{\nu }\left( p_{32}\right) +p_{31}^{\nu
}J_{2}\left( p_{32}\right) , \\
2p_{31}^{\alpha }J_{3\alpha }^{\nu } &=&-p_{31}^{2}J_{3}^{\nu }+J_{2}^{\nu
}\left( p_{21}\right) +J_{2}^{\nu }\left( p_{32}\right) +p_{31}^{\nu
}J_{2}\left( p_{32}\right) , \\
2J_{3\nu }^{\nu } &=&2m^{2}J_{3}+2J_{2}\left( p_{32}\right) +i\left( 4\pi
\right) ^{-2},  \label{J3uu}
\end{eqnarray}%
and vectors%
\begin{eqnarray}
2p_{21}^{\nu }J_{3\nu } &=&-p_{21}^{2}J_{3}+J_{2}\left( p_{31}\right)
-J_{2}\left( p_{32}\right) ,  \label{pJ} \\
2p_{31}^{\nu }J_{3\nu } &=&-p_{31}^{2}J_{3}+J_{2}\left( p_{21}\right)
-J_{2}\left( p_{32}\right) .  \label{qJ}
\end{eqnarray}

Although some reductions arise directly, other occurrences require further
algebraic manipulations. This circumstance manifests when one $J$-tensor
couples to the Levi-Civita symbol and rearranging indexes is necessary to
find momenta contractions. For vector integrals, we consider the identity $%
\varepsilon _{\lbrack \mu _{ab}\nu _{12}}p_{\nu _{3}]}J_{3}^{\nu _{1}}=0$ to
achieve the formula%
\begin{equation}
2\varepsilon _{\mu _{ab}\nu _{12}}\left[ p_{21}^{\nu _{2}}\left( p_{ij}\cdot
p_{31}\right) -p_{31}^{\nu _{2}}\left( p_{ij}\cdot p_{21}\right) \right]
J_{3}^{\nu _{1}}=-\varepsilon _{\mu _{ab}\nu _{23}}p_{21}^{\nu
_{2}}p_{31}^{\nu _{3}}\left[ 2p_{ij}^{\nu _{1}}J_{3\nu _{1}}\right] .
\label{SCH-2}
\end{equation}%
Similarly, we use $\varepsilon _{\lbrack \mu _{a}\nu _{123}}J_{3\mu
_{c}]}^{\nu _{1}}=0$ to reorganize terms involving the tensor integral 
\begin{eqnarray}
&&2\varepsilon _{\mu _{b}\nu _{123}}p_{21}^{\nu _{2}}p_{31}^{\nu
_{3}}J_{3\mu _{a}}^{\nu _{1}}-2\varepsilon _{\mu _{a}\nu _{123}}p_{21}^{\nu
_{2}}p_{31}^{\nu _{3}}J_{3\mu _{b}}^{\nu _{1}}  \notag \\
&=&\varepsilon _{\mu _{ab}\nu _{13}}p_{31}^{\nu _{3}}\left[ 2p_{21}^{\nu
_{2}}J_{3\nu _{2}}^{\nu _{1}}\right] -\varepsilon _{\mu _{ab}\nu
_{12}}p_{21}^{\nu _{2}}\left[ 2p_{31}^{\nu _{3}}J_{3\nu _{3}}^{\nu _{1}}%
\right] -\varepsilon _{\mu _{ab}\nu _{23}}p_{21}^{\nu _{2}}p_{31}^{\nu _{3}}%
\left[ 2J_{3\nu }^{\nu }\right] .  \label{SCH-3}
\end{eqnarray}

\textit{Axial} amplitudes have two structures: common tensors associated
with the version (\ref{C1})-(\ref{C3}) and subamplitudes. Starting with the
finite part of the tensor sector, let us explore the first version to
illustrate operations necessary for momenta contractions:%
\begin{equation}
C_{1\mu _{123}}^{\mathrm{finite}}=-2\varepsilon _{\mu _{13}\nu
_{12}}[p_{31}^{\nu _{2}}(J_{3\mu _{2}}^{\nu _{1}}+p_{21\mu _{2}}J_{3}^{\nu
_{1}})-p_{21}^{\nu _{2}}(J_{3\mu _{2}}^{\nu _{1}}+p_{31\mu _{2}}J_{3}^{\nu
_{1}})]-2\varepsilon _{\mu _{12}\nu _{12}}p_{21}^{\nu _{2}}(J_{3\mu
_{3}}^{\nu _{1}}+p_{31\mu _{3}}J_{3}^{\nu _{1}}).
\end{equation}%
The first parenthesis and $J$-vector contributions cancel out due to
symmetry properties; thus, the contraction with $p_{31}^{\mu _{1}}$ yields%
\begin{equation}
p_{31}^{\mu _{1}}C_{1\mu _{123}}^{\mathrm{finite}}=-2p_{21}^{\nu
_{2}}p_{31}^{\nu _{3}}(\varepsilon _{\mu _{3}\nu _{123}}J_{3\mu _{2}}^{\nu
_{1}}-\varepsilon _{\mu _{2}\nu _{123}}J_{3\mu _{3}}^{\nu _{1}}).
\end{equation}%
Since external momenta contract with the Levi-Civita symbol and not with $J$%
-integrals, one must permute indexes through the identity above (\ref{SCH-3}%
) to allow reductions of finite functions. This rearrangement implies the
presence of the trace $J_{3\nu }^{\nu }$ (\ref{J3uu}) and brings two
additional contributions: one proportional to squared mass and a numeric
factor. That differs from other contractions since their reductions are
immediate, only requiring the identification (\ref{SCH-2}). We extend this
analysis to other versions and cast all possible tensor contractions below,
stressing that additional terms accompany the contraction with the
version-defining index (in squared brackets). 
\begin{eqnarray}
p_{31}^{\mu _{1}}C_{1\mu _{123}}^{\mathrm{finite}} &=&\varepsilon _{\mu
_{23}\nu _{12}}\{\left( p_{31}^{\nu _{2}}p_{21}^{2}-p_{21}^{\nu
_{2}}p_{31}^{2}\right) J_{3}^{\nu _{1}}+p_{21}^{\nu _{1}}p_{31}^{\nu
_{2}}[2m^{2}J_{3}+J_{2}\left( p_{32}\right) +i(4\pi )^{-2}]\}  \label{cont1}
\\
p_{21}^{\mu _{2}}C_{1\mu _{123}}^{\mathrm{finite}} &=&\frac{1}{2}\varepsilon
_{\mu _{13}\nu _{12}}p_{32}^{\nu _{2}}\left\{ 2p_{21}^{2}\left( J_{3}^{\nu
_{1}}+p_{21}^{\nu _{1}}J_{3}\right) -p_{21}^{\nu _{1}}J_{2}\left(
p_{31}\right) \right\}  \label{cont2} \\
p_{32}^{\mu _{3}}C_{1\mu _{123}}^{\mathrm{finite}} &=&\frac{1}{2}\varepsilon
_{\mu _{12}\nu _{12}}p_{21}^{\nu _{2}}\left\{ -2p_{32}^{2}J_{3}^{\nu
_{1}}-p_{31}^{\nu _{1}}J_{2}\left( p_{31}\right) \right\}  \label{cont3}
\end{eqnarray}%
\begin{eqnarray}
p_{31}^{\mu _{1}}C_{2\mu _{123}}^{\mathrm{finite}} &=&\frac{1}{2}\varepsilon
_{\mu _{23}\nu _{12}}p_{32}^{\nu _{2}}\left\{ 2p_{31}^{2}\left( J_{3}^{\nu
_{1}}+p_{21}^{\nu _{1}}J_{3}\right) -p_{21}^{\nu _{1}}J_{2}\left(
p_{21}\right) \right\} \\
p_{21}^{\mu _{2}}C_{2\mu _{123}}^{\mathrm{finite}} &=&\varepsilon _{\mu
_{13}\nu _{12}}\{(p_{31}^{\nu _{2}}p_{21}^{2}-p_{21}^{\nu
_{2}}p_{31}^{2})J_{3}^{\nu _{1}}+p_{21}^{\nu _{1}}p_{31}^{\nu
_{2}}[2m^{2}J_{3}+J_{2}(p_{32})+i(4\pi )^{-2}]\} \\
p_{32}^{\mu _{3}}C_{2\mu _{123}}^{\mathrm{finite}} &=&\frac{1}{2}\varepsilon
_{\mu _{12}\nu _{12}}\left\{ 2p_{31}^{\nu _{2}}p_{32}^{2}J_{3}^{\nu
_{1}}+p_{21}^{\nu _{1}}p_{31}^{\nu _{2}}J_{2}\left( p_{21}\right) \right\}
\end{eqnarray}%
\begin{eqnarray}
p_{31}^{\mu _{1}}C_{3\mu _{123}}^{\mathrm{finite}} &=&\frac{1}{2}\varepsilon
_{\mu _{23}\nu _{12}}p_{21}^{\nu _{2}}\left\{ 2p_{31}^{2}J_{3}^{\nu
_{1}}+p_{31}^{\nu _{1}}J_{2}(p_{32})\right\} \\
p_{21}^{\mu _{2}}C_{3\mu _{123}}^{\mathrm{finite}} &=&\frac{1}{2}\varepsilon
_{\mu _{13}\nu _{12}}p_{31}^{\nu _{2}}\left\{ -2p_{21}^{2}J_{3}^{\nu
_{1}}-p_{21}^{\nu _{1}}J_{2}(p_{32})\right\} \\
p_{32}^{\mu _{3}}C_{3\mu _{123}}^{\mathrm{finite}} &=&\varepsilon _{\mu
_{12}\nu _{12}}\{\left( p_{21}^{\nu _{2}}p_{31}^{2}-p_{31}^{\nu
_{2}}p_{21}^{2}\right) J_{3}^{\nu _{1}}-p_{21}^{\nu _{1}}p_{31}^{\nu
_{2}}[2m^{2}J_{3}+J_{2}(p_{32})+i(4\pi )^{-2}]\}.  \label{cont9}
\end{eqnarray}

We have to sum contributions from subamplitudes to complete the finite
sector. That requires the same resources discussed above, but only vector
integrals remain, and again we identify (\ref{SCH-2}) to reduce them to
scalars. Terms proportional to the squared mass might arise from common
tensors and subamplitudes. They cancel out in vector contractions and
combine into the expected finite amplitudes (\ref{PVV})-(\ref{AAP}) in axial
contractions. That agrees with original expectations for momenta
contractions of all \textit{axial} amplitudes; however, we acknowledge one
additional numeric factor $i\left( 4\pi \right) ^{-2}$ when the contracted
index $\mu _{i}$ matches the $i$-th version.

To complete the analysis of RAGFs, we must perform momenta contractions over
divergent structures to identify differences between $AV$ amplitudes. Even
though different subamplitudes were identified, we showed that the divergent
sector is characteristic of the version (\ref{ST1})-(\ref{ST3}). They are
pure surface terms $S_{i}$ with the index $\mu _{i}$ appearing exclusively
within the Levi-Civita tensor and not in the actual surface term $\Delta
_{3} $. For all triangle amplitudes, identifications are automatic whenever
contractions consider another index $\mu _{j}$ with $i\neq j$. Nevertheless,
using the version-defining index ($i=j$) does not produce momenta
contractions with surface terms required for these identifications. Thus, in
parallel to the procedure for 2nd-order $J$-tensors, indexes are reorganized
through the identity%
\begin{equation}
\varepsilon _{\mu _{13}\nu _{12}}\Delta _{3\mu _{2}}^{\nu _{1}}-\varepsilon
_{\mu _{12}\nu _{12}}\Delta _{3\mu _{3}}^{\nu _{1}}=\varepsilon _{\mu
_{23}\nu _{12}}\Delta _{3\mu _{1}}^{\nu _{1}}+\varepsilon _{\mu _{123}\nu
_{1}}\Delta _{3\nu _{2}}^{\nu _{1}}-\varepsilon _{\mu _{123}\nu _{2}}\Delta
_{3\nu _{1}}^{\nu _{1}}.
\end{equation}

Again, let us approach the first version to exemplify. While relations from
indexes $\mu _{2}$ and $\mu _{3}$ are automatic, contracting $\mu _{1}$
demands the permutation introduced above. These operations yield (\ref%
{contS1}) after organizing momenta through $p_{ij}=P_{ir}-P_{jr}$. Besides
the expected contributions (\ref{AV(-)1})-(\ref{AV(-)3}), note the presence
of one additional term on the trace $\Delta _{3\alpha }^{\alpha }$
resembling what occurred for the finite part (in squared brackets). We cast
results for all versions in the sequence, so understanding this pattern is
possible. 
\begin{eqnarray}
p_{31}^{\mu _{1}}S_{1\mu _{123}} &=&2i\varepsilon _{\mu _{23}\nu
_{12}}\left( p_{12}^{\nu _{2}}P_{12}^{\nu _{3}}-p_{32}^{\nu _{2}}P_{32}^{\nu
_{3}}\right) \Delta _{3\nu _{3}}^{\nu _{1}}+\left[ 2i\varepsilon _{\mu
_{23}\nu _{23}}p_{21}^{\nu _{2}}p_{31}^{\nu _{3}}\Delta _{3\alpha }^{\alpha }%
\right]  \label{contS1} \\
p_{21}^{\mu _{2}}S_{1\mu _{123}} &=&2i\varepsilon _{\mu _{13}\nu
_{12}}\left( p_{31}^{\nu _{2}}P_{31}^{\nu _{3}}-p_{32}^{\nu _{2}}P_{32}^{\nu
_{3}}\right) \Delta _{3\nu _{3}}^{\nu _{1}} \\
p_{32}^{\mu _{3}}S_{1\mu _{123}} &=&2i\varepsilon _{\mu _{12}\nu
_{12}}\left( p_{21}^{\nu _{2}}P_{21}^{\nu _{3}}-p_{31}^{\nu _{2}}P_{31}^{\nu
_{3}}\right) \Delta _{3\nu _{3}}^{\nu _{1}}
\end{eqnarray}%
\begin{eqnarray}
p_{31}^{\mu _{1}}S_{2\mu _{123}} &=&2i\varepsilon _{\mu _{23}\nu
_{12}}\left( p_{12}^{\nu _{2}}P_{12}^{\nu _{3}}-p_{32}^{\nu _{2}}P_{32}^{\nu
_{3}}\right) \Delta _{3\nu _{3}}^{\nu _{1}} \\
p_{21}^{\mu _{2}}S_{2\mu _{123}} &=&2i\varepsilon _{\mu _{13}\nu
_{12}}\left( p_{31}^{\nu _{2}}P_{31}^{\nu _{3}}-p_{32}^{\nu _{2}}P_{32}^{\nu
_{3}}\right) \Delta _{3\nu _{3}}^{\nu _{1}}+\left[ 2i\varepsilon _{\mu
_{13}\nu _{23}}p_{21}^{\nu _{2}}p_{31}^{\nu _{3}}\Delta _{3\alpha }^{\alpha }%
\right]  \label{contS2} \\
p_{32}^{\mu _{3}}S_{2\mu _{123}} &=&2i\varepsilon _{\mu _{12}\nu
_{12}}\left( p_{21}^{\nu _{2}}P_{21}^{\nu _{3}}-p_{31}^{\nu _{2}}P_{31}^{\nu
_{3}}\right) \Delta _{3\nu _{3}}^{\nu _{1}}
\end{eqnarray}%
\begin{eqnarray}
p_{31}^{\mu _{1}}S_{3\mu _{123}} &=&2i\varepsilon _{\mu _{23}\nu
_{12}}\left( p_{12}^{\nu _{2}}P_{12}^{\nu _{3}}-p_{32}^{\nu _{2}}P_{32}^{\nu
_{3}}\right) \Delta _{3\nu _{3}}^{\nu _{1}} \\
p_{21}^{\mu _{2}}S_{3\mu _{123}} &=&2i\varepsilon _{\mu _{13}\nu
_{12}}\left( p_{31}^{\nu _{2}}P_{31}^{\nu _{3}}-p_{32}^{\nu _{2}}P_{32}^{\nu
_{3}}\right) \Delta _{3\nu _{3}}^{\nu _{1}} \\
p_{32}^{\mu _{3}}S_{3\mu _{123}} &=&2i\varepsilon _{\mu _{12}\nu
_{12}}\left( p_{21}^{\nu _{2}}P_{21}^{\nu _{3}}-p_{31}^{\nu _{2}}P_{31}^{\nu
_{3}}\right) \Delta _{3\nu _{3}}^{\nu _{1}}-\left[ 2i\varepsilon _{\mu
_{12}\nu _{23}}p_{21}^{\nu _{2}}p_{31}^{\nu _{3}}\Delta _{3\alpha }^{\alpha }%
\right]  \label{contS9}
\end{eqnarray}

With these properties known, let us explore the first $AVV$ version (\ref%
{AVV1}) to clarify the analysis of RAGFs. Using the contraction of the
subamplitude (\ref{TVPP})%
\begin{equation}
i\varepsilon _{\mu _{123}\nu }p_{31}^{\mu _{1}}(T_{\mathrm{finite}%
}^{VPP})^{\nu }=2i\varepsilon _{\mu _{23}\nu _{12}}p_{31}^{\nu _{2}}\left\{
2(p_{32}\cdot p_{21})J_{3}^{\nu _{1}}+p_{21}^{\nu
_{1}}[p_{31}^{2}J_{3}-J_{2}(p_{32})-J_{2}(p_{21})]\right\}
\end{equation}%
and of the common tensor $C_{1}$ (\ref{cont1}), we write the axial
contraction as follows: 
\begin{eqnarray}
p_{31}^{\mu _{1}}(T_{\mu _{123}}^{AVV})_{1} &=&p_{31}^{\mu _{1}}S_{1\mu
_{123}}+4i\varepsilon _{\mu _{23}\nu _{12}}p_{21}^{\nu _{1}}p_{31}^{\nu
_{2}}[2m^{2}J_{3}+i\left( 4\pi \right) ^{-2}]  \notag \\
&&-4i\varepsilon _{\mu _{23}\nu _{12}}\left[ p_{21}^{\nu
_{2}}p_{31}^{2}-p_{31}^{\nu _{2}}\left( p_{21}\cdot p_{31}\right) \right]
J_{3}^{\nu _{1}}  \notag \\
&&+2i\varepsilon _{\mu _{23}\nu _{12}}p_{21}^{\nu _{1}}p_{31}^{\nu
_{2}}[p_{31}^{2}J_{3}+J_{2}\left( p_{32}\right) -J_{2}\left( p_{21}\right) ].
\end{eqnarray}%
This organization emphasizes the second row as a variation of relation (\ref%
{SCH-2}), leading to reduction (\ref{qJ}) and ultimately canceling the third
row 
\begin{equation}
p_{31}^{\mu _{1}}(T_{\mu _{123}}^{AVV})_{1}=p_{31}^{\mu _{1}}S_{1\mu
_{123}}+4i\varepsilon _{\mu _{23}\nu _{12}}p_{21}^{\nu _{1}}p_{31}^{\nu
_{2}}[2m^{2}J_{3}+i\left( 4\pi \right) ^{-2}].
\end{equation}

At the end of this process, one identifies terms on the squared mass as the
finite amplitude $PVV$ (\ref{PVV}). On the other hand, it is direct to
identify the $AV$s from\ the contraction of the pure surface term $S_{1}$ (%
\ref{contS1}):%
\begin{equation}
p_{31}^{\mu _{1}}(T_{\mu _{123}}^{AVV})_{1}=T_{1\left( -\right) \mu
_{23}}^{AV}-2mT_{\mu _{23}}^{PVV}+2i\varepsilon _{\mu _{23}\nu
_{12}}p_{21}^{\nu _{1}}p_{31}^{\nu _{2}}[\Delta _{3\alpha }^{\alpha
}+2i\left( 4\pi \right) ^{-2}].
\end{equation}%
Terms in squared brackets appeared as a consequence of permutations within
2nd-order tensors ($J_{3}$ and $\Delta _{3}$), necessary when contracting
the version-defining index. Differently, vector RAGFs automatically apply
because they do not exhibit this feature. That occurs after reducing the
entire finite sector and identifying $AV$ amplitudes: 
\begin{eqnarray}
p_{21}^{\mu _{2}}(T_{\mu _{123}}^{AVV})_{1} &=&p_{21}^{\mu _{2}}S_{1\mu
_{123}}=T_{2\left( -\right) \mu _{13}}^{AV}, \\
p_{32}^{\mu _{3}}(T_{\mu _{123}}^{AVV})_{1} &=&p_{32}^{\mu _{3}}S_{1\mu
_{123}}=T_{3\left( -\right) \mu _{12}}^{AV}.
\end{eqnarray}

This pattern repeats for the first version of all \textit{axial} amplitudes (%
$AVV$, $VAV$, $VVA$, $AAA$). Whereas the first contraction exhibits the
additional term, other RAGFs are satisfied without conditions. The pattern
changes to the second and third versions, which show potentially-violating
terms in the version-defining index independently of the vertex nature as
axial or vector.

\begin{figure}[tbph]
\includegraphics[scale=0.8]{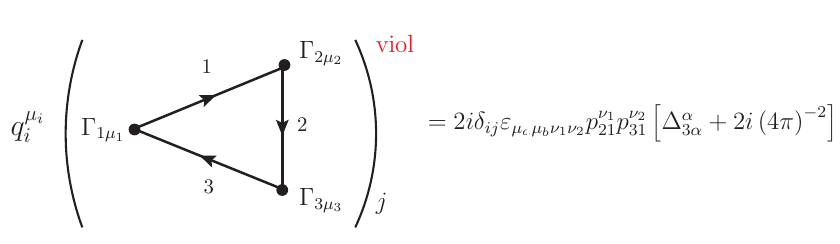}
\caption{The violation factor of the RAGF established for the contraction
with momenta $q_{i}^{\protect\mu _{1}}.$}
\label{diagviol}
\end{figure}

Following the developed steps, the equations below subsume all
potentially-offending terms emerging in contractions where the version is
defined%
\begin{equation}
\left\{ 
\begin{array}{c}
q_{1}^{\mu _{1}}(T_{\mu _{123}}^{\Gamma _{123}})_{1}^{\mathrm{viol}%
}=+2i\varepsilon _{\mu _{23}\nu _{12}}q_{2}^{\nu _{1}}q_{3}^{\nu
_{2}}[\Delta _{3\alpha }^{\alpha }+2i\left( 4\pi \right) ^{-2}] \\ 
q_{2}^{\mu _{2}}(T_{\mu _{123}}^{\Gamma _{123}})_{2}^{\mathrm{viol}%
}=+2i\varepsilon _{\mu _{13}\nu _{12}}q_{2}^{\nu _{1}}q_{3}^{\nu
_{2}}[\Delta _{3\alpha }^{\alpha }+2i\left( 4\pi \right) ^{-2}] \\ 
q_{3}^{\mu _{3}}(T_{\mu _{123}}^{\Gamma _{123}})_{3}^{\mathrm{viol}%
}=-2i\varepsilon _{\mu _{12}\nu _{12}}q_{2}^{\nu _{1}}q_{3}^{\nu
_{2}}[\Delta _{3\alpha }^{\alpha }+2i\left( 4\pi \right) ^{-2}].%
\end{array}%
\right.  \label{ragfViol4D}
\end{equation}%
We adopt the notation introduced in Figure \ref{diag1} to the routing
differences $q_{1}=p_{31}$, $q_{2}=p_{21}$, and $q_{3}=p_{32}$ to mark a
convention for first, second, and third vertices. In addition, the symbol $%
\Gamma _{123}\equiv \Gamma _{1}\Gamma _{2}\Gamma _{3}$ is an abbreviation
for all vertex configurations of \textit{axial} amplitudes.\ We consider
these relations nonautomatic since they depend on the value attributed to
surface terms, meaning they only apply under the constraint%
\begin{equation}
\Delta _{3\alpha }^{\alpha }=-2i\left( 4\pi \right) ^{-2}.  \label{Daa}
\end{equation}%
We offer the schematic graph in Figure \ref{diagviol} to visualize this
violation pattern. Other vertices (to each version) have their RAGFs
identically satisfied. 

From another perspective, all Ward identities would be valid by making
surface terms null if RAGFs apply identically (this works channel by
channel). Nevertheless, this outcome requires conflicting interpretations of
surface terms: zero for the momentum-space translational invariance and
nonzero for the linearity of integration. Thence, these properties do not
hold simultaneously. General tensor properties and the low-energy behavior
of finite amplitudes show these conclusions are inescapable in Subsection (%
\ref{LED4DSTS}). That is independent of any conceivable trace.

At this point, we explore differences among amplitude versions to understand
why the acknowledged results depend on the version-defining index.
Integrands of investigated versions are well-defined identical tensors.
However, after integration, the sampling of indexes makes finite parts and
surface terms different. We highlight differences among the three main
versions to elucidate this point:%
\begin{eqnarray}
(T_{\mu _{123}}^{\Gamma _{123}})_{1}-(T_{\mu _{123}}^{\Gamma _{123}})_{2}
&=&+2i\varepsilon _{\mu _{123}\nu }q_{3}^{\nu }[\Delta _{3\alpha }^{\alpha
}+2i\left( 4\pi \right) ^{-2}],  \label{v12} \\
(T_{\mu _{123}}^{\Gamma _{123}})_{1}-(T_{\mu _{123}}^{\Gamma _{123}})_{3}
&=&-2i\varepsilon _{\mu _{123}\nu }q_{2}^{\nu }[\Delta _{3\alpha }^{\alpha
}+2i\left( 4\pi \right) ^{-2}], \\
(T_{\mu _{123}}^{\Gamma _{123}})_{2}-(T_{\mu _{123}}^{\Gamma _{123}})_{3}
&=&-2i\varepsilon _{\mu _{123}\nu }q_{1}^{\nu }[\Delta _{3\alpha }^{\alpha
}+2i\left( 4\pi \right) ^{-2}].  \label{v23}
\end{eqnarray}%
After subtracting two versions, we reorganized indexes to identify
reductions of finite functions and recognize the same potentially-violating
term acknowledged in (\ref{ragfViol4D}).

Now, let us define the meaning of uniqueness adopted within this
investigation: any possible form to compute the same expression returns the
same result. This notion implies that an amplitude does not depend on Dirac
traces. Canceling the RHS of the equations above would be required to
achieve this property, which only happens when adopting the prescription $%
\Delta _{3\alpha }^{\alpha }=-2i\left( 4\pi \right) ^{-2}$. Meanwhile,
unlike in the two-dimensional context, the nonzero surface terms required by
this notion allow dependence on ambiguous combinations of arbitrary internal
momenta. In this sense, setting specific values for external momenta is
possible.

The trace of six matrices is the unique place\ where the amplitude versions
differ. Achieving traces different from those starting this argumentation is
possible through other identities involving the chiral matrix, Eq. (\ref%
{Chiral-Id}). Nonetheless, as detailed in Appendix \ref{Tr6G4D}, all
versions are linear combinations of those previously studied. That justifies
taking $(T_{\mu _{123}}^{\Gamma _{123}})_{i}$ as the basic versions;
moreover, \textit{they have the maximum number of RAGFs identically satisfied%
}, see Subsection (\ref{LED4DSTS}). Hence, we define a general form that
reproduces any accessible expression with the building-block versions 
\begin{equation}
\lbrack T_{\mu _{123}}^{\Gamma _{123}}]_{\left\{ r_{1}r_{2}r_{3}\right\} }=%
\frac{1}{r_{1}+r_{2}+r_{3}}\sum_{i=1}^{3}r_{i}(T_{\mu _{123}}^{\Gamma
_{123}})_{i},  \label{r123}
\end{equation}%
with weights $r_{1}+r_{2}+r_{3}\not=0$. It compiles all involved
arbitrariness, accounting for any choices regarding Dirac traces. From this
formula, assuming surface terms as zero after the integration, we identify
an infinity set of amplitudes that violate RAGFs by arbitrary amounts. That
allows obtaining different violation values found in the literature, e.g., 
\cite{Wu2006}.

We have shown how traces and surface terms interfere with linearity of
integration and uniqueness of the investigated tensors. In the subsequent
subsections, we demonstrate these properties are unavoidable since
conditions for RAGFs arise without explicit computations of the primary
amplitudes.

\subsection{Low-Energy Theorems I \label{LE4D}}

This subsection proposes a structure depending only on external momenta to
formulate a low-energy implication for a tensor representing three-point
amplitudes. That does not mean we ignore the possible presence of ambiguous
routing combinations because they can be transformed into linear covariant
combinations of physical momenta. The explored structure is a general
3rd-order tensor having odd parity:%
\begin{eqnarray}
F_{\mu _{123}} &=&\varepsilon _{\mu _{123}\nu }(q_{2}^{\nu }F_{1}+q_{3}^{\nu
}F_{2})+\varepsilon _{\mu _{12}\nu _{12}}q_{2}^{\nu _{1}}q_{3}^{\nu
_{2}}(q_{2\mu _{3}}G_{1}+q_{3\mu _{3}}G_{2})  \notag \\
&&+\varepsilon _{\mu _{13}\nu _{12}}q_{2}^{\nu _{1}}q_{3}^{\nu _{2}}\left(
q_{2\mu _{2}}G_{3}+q_{3\mu _{2}}G_{4}\right) +\varepsilon _{\mu _{23}\nu
_{12}}q_{2}^{\nu _{1}}q_{3}^{\nu _{2}}(q_{2\mu _{1}}G_{5}+q_{3\mu
_{1}}G_{6}).  \label{GForm}
\end{eqnarray}%
That is a function of two variables; namely, the incoming external momenta $%
q_{2}$ and $q_{3}$ associated with vertices $\Gamma _{2}$ and $\Gamma _{3}$
(following Figure \ref{diag1}). Conservation sets the relation $%
q_{1}=q_{2}+q_{3}$ with the outcoming momentum of the vertex $\Gamma _{1}$.

After performing momenta contractions, one identifies the arrangements $%
q_{i}^{\mu _{i}}F_{\mu _{123}}=\varepsilon _{\mu _{kl}\nu _{12}}q_{2}^{\nu
_{1}}q_{3}^{\nu _{2}}V_{i}$ with $k<l\not=i$. These operations lead to a set
of three functions written in terms of form factors belonging to the general
tensor%
\begin{equation}
\left\{ 
\begin{array}{l}
V_{1}=-F_{1}+F_{2}+\left( q_{1}\cdot q_{2}\right) G_{5}+\left( q_{1}\cdot
q_{3}\right) G_{6}, \\ 
V_{2}=-F_{2}+q_{2}^{2}G_{3}+\left( q_{2}\cdot q_{3}\right) G_{4}, \\ 
V_{3}=-F_{1}+q_{3}^{2}G_{2}+\left( q_{2}\cdot q_{3}\right) G_{1}.%
\end{array}%
\right.  \label{Vis}
\end{equation}%
Without any hypothesis about eventual symmetries nor restrictions over the
value of any of the quantities above, we construct an identity as follows%
\begin{equation}
V_{1}+V_{2}-V_{3}=q_{2}^{2}\left( G_{3}+G_{5}\right) +q_{3}^{2}\left(
G_{6}-G_{2}\right) +\left( q_{2}\cdot q_{3}\right) \left(
G_{4}+G_{5}+G_{6}-G_{1}\right) .
\end{equation}%
At the kinematical point where all bilinears are zero $q_{i}\cdot q_{j}=0$,
if the invariants do not have poles at this points, we derive a \textit{%
structural identity} among invariants%
\begin{equation}
V_{1}\left( 0\right) +V_{2}\left( 0\right) -V_{3}\left( 0\right) =0.
\label{VertsZero}
\end{equation}%
This relation contains information about symmetries or their violations at
the zero limit, even if no particular symmetry is needed for its deduction.
That occurs because it represents a constraint over three-point structures
arising on the RHS of proposed Ward identities (WIs).

Let us suppose that the $AVV$ axial contraction connects to\textbf{\ }the
amplitude coming from the pseudoscalar density to illustrate this resource: 
\begin{equation}
\varepsilon _{\mu _{23}\nu _{12}}q_{2}^{\nu _{1}}q_{3}^{\nu _{2}}V_{1}\left(
0\right) =-2mT_{\mu _{23}}^{PVV}\left( 0\right) =:\varepsilon _{\mu _{23}\nu
_{12}}q_{2}^{\nu _{1}}q_{3}^{\nu _{2}}\Omega _{1}^{PVV}\left( 0\right) ,
\end{equation}%
with the behavior (\ref{LEPVV}) leading to the value for the first invariant 
$V_{1}\left( 0\right) =(2\pi )^{-2}$. Since the constraint above prevents
the simultaneous vanishing of both other invariants $V_{2}\left( 0\right)
=V_{3}\left( 0\right) =0$, at least one vector WI is violated. On the other
hand, supposing that both vector WIs apply implies breaking the axial one.
That occurs because parameters defining the considered tensor and regularity
require the existence of an additional term $V_{1}\left( 0\right) =(2\pi
)^{-2}+\mathcal{A}$, the anomaly. Thus, $\mathcal{A}=-\Omega
_{1}^{PVV}\left( 0\right) $, relating a property of one finite amplitude and
the symmetry content of a rank-3 amplitude. Satisfying the symmetry at this
kinematical point does not guarantee invariance for all points; however, its
violation at zero implies symmetry violation.

That is the starting point of the violation pattern in anomalous amplitudes.
Numerical values presented above for invariants $V_{i}$ at zero represent
the preservation of corresponding WIs. Nevertheless, their simultaneous
occurrence implies a violation of the linear-algebra type solution given by
the \textit{structural identity} (\ref{VertsZero}). No tensor, independent
of its origin, can connect to the $PVV$ and have vanishing contractions
simultaneously with both momenta $q_{2}$ and $q_{3}$. Whenever an
axial-vertex contraction links to an amplitude coming from the pseudoscalar
density, there will be an anomaly in at least one vertex; the same
conclusion stands for other diagrams. These facts are known; however, the
form we raise is general.\textbf{\ }The low-energy theorem invoking vector
WIs is only one of the solutions, as in Section (4.2) of \cite{Treiman1985}.
The \textit{structural identity} is an exclusive and inviolable consequence
of properties assumed to the 3rd-order tensor\textbf{,} and symmetry
violations occur when the RHS terms of WIs do not behave accordingly.

The explicit computation of\textbf{\ }perturbative expressions corroborates
these assertions. Moreover, the RAGFs connect ultraviolet and infrared
features of amplitudes, namely $\Omega _{1}^{PVV}\left( 0\right) =2i\Delta
_{3\alpha }^{\alpha }$\label{IRUV}. That is the requirement for linearity
seen after evaluating the RAGFs, and it will be derived in the next
subsection.\textbf{\ }There\textbf{, }we assume the form $V_{i}=\Omega _{i}+%
\mathcal{A}_{i}$ and demonstrate\textbf{\ }the implication 
\begin{equation}
\Omega _{1}\left( 0\right) +\Omega _{2}\left( 0\right) -\Omega _{3}\left(
0\right) =(2\pi )^{-2},  \label{GamatZero}
\end{equation}%
where we suppress upper indexes in $\Omega _{i}$ coming from finite
functions (e.g., $PVV$-$PAA$), see (\ref{GR}). This equation holds to
classically non-conserved vector currents or amplitudes with three arbitrary
masses running in the loop. Although multiple-mass amplitudes are
complicated functions of these masses, the relation at the point zero is
ever the finite constant above.

Independently of divergent aspects, the last equation alone is incompatible
with the \textit{structural identity} (\ref{VertsZero}), characterizing
violations for rank-3 triangles under the form (\ref{GForm}). Hence,
anomalous terms from different vertices $\mathcal{A}_{i}$ obey the general
constraint%
\begin{equation}
\mathcal{A}_{1}+\mathcal{A}_{2}-\mathcal{A}_{3}=-(2\pi )^{-2},  \label{Ans}
\end{equation}%
This equation shows that by preserving two vector WIs (in $AVV$), the value
of its axial anomaly is unique. Likewise, any explicit tensor\footnote{%
This tensor can be obtained via regularization or not. See the approach of
G. Scharf (\cite{Scharf2010}) in Section 5.1, using causal perturbation
theory. The analogous to $PVV$ is not computed until the very end. Instead,
the authors study analogous differences between the contraction of $AVV$ and
the $PVV$ without Feynman diagrams.} having WIs violated by any quantity
obeys this equation if the $\mathcal{A}_{i}$ relate to finite amplitudes
coming from Feynman rules. The crossed channel of finite amplitudes only
brings a multiplicative factor of 2 in the last couple of equations.

It is possible to anticipate restrictions over surface terms based on the
general dependence that 3rd-order tensors have on such terms and preserving
the independence and arbitrariness of internal momenta sums. That is
achieved through the connection with $AV$\ functions via integration
linearity\textbf{.} In the next section, this reasoning leads to the
proposition $\Omega _{1}^{PVV}\left( 0\right) =2i\Delta _{3\alpha }^{\alpha
} $ and Eq. (\ref{GamatZero}).

\subsection{Low-Energy Theorems II \label{LED4DSTS}}

In Subsection (\ref{unique}), we performed explicit calculations related to
different amplitude versions. Without manipulating the integral expression
of the surface term, an additional term connecting it with a finite
contribution emerged in momenta contractions (\ref{ragfViol4D}). This
property implies that the Relation Among Green Functions (RAGF) from the
version-defining index is not automatic, bringing violating terms to the
corresponding Ward Identity (WI). Meanwhile, the previous subsection
established a low-energy implication from a general tensor of the external
momenta (\ref{GForm}). From this outset, a \textit{structural identity} (\ref%
{VertsZero}) shows that these violations are unavoidable.

Aiming for a clear argumentation, let us consider explicitly that the
involved expressions contain integrals exhibiting linear and logarithmic
power counting, i.e., the vector $\bar{J}_{2\mu }$ (\ref{J2u})\ and the
tensor $\bar{J}_{3\mu \nu }$ (\ref{J3uv}). They both depend on surface
terms, with the second one having ambiguous momenta as coefficients. In the
lack of\ translational invariance, routings parametrizing propagators are
independent and cannot be reduced to external momenta. Therefore, one must
consider this arbitrariness when building a general tensor to investigate
kinematic limits.

Considering this change of perspective, we will show that the low-energy
behavior of finite amplitudes precludes the simultaneous maintenance of
integration linearity and translational invariance. Ultimately, this
situation leads to anomalies since both these properties are requirements
for satisfying all WIs. This discussion emphasizes basic versions as those
that automatically satisfy the maximum number of RAGFs, albeit not all. We
advance that there is no need for computing anomalous amplitudes, so these
derivations are independent of specific trace versions.

Thereby, besides contributions on the external momenta (\ref{GForm}), the
general tensor must also consider the following terms%
\begin{eqnarray}
F_{\mu _{123}}^{\Delta } &=&\varepsilon _{\mu _{123}\nu _{1}}\left(
b_{1}P_{21}^{\nu _{2}}+b_{2}P_{31}^{\nu _{2}}+b_{3}P_{32}^{\nu _{2}}\right)
\Delta _{3\nu _{2}}^{\nu _{1}}  \notag \\
&&+\varepsilon _{\mu _{23}\nu _{12}}\left( a_{11}P_{21}^{\nu
_{2}}+a_{12}P_{31}^{\nu _{2}}+a_{13}P_{32}^{\nu _{2}}\right) \Delta _{3\mu
_{1}}^{\nu _{1}}  \notag \\
&&+\varepsilon _{\mu _{13}\nu _{12}}\left( a_{21}P_{21}^{\nu
_{2}}+a_{22}P_{31}^{\nu _{2}}+a_{23}P_{32}^{\nu _{2}}\right) \Delta _{3\mu
_{2}}^{\nu _{1}}  \notag \\
&&+\varepsilon _{\mu _{12}\nu _{12}}\left( a_{31}P_{21}^{\nu
_{2}}+a_{32}P_{31}^{\nu _{2}}+a_{33}P_{32}^{\nu _{2}}\right) \Delta _{3\mu
_{3}}^{\nu _{1}},  \label{Fdelta}
\end{eqnarray}%
with $P_{ij}=k_{i}+k_{j}$. The arbitrary constants $b_{j}$ and $a_{ij}$
summarize all degrees of freedom; thus, it is convenient to compact them
into the vectors $\boldsymbol{b}=\left( b_{1},b_{2},b_{3}\right) $ and $%
\boldsymbol{a}_{i}=\left( a_{i1},a_{i2},a_{i3}\right) $. The subindex $i$
links to the index $\mu _{i}$ associated with the vertex of amplitudes $%
T_{\mu _{123}}^{\Gamma _{123}}$. We will use the algebraic identity $%
\varepsilon _{\lbrack \mu _{1}\mu _{2}\mu _{3}\nu _{1}}\Delta _{3\nu
_{2}]}^{\nu _{2}}=0$ to simplify the study of relations with $AV$s when
expressing the tensor. It reduces the number of independent and arbitrary
parameters without losing information.

Contracting amplitudes with external momenta shows how finite amplitudes
determine surface terms. To clarify this idea, we propose one general
equation representing the satisfaction of all RAGFs%
\begin{equation}
q_{i}^{\mu _{i}}T_{\mu _{123}}^{\Gamma _{123}}=T_{i\left( -\right) \mu
_{kl}}^{AV}+\varepsilon _{\mu _{kl}\nu _{12}}q_{2}^{\nu _{1}}q_{3}^{\nu
_{2}}\Omega _{i}\left( q_{1},q_{2},q_{3}\right) ,  \label{GR}
\end{equation}%
with the index ordering constrained as $k<l\not=i\in \left\{ 1,2,3\right\} $%
. The second term on the RHS highlights the tensor structure while
symbolizing invariants linked to 2nd-rank finite amplitudes through $\Omega
_{i}$, with those generated in vector contractions being null. Anticipating
future comparisons, we rewrite $AV$ differences (\ref{AV(-)1})-(\ref{AV(-)3})%
\emph{\ }by expressing external momenta through sums:%
\begin{eqnarray}
T_{1\left( -\right) \mu _{23}}^{AV} &=&2i\varepsilon _{\mu _{23}\nu _{12}} 
\left[ P_{21}^{\nu _{2}}P_{32}^{\nu _{3}}-P_{31}^{\nu _{2}}\left(
P_{32}^{\nu _{3}}-P_{21}^{\nu _{3}}\right) -P_{32}^{\nu _{2}}P_{21}^{\nu
_{3}}\right] \Delta _{3\nu _{3}}^{\nu _{1}},  \label{1(-)} \\
T_{2\left( -\right) \mu _{13}}^{AV} &=&2i\varepsilon _{\mu _{13}\nu _{12}} 
\left[ -P_{21}^{\nu _{2}}\left( P_{31}^{\nu _{3}}-P_{32}^{\nu _{3}}\right)
-P_{31}^{\nu _{2}}P_{32}^{\nu _{3}}+P_{32}^{\nu _{2}}P_{31}^{\nu _{3}}\right]
\Delta _{3\nu _{3}}^{\nu _{1}},  \label{2(-)} \\
T_{3\left( -\right) \mu _{12}}^{AV} &=&2i\varepsilon _{\mu _{12}\nu _{12}} 
\left[ P_{21}^{\nu _{2}}P_{31}^{\nu _{3}}-P_{31}^{\nu _{2}}P_{21}^{\nu
_{3}}-P_{32}^{\nu _{2}}\left( P_{31}^{\nu _{3}}-P_{21}^{\nu _{3}}\right) %
\right] \Delta _{3\nu _{3}}^{\nu _{1}}.  \label{3(-)}
\end{eqnarray}

Performing contractions of the general structure (\ref{Fdelta}) is necessary
to verify the possibility of identifying the two-point functions above in
RAGFs. If this were to happen without additional conditions, they would be
simultaneously valid for any surface term values. Let us test this
possibility in the sequence.

We start by taking the first contraction and writing the result in terms of
the appropriate $P_{ij}$ combinations:%
\begin{eqnarray}
p_{31}^{\mu _{1}}F_{\mu _{123}}^{\Delta } &=&\varepsilon _{\mu _{23}\nu
_{12}}\Delta _{3\nu _{3}}^{\nu _{1}}[(a_{11}+b_{3})P_{21}^{\nu
_{2}}P_{32}^{\nu _{3}}+a_{12}P_{31}^{\nu _{2}}(P_{32}^{\nu _{3}}-P_{21}^{\nu
_{3}})-(a_{13}+b_{1})P_{32}^{\nu _{2}}P_{21}^{\nu _{3}}]  \notag \\
&&+\varepsilon _{\mu _{23}\nu _{12}}\Delta _{3\nu _{3}}^{\nu
_{1}}[-(a_{11}-b_{1})P_{21}^{\nu _{2}}P_{21}^{\nu
_{3}}+(a_{13}-b_{3})P_{32}^{\nu _{2}}P_{32}^{\nu _{3}}+b_{2}(P_{21}^{\nu
_{2}}-P_{32}^{\nu _{2}})P_{31}^{\nu _{3}}]  \notag \\
&&+\varepsilon _{\mu _{3}\nu _{123}}\Delta _{3\mu _{2}}^{\nu
_{1}}[-(a_{21}+a_{23})P_{21}^{\nu _{2}}P_{32}^{\nu _{3}}+a_{22}P_{31}^{\nu
_{2}}(P_{21}^{\nu _{3}}-P_{32}^{\nu _{3}})]  \notag \\
&&+\varepsilon _{\mu _{2}\nu _{123}}\Delta _{3\mu _{3}}^{\nu
_{1}}[-(a_{31}+a_{33})P_{21}^{\nu _{2}}P_{32}^{\nu _{3}}+a_{32}P_{31}^{\nu
_{2}}(P_{21}^{\nu _{3}}-P_{32}^{\nu _{3}})].
\end{eqnarray}%
After comparing this result with $AV$ amplitudes (\ref{1(-)}), we organized
non-zero terms in the first row. Vanishing the other rows sets most
coefficients directly, so one has to solve the remaining linear equations to
find $b_{3}=2i-b_{1}$ and $a_{12}=-2i$. By requiring the satisfaction of the
first RAGF, the original twelve parameters reduce to just three. Hence,
adopting a subindex corresponding to the considered contraction (with $q_{1}$%
), we organize this solution into the following matrix:%
\begin{equation}
(F_{\mu _{123}}^{\Delta })_{1}:\left( 
\begin{array}{c}
\boldsymbol{b} \\ 
\boldsymbol{a}_{1} \\ 
\boldsymbol{a}_{2} \\ 
\boldsymbol{a}_{3}%
\end{array}%
\right) =%
\begin{pmatrix}
b_{1} & 0 & 2i-b_{1} \\ 
b_{1} & -2i & 2i-b_{1} \\ 
-a_{23} & 0 & a_{23} \\ 
-a_{33} & 0 & a_{33}%
\end{pmatrix}%
.  \label{Sol1}
\end{equation}%
Extending this analysis to contractions with $q_{2}$ and $q_{3}$, we infer
the requirements to satisfy the corresponding relations. A comparison with
the differences between $AV$s establishes a system of linear equations whose
solutions follow%
\begin{equation}
(F_{\mu _{123}}^{\Delta })_{2}:%
\begin{pmatrix}
0 & b_{2} & 2i-b_{2} \\ 
0 & -a_{13} & a_{13} \\ 
2i & -b_{2} & b_{2}-2i \\ 
0 & -a_{33} & a_{33}%
\end{pmatrix}%
\text{;\quad }(F_{\mu _{123}}^{\Delta })_{3}:%
\begin{pmatrix}
b_{1} & 2i-b_{1} & 0 \\ 
a_{11} & -a_{11} & 0 \\ 
a_{21} & -a_{21} & 0 \\ 
b_{1} & 2i-b_{1} & -2i%
\end{pmatrix}%
.  \label{Sol2}
\end{equation}

Next, let us study the simultaneous satisfaction of two relations by putting
solutions together $(F_{\mu _{123}}^{\Delta })_{ij}=(F_{\mu _{123}}^{\Delta
})_{i}\cap (F_{\mu _{123}}^{\Delta })_{j}$. The intersection of the first
two sets determines all coefficients without recurring to further conditions
regarding surface terms. In other words, the hypothesis of satisfaction of
the first and second RAGFs constrains the general tensor to%
\begin{equation}
(F_{\mu _{123}}^{\Delta })_{12}=2i[\varepsilon _{\mu _{13}\nu
_{12}}(P_{21}^{\nu _{2}}-P_{32}^{\nu _{2}})\Delta _{3\mu _{2}}^{\nu
_{1}}+\varepsilon _{\mu _{23}\nu _{12}}(P_{32}^{\nu _{2}}-P_{31}^{\nu
_{2}})\Delta _{3\mu _{1}}^{\nu _{1}}+\varepsilon _{\mu _{123}\nu
_{1}}P_{32}^{\nu _{2}}\Delta _{3\nu _{2}}^{\nu _{1}}],
\end{equation}%
which is incompatible with the coefficients of the third set. We observe the
same circumstances when combining other solutions. Single-solutions depend
on three independent parameters and are compatible in pairs, which means
that coefficients are unique once one pair of RAGFs is determined.
Therefore, the complementary contraction always leads to an incompatible
solution.

Now, identifying $p_{ij}=P_{il}-P_{jl}$, the achieved tensors correspond to
the divergent sector of amplitude versions computed explicitly (\ref{ST1})-(%
\ref{ST3}):%
\begin{equation}
(F_{\mu _{123}}^{\Delta })_{23}=S_{1\mu _{123}};\quad (F_{\mu
_{123}}^{\Delta })_{13}=S_{2\mu _{123}};\quad (F_{\mu _{123}}^{\Delta
})_{12}=S_{3\mu _{123}}.
\end{equation}%
As a consequence, their contractions also follow the properties (\ref{contS1}%
)-(\ref{contS9}). We stress that these results come from the analysis of the
divergent structure of a general rank-3 tensor of mass-dimension one (\ref%
{Fdelta}), independently of the explicit approach developed at the outset of
this section.

Let us resume the discussion about low-energy implications by considering
this new information. For instance, in the hypothesis of satisfying both
vector RAGFs, the complete tensor structure of any \textit{anomalous}
amplitude ($AVV$, $VAV$, $VVA$, $AAA$) assumes the form:%
\begin{equation}
T_{\mu _{123}}^{\Gamma _{123}}=(F_{\mu _{123}}^{\Delta })_{23}+\hat{F}_{\mu
_{123}}=S_{1\mu _{123}}+\hat{F}_{\mu _{123}}.
\end{equation}%
Differently from the original context (\ref{GForm}), the term $\hat{F}_{\mu
_{123}}$ represents strictly finite parts this time, justifying the adoption
of the "hats" notation. In that sense, note that all considerations from the
previous subsection extend to this analysis. Momenta contractions of these
finite contributions lead to $q_{i}^{\mu _{i}}\hat{F}_{\mu
_{123}}=\varepsilon _{\mu _{kl}\nu _{12}}q_{2}^{\nu _{1}}q_{3}^{\nu _{2}}%
\hat{V}_{i}$, linking to invariants $\hat{V}_{i}$ that are functions of form
factors belonging to the general tensor. We cast these results in the
sequence 
\begin{eqnarray}
q_{1}^{\mu _{1}}T_{\mu _{123}}^{\Gamma _{123}}-T_{1\left( -\right) \mu
_{23}}^{AV} &=&\varepsilon _{\mu _{23}\nu _{12}}q_{2}^{\nu _{1}}q_{3}^{\nu
_{2}}(\hat{V}_{1}+2i\Delta _{3\alpha }^{\alpha })=\varepsilon _{\mu _{23}\nu
_{12}}q_{2}^{\nu _{1}}q_{3}^{\nu _{2}}\Omega _{1}, \\
q_{2}^{\mu _{2}}T_{\mu _{123}}^{\Gamma _{123}}-T_{2\left( -\right) \mu
_{13}}^{AV} &=&\varepsilon _{\mu _{13}\nu _{12}}q_{2}^{\nu _{1}}q_{3}^{\nu
_{2}}\hat{V}_{2}=\varepsilon _{\mu _{13}\nu _{12}}q_{2}^{\nu _{1}}q_{3}^{\nu
_{2}}\Omega _{2}, \\
q_{3}^{\mu _{3}}T_{\mu _{123}}^{\Gamma _{123}}-T_{3\left( -\right) \mu
_{12}}^{AV} &=&\varepsilon _{\mu _{12}\nu _{12}}q_{2}^{\nu _{1}}q_{3}^{\nu
_{2}}\hat{V}_{3}=\varepsilon _{\mu _{12}\nu _{12}}q_{2}^{\nu _{1}}q_{3}^{\nu
_{2}}\Omega _{3},
\end{eqnarray}%
where $\Omega _{i}$ are the invariants associated with finite amplitudes (%
\ref{GR}). The additional object $\Delta _{3\alpha }^{\alpha }$ emerged from
index permutations within the $S_{1}$-contraction, as acknowledged in the
analysis of RAGFs (\ref{contS1}). Its presence characterizes the
corresponding relation as non-automatic since the equality between $\hat{V}%
_{1}$ and $\Omega _{1}$ is not direct.

Given the analogy with the previous subsection, the \textit{structural
identity }(\ref{VertsZero}) assumes the form 
\begin{equation}
\hat{V}_{1}\left( 0\right) +\hat{V}_{2}\left( 0\right) -\hat{V}_{3}\left(
0\right) =0\Rightarrow 2i\Delta _{3\alpha }^{\alpha }=\Omega _{1}\left(
0\right) +\Omega _{2}\left( 0\right) -\Omega _{3}\left( 0\right) .
\end{equation}%
This equation is true regardless of the RAGFs satisfied by hypotheses.
Changing the tensor sector to $S_{2}$ or $S_{3}$ changes the contraction
originating $\Delta _{3\alpha }^{\alpha }$, which does not affect the
achieved result. We obtained a proper relation connecting surface terms with
a kinematical property of finite functions, generalizing the particular
occurrence $\Omega _{1}^{PVV}\left( 0\right) =2i\Delta _{3\alpha }^{\alpha }$
(\ref{IRUV}). The outset was a tensor with two RAGFs satisfied without
restriction, connected to $AV$ differences and finite amplitudes. Explicit
computations from Subsection (\ref{unique}) corroborate the result above.

On the other hand, examining the low-energy behavior of finite amplitudes (%
\ref{LEPVV})-(\ref{LEPAA}) allows for assessing the numeric value of the
expression above. In the case of the $AVV$ amplitude and vertex
permutations, two form factors are zero, and the other yields the value%
\begin{equation}
\Omega ^{PVV}=\Omega ^{VPV}=-\Omega ^{VVP}=(2\pi )^{-2}.
\end{equation}%
The same value manifests when analyzing the $AAA$ with its three non-zero
contributions\footnote{%
Our discussion applies to theories involving different masses. Since all $%
\Omega $s are non-zero under these circumstances, calculations would be
similar to those for the $AAA$ amplitude. Even though each $\Omega $ is
mass-dependent, the combination dictated by the \textit{structural identity}
is a constant.}:%
\begin{equation}
\Omega _{1}^{PAA}\left( 0\right) +\Omega _{2}^{APA}\left( 0\right) -\Omega
_{3}^{AAP}\left( 0\right) =(2\pi )^{-2}.  \label{omega(AAA)}
\end{equation}%
These kinematical properties set the value of the surface terms, producing
the same condition necessary to find unique amplitudes that satisfy all
RAGFs: 
\begin{equation}
\text{RAGF}\Leftrightarrow 2i\Delta _{3\alpha }^{\alpha }=(2\pi )^{-2}.
\label{LiUni4D}
\end{equation}

In the previous subsection, we deduced a \textit{structural identity} from
scalar invariants $V_{i}$ of a general third-rank tensor (\ref{VertsZero}).
This equation applies as long as there are no poles at zero, and it does not
require any hypothesis about symmetries. When identifying the result of
momenta contractions with amplitudes coming from WIs, we get a device to
anticipate the impossibility of realizing all WIs. As these amplitudes are
finite and immune to ambiguities, this analysis does not depend on the
scheme used to compute divergences. This competition involving symmetries
materializes into the invariants $\hat{V}_{i}=\Omega _{i}+\mathcal{A}_{i}$,
which produce anomalous factors $\mathcal{A}_{i}$ to maintain the \textit{%
structural identity}: 
\begin{equation}
(\hat{V}_{1}+\hat{V}_{2}-\hat{V}_{3})|_{0}=\left( \Omega _{1}+\Omega
_{2}-\Omega _{3}\right) |_{0}+\mathcal{A}_{1}+\mathcal{A}_{2}-\mathcal{A}%
_{3}=0.
\end{equation}

Meanwhile, by preserving the arbitrariness of internal momenta and surface
terms, we observed that the low-energy behavior of these finite amplitudes
links to the numerical value of surface terms. This value is the same that
guarantees the uniqueness of axial amplitudes while satisfying all RAGFs. 
\textit{For these perturbative amplitudes, shift-invariance is lost when the
linearity of integration is obeyed and vice-versa. }Hence, kinematical
limits of finite amplitudes are incompatible with the whole set of WIs, as
already established in two dimensions. The main counterpoint of this work is
that anomalies originate in finite functions, differing from the literature
and its focus on regularization properties. We extend this argumentation to
extra dimensions in the ensuing section.

\section{Six-Dimensional Amplitudes}

\label{6Dim4Pt}This section explores the six-dimensional box to illustrate
the generality of results seen previously. The integrand of this amplitude (%
\ref{tt}) contains traces involving the chiral matrix, with the only nonzero
contributions being the following:%
\begin{eqnarray}
t_{\mu _{1234}}^{AVVV} &=&K_{1234}^{\nu _{1234}}\mathrm{tr}(\gamma _{\ast
\mu _{1}\nu _{1}\mu _{2}\nu _{2}\mu _{3}\nu _{3}\mu _{4}\nu _{4}})\frac{1}{%
D_{1234}} \\
&&-m^{2}\mathrm{tr}(\gamma _{\ast \mu _{1234}\nu _{12}})(K_{12}^{\nu
_{12}}-K_{13}^{\nu _{12}}+K_{14}^{\nu _{12}}+K_{23}^{\nu _{12}}-K_{24}^{\nu
_{12}}+K_{34}^{\nu _{12}})\frac{1}{D_{1234}}.  \notag
\end{eqnarray}%
Our focus here is on the trace with eight Dirac matrices, which yields a
combination of products between the Levi-Civita symbol and the metric.
Different ways to compute this object lead to different tensor arrangements.
Although they compound identities, such a connection is not straightforward
when comparing integrated amplitudes.

Here, we introduce two versions defined by replacing the chiral matrix
definition adjacent to the first and second vertices (labeled through
subindexes 1 and 2). By prioritizing one vertex, its index appears
exclusively in the Levi-Civita symbol. Such a feature is transparent in the
organization achieved after integration%
\begin{eqnarray}
(T_{\mu _{1234}}^{AVVV})_{1} &=&-8\left[ \varepsilon _{\mu _{134}\nu
_{123}}T_{\mu _{2}}^{\left( +-+\right) \nu _{123}}-\varepsilon _{\mu
_{124}\nu _{123}}T_{\mu _{3}}^{\left( --+\right) \nu _{123}}+\varepsilon
_{\mu _{123}\nu _{123}}T_{\mu _{4}}^{\left( -++\right) \nu _{123}}\right] 
\label{6D1} \\
&&-\frac{1}{2}\varepsilon _{\mu _{1234}}^{\qquad \nu _{12}}T_{\nu _{12}}^{%
\tilde{T}PPP},  \notag \\
(T_{\mu _{1234}}^{AVVV})_{2} &=&-8\left[ \varepsilon _{\mu _{234}\nu
_{123}}T_{\mu _{1}}^{\left( +-+\right) \nu _{123}}-\varepsilon _{\mu
_{214}\nu _{123}}T_{\mu _{3}}^{\left( ++-\right) \nu _{123}}+\varepsilon
_{\mu _{213}\nu _{123}}T_{\mu _{4}}^{\left( +--\right) \nu _{123}}\right] 
\label{6D2} \\
&&-\frac{1}{2}\varepsilon _{\mu _{1234}}^{\qquad \nu _{12}}T_{\nu
_{12}}^{STPP},  \notag
\end{eqnarray}%
where tensor and pseudotensor vertices arise naturally (\ref{SetofVertexes}%
). Consult the expressions attributed to these substructures in Appendix \ref%
{sixdamp}. We emphasize that the tensor $T^{\left( -+-\right) }$ corresponds
to vanishing integrals, as it occurred for similar objects in other
dimensions.

The discussion above reflects on the study of RAGFs; consult Eqs. (\ref%
{RAGF1}) and (\ref{RAGF2}). Their verifications employ two new amplitudes,
both having simple structures since their only non-zero contributions are
traces involving six Dirac matrices plus the chiral one. The integrated
three-point function corresponds to a pure surface term%
\begin{equation}
T_{\mu _{1}\mu _{2}\mu _{3}}^{AVV}\left( i,j,l\right) =\frac{8}{3}%
\varepsilon _{\mu _{123}\nu _{123}}p_{ji}^{\nu _{2}}p_{li}^{\nu
_{3}}P_{ijl}^{\nu _{4}}\Delta _{4\nu _{4}}^{\nu _{1}},  \label{6Dsur}
\end{equation}%
while the box arising from the axial contraction leads to a finite integral%
\begin{equation}
T_{\mu _{234}}^{PVVV}=-8m\varepsilon _{\mu _{234}\nu _{123}}p_{21}^{\nu
_{1}}p_{32}^{\nu _{2}}p_{43}^{\nu _{3}}J_{4}.  \label{6Dfin}
\end{equation}

Thus, inquiring about one RAGF requires contracting the original amplitude
and identifying the objects above. External momenta couple directly with
finite tensors $J_{4}$ and surface terms $\Delta _{4}$ for most cases, so
relations apply without further conditions. Nonetheless, that does not occur
for the contraction with the vertex-defining index, whose satisfaction is
not automatic.

For instance, the first $AVVV$ version used a trace prioritizing the index $%
\mu _{1}$ (axial vertex), so this index only appears within the Levi-Civita
symbol (\ref{6D1}). When performing the corresponding contraction, this
tensor arrangement is inadequate for manipulations, and permutations are
necessary. Although that allows identifying amplitudes, it brings
potentially-violating contributions:%
\begin{eqnarray}
p_{41}^{\mu _{1}}(T_{\mu _{1234}}^{AVVV})_{1} &=&T_{\mu _{4}\mu _{2}\mu
_{3}}^{AVV}\left( 1,2,3\right) -T_{\mu _{2}\mu _{3}\mu _{4}}^{AVV}\left(
2,3,4\right) -2mT_{\mu _{234}}^{PVVV} \\
&&+\frac{8}{3}\varepsilon _{\mu _{234}\nu _{123}}p_{21}^{\nu
_{1}}p_{32}^{\nu _{2}}p_{43}^{\nu _{3}}[\Delta _{4\rho }^{\rho }+i\left(
4\pi \right) ^{-3}].  \notag
\end{eqnarray}%
When considering the second box version (\ref{6D2}), an analogous situation
manifests in the following vector contraction:%
\begin{eqnarray}
p_{21}^{\mu _{2}}(T_{\mu _{1234}}^{AVVV})_{2} &=&T_{\mu _{134}}^{AVV}\left(
1,3,4\right) -T_{\mu _{134}}^{AVV}\left( 2,3,4\right)  \notag \\
&&+\frac{8}{3}\varepsilon _{\mu _{134}\nu _{123}}p_{21}^{\nu
_{1}}p_{32}^{\nu _{2}}p_{43}^{\nu _{3}}[\Delta _{4\rho }^{\rho }+i\left(
4\pi \right) ^{-3}].
\end{eqnarray}

Again, a similar outcome manifests if one compares both versions directly:%
\begin{equation}
(T_{\mu _{1234}}^{AVVV})_{1}-(T_{\mu _{1234}}^{AVVV})_{2}=-\frac{8}{3}%
\varepsilon _{\mu _{1234}\nu _{12}}p_{32}^{\nu _{1}}p_{43}^{\nu _{2}}[\Delta
_{4\rho }^{\rho }+i\left( 4\pi \right) ^{-3}].
\end{equation}%
That clarifies the connection between linearity and uniqueness in the sense
we posed. Different formulae to the traces do not deliver identical tensors,
and their equivalence depends on the precise value attributed to the surface
term. Under the condition of canceling the object between squared brackets,
these tensors coincide, and all RAGFs apply.

Momenta contractions also link to WIs, exhibiting the same features seen in
other dimensions. To clarify this aspect, one follows the analysis developed
in Subsection (\ref{LE4D}) and writes the box amplitude through a general
tensor. Thus, properties relating form factors with invariants $V_{i}$ arise
when performing momenta contractions over this general structure $q_{i}^{\mu
_{i}}F_{\mu _{1234}}=\varepsilon _{\mu _{1\cdots \hat{\imath}\cdots 4}\nu
_{123}}q_{2}^{\nu _{1}}q_{3}^{\nu _{2}}q_{4}^{\nu _{3}}V_{i}$. Without
assuming any hypothesis about symmetries, putting these pieces of
information together allows deriving a \textit{structural identity} among
invariants at the kinematical point $q_{i}\cdot q_{j}=0$:%
\begin{equation}
V_{1}\left( 0\right) +V_{2}\left( 0\right) -V_{3}\left( 0\right)
+V_{4}\left( 0\right) =0.  \label{Vis6}
\end{equation}

We saw that each invariant might contain two parts, one associated with a
finite function and the other corresponding to anomalous contributions
arising in contractions. For the investigated case, only the axial
contraction leads to a finite structure linked to the amplitude $PVVV$.
Taking its explicit form (\ref{6Dfin}), we replace the finite object
definition $J_{4}=i\left( 4\pi \right) ^{-3}Z_{000}^{\left( -1\right)
}\left( p,q,r\right) $ and use its limit (\ref{LetZ6D}) to evaluate the
low-energy behavior of this amplitude at the point where all bilinears are
zero:%
\begin{equation}
-2mT_{\mu _{234}}^{PVVV}=:\varepsilon _{\mu _{234}\nu _{123}}p_{21}^{\nu
_{1}}p_{32}^{\nu _{2}}p_{43}^{\nu _{3}}\Omega ^{PVVV}\left( 0\right) =-\frac{%
8i}{3\left( 4\pi \right) ^{3}}\varepsilon _{\mu _{234}\nu _{123}}p_{21}^{\nu
_{1}}p_{32}^{\nu _{2}}p_{43}^{\nu _{3}}\not=0
\end{equation}%
Since this outcome differs from zero, this equation states that at least one
WI must be violated as compensation.

Usually, the literature opts for preserving all vector identities by letting
the axial one broken. This scenario is accomplished by the first amplitude
version (when surface terms vanish), which prioritizes the index
corresponding to the axial vertex. One anomalous contribution arises to the
axial contraction under these circumstances, so the \textit{structural
identity} yields%
\begin{equation}
V_{1}\left( 0\right) =\Omega ^{PVVV}\left( 0\right) +\mathcal{A}_{1}=0.
\end{equation}%
Alternatively, we cast one case of preserving the axial identity by
exploring the second amplitude version. The anomalous contribution appears
for the second vertex as represented in the \textit{structural identity}:%
\begin{equation}
V_{1}\left( 0\right) +V_{2}\left( 0\right) =\Omega ^{PVVV}\left( 0\right) +%
\mathcal{A}_{2}=0.
\end{equation}%
This perspective shows that the value assumed by an anomaly comes from the
kinematic behavior of finite functions and not from divergences.

\section{Final Remarks and Perspectives}

\label{finalremarks}This investigation looks for a better understanding of
anomalies by approaching $(n+1)$-point perturbative amplitudes in a $2n$%
-dimensional setting. They combine axial and vector vertices to form odd
tensors, whose Dirac trace of the highest order contains two gamma matrices
beyond the space-time dimension. This structure allows different
expressions, considered identities at the integrand level. Nevertheless,
connecting them is not automatic after loop integration since the divergent
character of amplitudes implies the presence of surface terms.

The IReg strategy was crucial to this exploration because it avoids
evaluating divergent objects initially. That maintains the connection among
all expressions attributed to the same object, allowing a clear view of the\
consequences of trace choices. As results are analogous in different
dimensions, consult the two-dimensional case for a simpler view (\ref{oddav}%
)-(\ref{oddva}). By replacing the chiral matrix definition adjacent to one
vertex, we limit the occurrence of this \textit{version-defining} index
solely to the Levi-Civita symbol. We stress this tensor structure is
unrelated to the nature of the vertex as axial or vector.

Such a feature affects momenta contractions embodied in Relations Among
Green Functions (RAGFs). Notwithstanding these constraints originate from
algebraic operations, potentially-violating terms arise after integration
for contractions with the version-defining index (\ref{ragfViol4D}). These
terms also distinguish amplitude versions achieved through different trace
choices (\ref{v12})-(\ref{v23}). From these results, it is possible to
obtain unique perturbative solutions that satisfy all RAGFs by choosing
specific finite values for surface terms (\ref{Daa}). That preserves the
linearity of integration in this context; however, it breaks all symmetry
expectations for odd and even correlators.

At the same time, symmetry implications arise from momenta contractions
through Ward identities (WIs). Under the hypothesis that RAGFs apply,
translational invariance would be sufficient to ensure the validity of both
axial and vector WIs. This invariance imposes the vanishing of lower-point
amplitudes inside these relations, leading to the cancellation of surface
terms. Nevertheless, that is not enough to maintain the RAGF with
potentially-violating terms. Even by imposing translational invariance, one
anomalous contribution emerges from the finite sector of the amplitudes.

The result above agrees with the recognized competition between gauge and
chiral symmetries; however, we propose a broader perspective. By
investigating strategies to take Dirac traces, we derived distinct
expressions for an amplitude (\ref{r123}). They are combinations of the most
fundamental ones (called version-defining) and carry violations in more
contractions. Under this reasoning, preserving the vector symmetry is only
one possibility. That is the case of the first $AVV$ version, which
prioritizes the index of the axial vertex and violates the corresponding WI.
Table \ref{tab2d} casts the two-dimensional cases, emphasizing the
version-defining occurrences and one of their combinations (third version).

Further explorations on this subject do not concern this work, but we aim to
publish them soon. They include a complete analysis of trace operations
within triangle amplitudes, showing a route to significant simplifications
in perturbative calculations. Following our perspecticve on ambiguities and
exclusive manipulation of finite integrals, we intend to study trace
anomalies for Weyl fermions in four-dimensional three-point correlators.
Such a theme appears in recent debates about the contributions of Pontryagin
density to these anomalies \cite{Bonora2014}-\cite{Bonora2022}, \cite%
{Liu2022} and \cite{Bastianelli2016}-\cite{Larue2023}.

Here, we also proposed a general tensor form for amplitudes to investigate
low-energy theorems, clarifying the opposition between translational
invariance and linearity of integration. First, supposing coefficients on
external momenta, \textit{structural identities} involving invariants arise
in different dimensions: (\ref{id2D}), (\ref{VertsZero}), and (\ref{Vis6}).
They contain kinematical limits of finite functions that should be zero but
assume another value instead. Hence, the finite content $\Omega $ demands
anomalous contributions $\mathcal{A}$ to satisfy these identities%
\begin{equation}
\Omega _{1}\left( 0\right) +\sum_{i=1}^{n+1}\left( -1\right) ^{i}\Omega
_{i}\left( 0\right) +\mathcal{A}_{1}+\sum_{i=2}^{n+1}\left( -1\right) ^{i}%
\mathcal{A}_{i}=0,
\end{equation}%
showing that violations are unavoidable and have a fixed value. Nonetheless,
the distribution of anomalous contributions still depends on trace choices.

Second, we admit the dependence on arbitrary routings that break
translational invariance. That allows deriving the structure of surface
terms without computing amplitudes, emphasizing the impossibility of
automatic satisfaction of all RAGFs. Meanwhile, \textit{structural
identities }still apply and associate the surface term value with the
kinematical limits of finite functions%
\begin{equation}
\Omega _{1}\left( 0\right) +\sum_{i=2}^{n+1}\left( -1\right) ^{i}\Omega
_{i}\left( 0\right) =\frac{2^{n}i^{n-1}}{n}\Delta _{n+1;\alpha }^{\alpha },
\end{equation}%
reproducing the condition for linearity maintenance%
\begin{equation}
\Delta _{n+1;\alpha }^{\alpha }=-\frac{2}{\left( n-1\right) !}\frac{i}{%
\left( 4\pi \right) ^{n}}.
\end{equation}%
All explored facets apply to amplitudes in other even dimensions, with the
final equations being general. They are also valid for propagators featuring
arbitrary masses, so we aim to elaborate on this discussion in the future.

\appendix

\section{Two-Dimensional Feynman Integrals}

\label{AppInt2D}\textbf{One-propagator integrals}%
\begin{eqnarray}
\bar{J}_{1}\left( k_{i}\right) &=&I_{\log }^{\left( 2\right) } \\
\bar{J}_{1}^{\mu }\left( k_{i}\right) &=&-k_{i}^{\nu }\Delta _{2\nu
}^{\left( 2\right) \mu }
\end{eqnarray}

\textbf{Two-propagator integrals}%
\begin{eqnarray}
\overline{J}_{2} &=&J_{2}=i\left( 4\pi \right) ^{-1}[Z_{0}^{\left( -1\right)
}\left( p^{2},m^{2}\right) ]  \label{2DJ2} \\
\overline{J}_{2}^{\mu _{1}} &=&J_{2}^{\mu _{1}}=i\left( 4\pi \right)
^{-1}[-q^{\mu _{1}}Z_{1}^{\left( -1\right) }]  \label{2DJ2mu1} \\
\bar{J}_{2}^{\mu _{12}} &=&J_{2}^{\mu _{12}}+(\Delta _{2}^{\left( 2\right)
\mu _{12}}+g^{\mu _{12}}I_{\log }^{\left( 2\right) })/2 \\
J_{2}^{\mu _{12}} &=&i\left( 4\pi \right) ^{-1}[-g^{\mu _{12}}Z_{0}^{\left(
0\right) }/2+q^{\mu _{12}}Z_{2}^{\left( -1\right) }]
\end{eqnarray}

\textbf{Reductions of finite functions}%
\begin{eqnarray}
Z_{0}^{\left( 0\right) } &=&2q^{2}Z_{2}^{\left( -1\right)
}-q^{2}Z_{1}^{\left( -1\right) }\text{, }2Z_{1}^{\left( -1\right)
}=Z_{0}^{\left( -1\right) } \\
q^{2}Z_{n+2}^{\left( -1\right) } &=&q^{2}Z_{n+1}^{\left( -1\right)
}-m^{2}Z_{n}^{\left( -1\right) }-\left( n+1\right) ^{-1}\text{ with }%
n=0,1,3,\cdots
\end{eqnarray}

\textbf{Reductions of tensors} 
\begin{eqnarray}
2J_{2}^{\mu _{1}} &=&-q^{\mu _{1}}J_{2}\text{ and }2q_{\mu _{1}}J_{2}^{\mu
_{1}}=-q^{2}J_{2} \\
2q_{\mu _{1}}J_{2}^{\mu _{12}} &=&-q^{2}J_{2}^{\mu _{2}}\text{ and }g_{\mu
_{12}}J_{2}^{\mu _{12}}=m^{2}J_{2}+i\left( 4\pi \right) ^{-1}
\end{eqnarray}

\section{Four-Dimensional Feynman Integrals}

\label{AppInt4D}\textbf{Two-propagator integrals\ }%
\begin{eqnarray}
\bar{J}_{2} &=&J_{2}\left( p_{ij}\right) +I_{\log }^{\left( 4\right) }\text{
with }J_{2}\left( p_{ij}\right) =i\left( 4\pi \right) ^{-2}[-Z_{0}^{\left(
0\right) }\left( p_{ij}^{2},m^{2}\right) ] \\
\bar{J}_{2\mu } &=&J_{2\mu }\left( p_{ij}\right) -(P_{ij}^{\nu }\Delta
_{3\mu \nu }^{\left( 4\right) }+p_{ji\mu }I_{\log }^{\left( 4\right) })/2
\label{J2u} \\
J_{2\mu }\left( p_{ij}\right) &=&i\left( 4\pi \right) ^{-2}[p_{ij\mu
}Z_{1}^{\left( 0\right) }\left( p_{ij}^{2},m^{2}\right) ]
\end{eqnarray}

\textbf{Three-propagator integrals }using general variables $p$ and $q$%
\textbf{\ }%
\begin{eqnarray}
\bar{J}_{3} &=&J_{3}=i\left( 4\pi \right) ^{-2}[Z_{00}^{\left( -1\right)
}\left( p,q\right) ] \\
\bar{J}_{3\mu } &=&J_{3\mu }=i\left( 4\pi \right) ^{-2}[-p_{\mu
}Z_{10}^{\left( -1\right) }-q_{\mu }Z_{01}^{\left( -1\right) }] \\
\bar{J}_{3\mu _{12}} &=&J_{3\mu _{12}}+(\Delta _{3\mu _{12}}^{\left(
4\right) }+g_{\mu _{12}}I_{\log }^{\left( 4\right) })/4  \label{J3uv} \\
J_{3\mu _{12}} &=&i\left( 4\pi \right) ^{-2}[p_{\mu _{12}}Z_{20}^{\left(
-1\right) }+q_{\mu _{12}}Z_{02}^{\left( -1\right) }+p_{(\mu _{1}}q_{\mu
_{2})}Z_{11}^{\left( -1\right) }-\frac{1}{2}g_{\mu _{12}}Z_{00}^{\left(
0\right) }]
\end{eqnarray}

\textbf{Reductions of finite functions} using $2Z_{1}^{\left( 0\right)
}=Z_{0}^{\left( 0\right) }$ and the Kronecker symbol $\delta _{n0}$ 
\begin{eqnarray}
&&2[p^{2}Z_{n+1;m}^{\left( -1\right) }+\left( p\cdot q\right)
Z_{n;m+1}^{\left( -1\right) }] \\
&=&p^{2}Z_{n;m}^{\left( -1\right) }+\left( 1-\delta _{n0}\right)
nZ_{n-1,m}^{\left( 0\right) }+\delta _{n0}Z_{m}^{\left( 0\right) }\left(
p_{31}\right) -\sum_{s=0}^{m}\left( -1\right) ^{s}\binom{m}{s}%
Z_{n+s}^{\left( 0\right) }\left( p_{32}\right)  \notag \\
&&2[q^{2}Z_{n;m+1}^{\left( -1\right) }+\left( p\cdot q\right)
Z_{n+1;m}^{\left( -1\right) }] \\
&=&q^{2}Z_{n;m}^{\left( -1\right) }+\left( 1-\delta _{m0}\right)
mZ_{n;m-1}^{\left( 0\right) }+\delta _{m0}Z_{n}^{\left( 0\right) }\left(
p_{21}\right) -\sum_{s=0}^{m}\left( -1\right) ^{s}\binom{m}{s}%
Z_{n+s}^{\left( 0\right) }\left( p_{32}\right)  \notag
\end{eqnarray}%
\begin{equation}
2Z_{00}^{\left( 0\right) }=[p^{2}Z_{10}^{\left( -1\right)
}+q^{2}Z_{01}^{\left( -1\right) }]-2m^{2}Z_{00}^{\left( -1\right)
}+2Z_{1}^{\left( 0\right) }\left( q-p\right) -1  \label{Z00^0}
\end{equation}

\textbf{Reductions of tensors}%
\begin{eqnarray*}
2p^{\mu _{1}}J_{3\mu _{1}} &=&-p^{2}J_{3}+[J_{2}\left( q\right) -J_{2}\left(
q-p\right) ] \\
2q^{\mu _{1}}J_{3\mu _{1}} &=&-q^{2}J_{3}+[J_{2}\left( p\right) -J_{2}\left(
q-p\right) ]
\end{eqnarray*}%
\begin{eqnarray*}
2p^{\mu _{1}}J_{3\mu _{12}} &=&-p^{2}J_{3\mu _{2}}+[J_{2\mu _{2}}\left(
q\right) +J_{2\mu _{2}}\left( q-p\right) +q_{\mu _{2}}J_{2}\left( q-p\right)
] \\
2q^{\mu _{1}}J_{3\mu _{12}} &=&-q^{2}J_{3\mu _{2}}+[J_{2\mu _{2}}\left(
p\right) +J_{2\mu _{2}}\left( q-p\right) +q_{\mu _{2}}J_{2}\left( q-p\right)
]
\end{eqnarray*}%
\begin{equation}
g^{\mu _{12}}J_{3\mu _{12}}=m^{2}J_{3}+J_{2}\left( q-p\right) +i\left[
2\left( 4\pi \right) ^{2}\right] ^{-1}  \label{trJ3}
\end{equation}

\section{Six-Dimensional Feynman Integrals}

\label{AppInt6D}\textbf{Three-propagator integrals}%
\begin{eqnarray*}
\bar{J}_{3} &=&J_{3}+I_{\log }^{\left( 6\right) }\text{ with }J_{3}\left(
p,q\right) =i\left( 4\pi \right) ^{-3}[-Z_{00}^{\left( 0\right) }\left(
p,q\right) ] \\
\bar{J}_{3}^{\mu _{1}}\left( k_{1},k_{2},k_{3}\right) &=&J_{3}^{\mu
_{1}}\left( k_{1},k_{2},k_{3}\right) -l^{\nu _{1}}\Delta _{4\nu
_{1}}^{\left( 6\right) \mu _{1}}/3-\left( p_{21}^{\mu _{1}}+p_{31}^{\mu
_{1}}\right) I_{\log }^{\left( 6\right) }/3 \\
J_{3}^{\mu _{1}}\left( k_{1},k_{2},k_{3}\right) &=&i\left( 4\pi \right)
^{-3}[p_{21}^{\mu _{1}}Z_{10}^{\left( 0\right) }+p_{31}^{\mu
_{1}}Z_{01}^{\left( 0\right) }]
\end{eqnarray*}

\textbf{Four-propagator integrals}%
\begin{eqnarray*}
\overline{J}_{4} &=&J_{4}=i\left( 4\pi \right) ^{-3}[Z_{000}^{\left(
-1\right) }\left( p,q,r\right) ] \\
\overline{J}_{4\mu _{1}} &=&J_{4\mu _{1}}=i\left( 4\pi \right) ^{-3}[-p_{\mu
_{1}}Z_{100}^{\left( -1\right) }-q_{\mu _{1}}Z_{010}^{\left( -1\right)
}-r_{\mu _{1}}Z_{001}^{\left( -1\right) }] \\
\bar{J}_{4\mu _{12}} &=&J_{4\mu _{12}}+(\Delta _{4\mu _{12}}^{\left(
6\right) }+g_{\mu _{12}}I_{\log }^{\left( 6\right) })/6 \\
J_{4\mu _{1}\mu _{2}} &=&i\left( 4\pi \right) ^{-3}[-g_{\mu
_{12}}Z_{000}^{\left( 0\right) }/2+p_{\mu _{12}}Z_{200}^{\left( -1\right)
}+q_{\mu _{12}}Z_{020}^{\left( -1\right) }+r_{\mu _{12}}Z_{002}^{\left(
-1\right) } \\
&&+p_{(\mu _{1}}q_{\mu _{2})}Z_{110}^{\left( -1\right) }+p_{(\mu _{1}}r_{\mu
_{2})}Z_{101}^{\left( -1\right) }+q_{(\mu _{1}}r_{\mu _{2})}Z_{011}^{\left(
-1\right) }]
\end{eqnarray*}

\textbf{Reductions of finite functions }using the binomial coefficient $%
C_{s}^{k}=\binom{k}{s}$%
\begin{eqnarray*}
&&2[p^{2}Z_{n+1;m;k}^{\left( -1\right) }+\left( p\cdot q\right)
Z_{n;m+1;k}^{\left( -1\right) }+\left( p\cdot r\right) Z_{n;m;k+1}^{\left(
-1\right) }] \\
&=&p^{2}Z_{n;m;k}^{\left( -1\right) }+\left( 1-\delta _{n0}\right)
nZ_{n-1;m;k}^{\left( 0\right) }+\delta _{n0}Z_{m;k}^{\left( 0\right) }\left(
q,r\right) -\sum_{s_{1}=0}^{k}\sum_{s_{2}=0}^{s_{1}}\left( -1\right)
^{s_{1}}C_{s_{1}}^{k}C_{s_{2}}^{s_{1}}Z_{n+s_{1}-s_{2};m+s_{2}}^{\left(
0\right) }\left( p_{42},p_{43}\right)
\end{eqnarray*}%
\begin{eqnarray*}
&&2[q^{2}Z_{n;m+1;k}^{\left( -1\right) }+\left( p\cdot q\right)
Z_{n+1;m;k}^{\left( -1\right) }+\left( q\cdot r\right) Z_{n;m;k+1}^{\left(
-1\right) }] \\
&=&q^{2}Z_{n;m;k}^{\left( -1\right) }+\left( 1-\delta _{m0}\right)
mZ_{n;m-1;k}^{\left( 0\right) }+\delta _{m0}Z_{n;k}^{\left( 0\right) }\left(
p,r\right) -\sum_{s_{1}=0}^{k}\sum_{s_{2}=0}^{s_{1}}\left( -1\right)
^{s_{1}}C_{s_{1}}^{k}C_{s_{2}}^{s_{1}}Z_{n+s_{1}-s_{2};m+s_{2}}^{\left(
0\right) }\left( p_{42},p_{43}\right)
\end{eqnarray*}%
\begin{eqnarray*}
&&2[r^{2}Z_{n;m;k+1}^{\left( -1\right) }+\left( p\cdot r\right)
Z_{n+1;m;k}^{\left( -1\right) }+\left( q\cdot r\right) Z_{n;m+1;k}^{\left(
-1\right) }] \\
&=&r^{2}Z_{n;m;k}^{\left( -1\right) }+\left( 1-\delta _{k0}\right)
kZ_{n;m;k-1}^{\left( 0\right) }+\delta _{k0}Z_{n;m}^{\left( 0\right) }\left(
p,q\right) -\sum_{s_{1}=0}^{k}\sum_{s_{2}=0}^{s_{1}}\left( -1\right)
^{s_{1}}C_{s_{1}}^{k}C_{s_{2}}^{s_{1}}Z_{n+s_{1}-s_{2};m+s_{2}}^{\left(
0\right) }\left( p_{42},p_{43}\right)
\end{eqnarray*}%
\begin{equation*}
-3Z_{000}^{\left( 0\right) }=2m^{2}Z_{000}^{\left( -1\right) }+\frac{1}{3}%
-[p^{2}Z_{100}^{\left( -1\right) }+q^{2}Z_{010}^{\left( -1\right)
}+r^{2}Z_{001}^{\left( -1\right) }]-Z_{00}^{\left( 0\right) }\left(
p_{42},p_{43}\right)
\end{equation*}

\textbf{Reductions of tensors} using $p=p_{21}$, $q=p_{31}$, and $r=p_{41}$%
\begin{eqnarray*}
2p^{\mu _{1}}J_{4\mu _{1}} &=&-p^{2}J_{4}+J_{3}\left( q,r\right)
-J_{3}\left( r-p,r-q\right) \\
2q^{\mu _{1}}J_{4\mu _{1}} &=&-q^{2}J_{4}+J_{3}\left( p,r\right)
-J_{3}\left( r-p,r-q\right) \\
2r^{\mu _{1}}J_{4\mu _{1}} &=&-r^{2}J_{4}+J_{3}\left( p,q\right)
-J_{3}\left( r-p,r-q\right)
\end{eqnarray*}%
\begin{eqnarray*}
2p^{\mu _{1}}J_{4\mu _{1}\mu _{2}} &=&-p^{2}J_{4\mu _{2}}+J_{3\mu
_{2}}\left( p_{42},p_{43}\right) +J_{3\mu _{2}}\left( p_{31},p_{41}\right)
+p_{41\mu _{2}}J_{3}\left( p_{42},p_{43}\right) \\
2q^{\mu _{1}}J_{4\mu _{1}\mu _{2}} &=&-q^{2}J_{4\mu _{2}}+J_{3\mu
_{2}}\left( p_{42},p_{43}\right) +J_{3\mu _{2}}\left( p_{21},p_{41}\right)
+p_{41\mu _{2}}J_{3}\left( p_{42},p_{43}\right) \\
2r^{\mu _{1}}J_{4\mu _{1}\mu _{2}} &=&-r^{2}J_{4\mu _{2}}+J_{3\mu
_{2}}\left( p_{42},p_{43}\right) +J_{3\mu _{2}}\left( p_{21},p_{31}\right)
+p_{41\mu _{2}}J_{3}\left( p_{42},p_{43}\right)
\end{eqnarray*}%
\begin{equation*}
2g^{\mu _{12}}J_{4\mu _{1}\mu _{2}}=i[3\left( 4\pi \right)
^{3}]^{-1}+2m^{2}J_{4}+2J_{3}\left( p_{42},p_{43}\right)
\end{equation*}

\section{Four-Dimensional Subamplitudes}

\label{AppSub}We cast vector subamplitudes in this appendix. They are
ordered following the amplitudes that originate them ($AVV$, $VAV$, $VVA$,
and $AAA$) and then grouped according to the version. That emphasizes
patterns attributed to each version and additional terms depending on the
squared mass.

\textbf{First version} 
\begin{eqnarray*}
(T^{VPP})^{\nu _{1}} &=&2\left[ P_{31}^{\nu _{2}}\Delta _{3\nu _{2}}^{\nu
_{1}}+(p_{21}^{\nu _{1}}-p_{32}^{\nu _{1}})I_{\log }\right] -4\left(
p_{21}\cdot p_{32}\right) J_{3}^{\nu _{1}} \\
&&+2\left[ (p_{31}^{\nu _{1}}p_{21}^{2}-p_{21}^{\nu
_{1}}p_{31}^{2})J_{3}+p_{21}^{\nu _{1}}J_{2}\left( p_{21}\right)
-p_{32}^{\nu _{1}}J_{2}\left( p_{32}\right) \right] \\
\left( T^{ASP}\right) ^{\nu _{1}} &=&2\left[ P_{31}^{\nu _{2}}\Delta _{3\nu
_{2}}^{\nu _{1}}+\left( p_{21}^{\nu _{1}}-p_{32}^{\nu _{1}}\right) I_{\log }%
\right] -4\left( p_{21}\cdot p_{32}\right) J_{3}^{\nu _{1}} \\
&&+2\left[ \left( p_{31}^{\nu _{1}}p_{21}^{2}-p_{21}^{\nu
_{1}}p_{31}^{2}-4m^{2}p_{32}^{\nu _{1}}\right) J_{3}+p_{21}^{\nu
_{1}}J_{2}\left( p_{21}\right) -p_{32}^{\nu _{1}}J_{2}\left( p_{32}\right) %
\right] \\
-\left( T^{APS}\right) ^{\nu _{1}} &=&2\left[ P_{31}^{\nu _{2}}\Delta _{3\nu
_{2}}^{\nu _{1}}+(p_{21}^{\nu _{1}}-p_{32}^{\nu _{1}})I_{\log }\right]
-4\left( p_{21}\cdot p_{32}\right) J_{3}^{\nu _{1}} \\
&&+2\left[ \left( p_{31}^{\nu _{1}}p_{21}^{2}-p_{21}^{\nu
_{1}}p_{31}^{2}+4m^{2}p_{21}^{\nu _{1}}\right) J_{3}+p_{21}^{\nu
_{1}}J_{2}\left( p_{21}\right) -p_{32}^{\nu _{1}}J_{2}\left( p_{32}\right) %
\right] \\
-\left( T^{VSS}\right) ^{\nu _{1}} &=&2\left[ P_{31}^{\nu _{2}}\Delta _{3\nu
_{2}}^{\nu _{1}}+(p_{21}^{\nu _{1}}-p_{32}^{\nu _{1}})I_{\log }\right]
-4\left( p_{21}\cdot p_{32}+4m^{2}\right) J_{3}^{\nu _{1}} \\
&&+2\left[ \left( p_{31}^{\nu _{1}}p_{21}^{2}-p_{21}^{\nu
_{1}}p_{31}^{2}-4m^{2}p_{31}^{\nu _{1}}\right) J_{3}+p_{21}^{\nu
_{1}}J_{2}\left( p_{21}\right) -p_{32}^{\nu _{1}}J_{2}\left( p_{32}\right) %
\right]
\end{eqnarray*}

\textbf{Second version}%
\begin{eqnarray*}
-\left( T^{SAP}\right) ^{\nu _{1}} &=&2\left[ P_{21}^{\nu _{2}}\Delta _{3\nu
_{2}}^{\nu _{1}}+\left( p_{32}^{\nu _{1}}+p_{31}^{\nu _{1}}\right) I_{\log }%
\right] +4\left( p_{32}\cdot p_{31}\right) J_{3}^{\nu _{1}} \\
&&+2\left[ \left( p_{21}^{\nu _{1}}p_{31}^{2}-p_{31}^{\nu
_{1}}p_{21}^{2}+4m^{2}p_{32}^{\nu _{1}}\right) J_{3}+p_{32}^{\nu
_{1}}J_{2}\left( p_{32}\right) +p_{31}^{\nu _{1}}J_{2}\left( p_{31}\right) %
\right] \\
\left( T^{PVP}\right) ^{\nu _{1}} &=&2\left[ P_{21}^{\nu _{2}}\Delta _{3\nu
_{2}}^{\nu _{1}}+\left( p_{32}^{\nu _{1}}+p_{31}^{\nu _{1}}\right) I_{\log }%
\right] +4\left( p_{32}\cdot p_{31}\right) J_{3}^{\nu _{1}} \\
&&+2\left[ \left( p_{21}^{\nu _{1}}p_{31}^{2}-p_{31}^{\nu
_{1}}p_{21}^{2}\right) J_{3}+p_{32}^{\nu _{1}}J_{2}\left( p_{32}\right)
+p_{31}^{\nu _{1}}J_{2}\left( p_{31}\right) \right] \\
\left( T^{PAS}\right) ^{\nu _{1}} &=&2\left[ P_{21}^{\nu _{2}}\Delta _{3\nu
_{2}}^{\nu _{1}}+\left( p_{32}^{\nu _{1}}+p_{31}^{\nu _{1}}\right) I_{\log }%
\right] +4\left( p_{32}\cdot p_{31}\right) J_{3}^{\nu _{1}} \\
&&+2\left[ \left( p_{21}^{\nu _{1}}p_{31}^{2}-p_{31}^{\nu
_{1}}p_{21}^{2}+4m^{2}p_{31}^{\nu _{1}}\right) J_{3}+p_{32}^{\nu
_{1}}J_{2}\left( p_{32}\right) +p_{31}^{\nu _{1}}J_{2}\left( p_{31}\right) %
\right] \\
-\left( T^{SVS}\right) ^{\nu _{1}} &=&2\left[ P_{21}^{\nu _{2}}\Delta _{3\nu
_{2}}^{\nu _{1}}+\left( p_{32}^{\nu _{1}}+p_{31}^{\nu _{1}}\right) I_{\log }%
\right] +4\left( p_{32}\cdot p_{31}-4m^{2}\right) J_{3}^{\nu _{1}} \\
&&+2\left[ \left( p_{21}^{\nu _{1}}p_{31}^{2}-p_{31}^{\nu
_{1}}p_{21}^{2}-4m^{2}p_{21}^{\nu _{1}}\right) J_{3}+p_{32}^{\nu
_{1}}J_{2}\left( p_{32}\right) +p_{31}^{\nu _{1}}J_{2}\left( p_{31}\right) %
\right]
\end{eqnarray*}

\textbf{Third version}%
\begin{eqnarray*}
\left( T^{SPA}\right) ^{\nu _{1}} &=&2\left[ P_{32}^{\nu _{2}}\Delta _{3\nu
_{2}}^{\nu _{1}}-\left( p_{21}^{\nu _{1}}+p_{31}^{\nu _{1}}\right) I_{\log }%
\right] +4\left( p_{21}\cdot p_{31}\right) J_{3}^{\nu _{1}} \\
&&+2\left[ \left( p_{31}^{\nu _{1}}p_{21}^{2}+p_{21}^{\nu
_{1}}p_{31}^{2}-4m^{2}p_{21}^{\nu _{1}}\right) J_{3}-p_{21}^{\nu
_{1}}J_{2}\left( p_{21}\right) -p_{31}^{\nu _{1}}J_{2}\left( p_{31}\right) %
\right] \\
-\left( T^{PSA}\right) ^{\nu _{1}} &=&2\left[ P_{32}^{\nu _{2}}\Delta _{3\nu
_{2}}^{\nu _{1}}-\left( p_{21}^{\nu _{1}}+p_{31}^{\nu _{1}}\right) I_{\log }%
\right] +4\left( p_{21}\cdot p_{31}\right) J_{3}^{\nu _{1}} \\
&&+2\left[ \left( p_{31}^{\nu _{1}}p_{21}^{2}+p_{21}^{\nu
_{1}}p_{31}^{2}-4m^{2}p_{31}^{\nu _{1}}\right) J_{3}-p_{21}^{\nu
_{1}}J_{2}\left( p_{21}\right) -p_{31}^{\nu _{1}}J_{2}\left( p_{31}\right) %
\right] \\
\left( T^{PPV}\right) ^{\nu _{1}} &=&2\left[ P_{32}^{\nu _{2}}\Delta _{3\nu
_{2}}^{\nu _{1}}-\left( p_{21}^{\nu _{1}}+p_{31}^{\nu _{1}}\right) I_{\log }%
\right] +4\left( p_{21}\cdot p_{31}\right) J_{3}^{\nu _{1}} \\
&&+2\left[ \left( p_{31}^{\nu _{1}}p_{21}^{2}+p_{21}^{\nu
_{1}}p_{31}^{2}\right) J_{3}-p_{21}^{\nu _{1}}J_{2}\left( p_{21}\right)
-p_{31}^{\nu _{1}}J_{2}\left( p_{31}\right) \right] \\
-\left( T^{SSV}\right) ^{\nu _{1}} &=&2\left[ P_{32}^{\nu _{2}}\Delta _{3\nu
_{2}}^{\nu _{1}}-\left( p_{21}^{\nu _{1}}+p_{31}^{\nu _{1}}\right) I_{\log }%
\right] +4\left( p_{21}\cdot p_{31}-4m^{2}\right) J_{3}^{\nu _{1}} \\
&&+2\left[ \left( p_{31}^{\nu _{1}}p_{21}^{2}+p_{21}^{\nu
_{1}}p_{31}^{2}-4m^{2}\left( p_{21}^{\nu _{1}}+p_{31}^{\nu _{1}}\right)
\right) J_{3}-p_{21}^{\nu _{1}}J_{2}\left( p_{21}\right) -p_{31}^{\nu
_{1}}J_{2}\left( p_{31}\right) \right]
\end{eqnarray*}

\section{Four-Dimensional Trace Versions}

\label{Tr6G4D}One uses the following identities to insert the Levi-Civita
tensor in traces with the chiral matrix%
\begin{equation*}
\gamma _{\ast }\gamma _{\left[ \mu _{1}\cdots \mu _{r}\right] }=\frac{%
i^{n-1+r\left( r+1\right) }}{\left( 2n-r\right) !}\varepsilon _{\mu
_{1}\cdots \mu _{r}\nu _{r+1}\cdots \nu _{2n}}\gamma ^{\left[ \nu
_{r+1}\cdots \nu _{2n}\right] },
\end{equation*}%
where the notation $\gamma _{\left[ \mu _{1}\cdots \mu _{r}\right] }$
indicates antisymmetrized products of gammas and the investigated dimension
is $2n=4$. This appendix uses this resource to achieve different trace
expressions and explore their relations.

\textbf{Trace using the definition }$\gamma _{\ast }=i\varepsilon _{\nu
_{1}\nu _{2}\nu _{3}\nu _{4}}\gamma ^{\nu _{1}\nu _{2}\nu _{3}\nu _{4}}/4!$%
\textbf{\ - }The three leading positions to substitute the definition are
around vertices $\Gamma _{1}$, $\Gamma _{2}$, and $\Gamma _{3}$. Even if
that brings six options, the same integrated expressions arise regardless of
replacing at the left or right. Thus, we cast the possibilities in the
sequence 
\begin{eqnarray*}
t_{1} &=&\mathrm{tr}(\gamma _{\ast }\gamma _{\mu _{1}\nu _{1}\mu _{2}\nu
_{2}\mu _{3}\nu _{3}})=i\varepsilon ^{\alpha _{1}\alpha _{2}\alpha
_{3}\alpha _{4}}\mathrm{tr}(\gamma _{\alpha _{1}\alpha _{2}\alpha _{3}\alpha
_{4}}\gamma _{\mu _{1}\nu _{1}\mu _{2}\nu _{2}\mu _{3}\nu _{3}})/4! \\
&=&+g_{\mu _{1}\nu _{1}}\varepsilon _{\mu _{2}\nu _{2}\mu _{3}\nu
_{3}}-g_{\mu _{1}\mu _{2}}\varepsilon _{\nu _{1}\nu _{2}\mu _{3}\nu
_{3}}+g_{\mu _{1}\nu _{2}}\varepsilon _{\nu _{1}\mu _{2}\mu _{3}\nu
_{3}}-g_{\mu _{1}\mu _{3}}\varepsilon _{\nu _{1}\mu _{2}\nu _{2}\nu
_{3}}+g_{\mu _{1}\nu _{3}}\varepsilon _{\nu _{1}\mu _{2}\nu _{2}\mu _{3}} \\
&&+g_{\nu _{1}\mu _{2}}\varepsilon _{\mu _{1}\nu _{2}\mu _{3}\nu
_{3}}-g_{\nu _{1}\nu _{2}}\varepsilon _{\mu _{1}\mu _{2}\mu _{3}\nu
_{3}}+g_{\nu _{1}\mu _{3}}\varepsilon _{\mu _{1}\mu _{2}\nu _{2}\nu
_{3}}-g_{\nu _{1}\nu _{3}}\varepsilon _{\mu _{1}\mu _{2}\nu _{2}\mu
_{3}}+g_{\mu _{2}\nu _{2}}\varepsilon _{\mu _{1}\nu _{1}\mu _{3}\nu _{3}} \\
&&-g_{\mu _{2}\mu _{3}}\varepsilon _{\mu _{1}\nu _{1}\nu _{2}\nu
_{3}}+g_{\mu _{2}\nu _{3}}\varepsilon _{\mu _{1}\nu _{1}\nu _{2}\mu
_{3}}+g_{\nu _{2}\mu _{3}}\varepsilon _{\mu _{1}\nu _{1}\mu _{2}\nu
_{3}}-g_{\nu _{2}\nu _{3}}\varepsilon _{\mu _{1}\nu _{1}\mu _{2}\mu
_{3}}+g_{\mu _{3}\nu _{3}}\varepsilon _{\mu _{1}\nu _{1}\mu _{2}\nu _{2}},
\end{eqnarray*}%
\begin{eqnarray*}
t_{2} &=&\mathrm{tr}(\gamma _{\mu _{1}\nu _{1}}\gamma _{\ast }\gamma _{\mu
_{2}\nu _{2}\mu _{3}\nu _{3}})=i\varepsilon ^{\alpha _{1}\alpha _{2}\alpha
_{3}\alpha _{4}}\mathrm{tr}(\gamma _{\mu _{1}\nu _{1}}\gamma _{\alpha
_{1}\alpha _{2}\alpha _{3}\alpha _{4}}\gamma _{\mu _{2}\nu _{2}\mu _{3}\nu
_{3}})/4! \\
&=&+g_{\mu _{1}\nu _{1}}\varepsilon _{\mu _{2}\nu _{2}\mu _{3}\nu
_{3}}+g_{\mu _{1}\mu _{2}}\varepsilon _{\nu _{1}\nu _{2}\mu _{3}\nu
_{3}}-g_{\mu _{1}\nu _{2}}\varepsilon _{\mu _{2}\nu _{2}\mu _{3}\nu
_{3}}+g_{\mu _{1}\mu _{3}}\varepsilon _{\nu _{1}\mu _{2}\nu _{2}\nu
_{3}}-g_{\mu _{1}\nu _{3}}\varepsilon _{\nu _{1}\mu _{2}\nu _{2}\mu _{3}} \\
&&-g_{\nu _{1}\mu _{2}}\varepsilon _{\mu _{1}\nu _{2}\mu _{3}\nu
_{3}}+g_{\nu _{1}\nu _{2}}\varepsilon _{\mu _{1}\mu _{2}\mu _{3}\nu
_{3}}-g_{\nu _{1}\mu _{3}}\varepsilon _{\mu _{1}\mu _{2}\nu _{2}\nu
_{3}}+g_{\nu _{1}\nu _{3}}\varepsilon _{\mu _{1}\mu _{2}\nu _{2}\mu
_{3}}+g_{\mu _{2}\nu _{2}}\varepsilon _{\mu _{1}\nu _{1}\mu _{3}\nu _{3}} \\
&&-g_{\mu _{2}\mu _{3}}\varepsilon _{\mu _{1}\nu _{1}\nu _{2}\nu
_{3}}+g_{\mu _{2}\nu _{3}}\varepsilon _{\mu _{1}\nu _{1}\nu _{2}\mu
_{3}}+g_{\nu _{2}\mu _{3}}\varepsilon _{\mu _{1}\nu _{1}\mu _{2}\nu
_{3}}-g_{\nu _{2}\nu _{3}}\varepsilon _{\mu _{1}\nu _{1}\mu _{2}\mu
_{3}}+g_{\mu _{3}\nu _{3}}\varepsilon _{\mu _{1}\nu _{1}\mu _{2}\nu _{2}},
\end{eqnarray*}%
\begin{eqnarray*}
t_{3} &=&\mathrm{tr}(\gamma _{\mu _{1}\nu _{1}\mu _{2}\nu _{2}}\gamma _{\ast
}\gamma _{\mu _{3}\nu _{3}})=i\varepsilon ^{\alpha _{1}\alpha _{2}\alpha
_{3}\alpha _{4}}\mathrm{tr}(\gamma _{\mu _{1}\nu _{1}\mu _{2}\nu _{2}}\gamma
_{\alpha _{1}\alpha _{2}\alpha _{3}\alpha _{4}}\gamma _{\mu _{3}\nu _{3}})/4!
\\
&=&+g_{\mu _{1}\nu _{1}}\varepsilon _{\mu _{2}\nu _{2}\mu _{3}\nu
_{3}}-g_{\mu _{1}\mu _{2}}\varepsilon _{\nu _{1}\nu _{2}\mu _{3}\nu
_{3}}+g_{\mu _{1}\nu _{2}}\varepsilon _{\mu _{2}\nu _{2}\mu _{3}\nu
_{3}}+g_{\mu _{1}\mu _{3}}\varepsilon _{\nu _{1}\mu _{2}\nu _{2}\nu
_{3}}-g_{\mu _{1}\nu _{3}}\varepsilon _{\nu _{1}\mu _{2}\nu _{2}\mu _{3}} \\
&&+g_{\nu _{1}\mu _{2}}\varepsilon _{\mu _{1}\nu _{2}\mu _{3}\nu
_{3}}-g_{\nu _{1}\nu _{2}}\varepsilon _{\mu _{1}\mu _{2}\mu _{3}\nu
_{3}}-g_{\nu _{1}\mu _{3}}\varepsilon _{\mu _{1}\mu _{2}\nu _{2}\nu
_{3}}+g_{\nu _{1}\nu _{3}}\varepsilon _{\mu _{1}\mu _{2}\nu _{2}\mu
_{3}}+g_{\mu _{2}\nu _{2}}\varepsilon _{\mu _{1}\nu _{1}\mu _{3}\nu _{3}} \\
&&+g_{\mu _{2}\mu _{3}}\varepsilon _{\mu _{1}\nu _{1}\nu _{2}\nu
_{3}}-g_{\mu _{2}\nu _{3}}\varepsilon _{\mu _{1}\nu _{1}\nu _{2}\mu
_{3}}-g_{\nu _{2}\mu _{3}}\varepsilon _{\mu _{1}\nu _{1}\mu _{2}\nu
_{3}}+g_{\nu _{2}\nu _{3}}\varepsilon _{\mu _{1}\nu _{1}\mu _{2}\mu
_{3}}+g_{\mu _{3}\nu _{3}}\varepsilon _{\mu _{1}\nu _{1}\mu _{2}\nu _{2}},
\end{eqnarray*}%
where we omit the global factor $4i$. Since each expression contains fifteen
monomials featuring all index configurations, signs are the unique
distinguishing factor among them. That is also the reason why references
often name them symmetric or democratic \cite{AguilaVictoria1998, Wu2006,
Viglioni2016}.

These (main) versions play fundamental roles in this investigation as they
are enough to obtain any other result. That is evident for \textbf{traces
employing} $\gamma _{\ast }\gamma _{a}=-i\varepsilon _{a\alpha _{1}\alpha
_{2}\alpha _{3}}\gamma ^{\alpha _{1}\alpha _{2}\alpha _{3}}/3!$. After using
this identity for the chiral matrix and the first gamma, we write this trace
through ten monomials. Although some index configurations are absent, the
integrated expression coincides with the $a$-th main version. That occurs
because extra terms vanish in the integration of finite null integrals
embodied into the $t^{\left( -+\right) }$ tensor (\ref{T-+}) and the $ASS$
amplitude (\ref{ASS}). Following these specific choices brings
simplifications while maintaining the complete organization adopted
throughout this work.%
\begin{equation*}
\eta _{1}\left( a\right) =\mathrm{tr}\left( \gamma _{\ast }\gamma
_{abcdef}\right) =-i\varepsilon _{a}^{\quad \alpha _{1}\alpha _{2}\alpha
_{3}}\mathrm{tr}\left( \gamma _{\alpha _{1}\alpha _{2}\alpha _{3}}\gamma
_{bcdef}\right) /6
\end{equation*}%
\begin{eqnarray*}
\eta _{1}\left( a\right) &=&g_{\text{$b$}c}\varepsilon _{\text{$a$}def}-g_{%
\text{$b$}d}\varepsilon _{\text{$a$}cef}+g_{\text{$b$}e}\varepsilon _{\text{$%
a$}cdf}-g_{\text{$b$}f}\varepsilon _{\text{$a$}cde}+g_{\text{$c$}%
d}\varepsilon _{\text{$a$}bef} \\
&&-g_{\text{$c$}e}\varepsilon _{\text{$a$}bdf}+g_{\text{$c$}f}\varepsilon _{%
\text{$a$}bde}+g_{de}\varepsilon _{abcf}+g_{ef}\varepsilon
_{abcd}-g_{df}\varepsilon _{abce}
\end{eqnarray*}

If we use any other identity\ involving the chiral matrix, trace expressions
relate to linear combinations of the main versions (\ref{r123}).\ We
approach some of these possibilities in the sequence while highlighting
relations at the integrand level. Other associations apply only after
integration.

\textbf{Trace using} $\gamma _{\ast }\gamma _{\left[ ab\right]
}=-i\varepsilon _{ab\nu _{1}\nu _{2}}\gamma ^{\nu _{1}\nu _{2}}/2!$ - This
case requires expressing the ordinary product in terms of the
antisymmetrized one. We find seven monomials after taking the traces.%
\begin{equation*}
\gamma _{\ast }\gamma _{ab}=-\frac{1}{2}i\varepsilon _{ab\alpha _{1}\alpha
_{2}}\gamma ^{\alpha _{1}\alpha _{2}}+g_{ab}\gamma _{\ast }
\end{equation*}%
\begin{eqnarray*}
\eta _{2}\left( ab\right) =\mathrm{tr}\left( \gamma _{\ast }\gamma
_{abcdef}\right) &=&g_{ab}\varepsilon _{cdef}+g_{cd}\varepsilon
_{abef}-g_{ce}\varepsilon _{abdf}+g_{cf}\varepsilon _{abde} \\
&&+g_{d\text{$e$}}\varepsilon _{abcf}-g_{df}\varepsilon
_{abce}+g_{ef}\varepsilon _{abcd}
\end{eqnarray*}%
\begin{equation*}
t_{1}+t_{2}=2\eta _{2}\left( \mu _{1}\nu _{1}\right)
\end{equation*}

\textbf{Trace using} $\gamma _{\ast }\gamma _{\left[ abc\right]
}=i\varepsilon _{abc\nu }\gamma ^{\nu }$ - Following a similar procedure we
find six monomials.%
\begin{equation*}
\gamma _{\ast }\gamma _{abc}=i\varepsilon _{abc\nu }\gamma ^{\nu }+\gamma
_{\ast }\left( g_{bc}\gamma _{a}-g_{ac}\gamma _{b}+g_{ab}\gamma _{c}\right)
\end{equation*}%
\begin{equation*}
\eta _{3}\left( abc\right) =\mathrm{tr}\left( \gamma _{\ast }\gamma
_{abcdef}\right) =g_{ab}\varepsilon _{cdef}-g_{ac}\varepsilon
_{bdef}+g_{bc}\varepsilon _{adef}+g_{de}\varepsilon
_{abcf}-g_{df}\varepsilon _{abce}+g_{ef}\varepsilon _{abcd}
\end{equation*}

\textbf{Trace using} $\gamma _{\ast }\gamma _{\left[ abcd\right]
}=i\varepsilon _{abcd}$ - This case also generates seven monomials. 
\begin{eqnarray*}
\gamma _{\ast }\gamma _{abcd} &=&i\varepsilon _{abcd}\mathbf{1}+g_{ab}\gamma
_{\ast }\gamma _{\left[ cd\right] }-g_{ac}\gamma _{\ast }\gamma _{\left[ bd%
\right] }+g_{ad}\gamma _{\ast }\gamma _{\left[ bc\right] } \\
&&+g_{bc}\gamma _{\ast }\gamma _{\left[ ad\right] }-g_{bd}\gamma _{\ast
}\gamma _{\left[ ac\right] }+g_{cd}\gamma _{\ast }\gamma _{\left[ ab\right]
}+\left( g_{ab}g_{cd}-g_{ac}g_{bd}+g_{ad}g_{bc}\right) \gamma _{\ast }
\end{eqnarray*}%
\begin{eqnarray*}
\eta _{4}\left( abcd\right) =\mathrm{tr}\left( \gamma _{\ast }\gamma
_{abcdef}\right) &=&g_{ab}\varepsilon _{cdef}-g_{ac}\varepsilon
_{bdef}+g_{ad}\varepsilon _{bcef}+g_{bc}\varepsilon _{adef} \\
&&-g_{bd}\varepsilon _{acef}+g_{cd}\varepsilon _{abef}+g_{ef}\varepsilon
_{abcd}
\end{eqnarray*}%
\begin{equation*}
t_{2}+t_{3}=2\eta _{4}\left( \mu _{3}\nu _{3}\mu _{1}\nu _{1}\right)
\end{equation*}

\section{Six-Dimensional Substructures}

\label{sixdamp}Following the pattern acknowledged in all studied dimensions,
these amplitudes contain standard tensors and subamplitudes. Below, we
introduce their finite content using the "hats" notation. Both sectors
exhibit irreducible divergent objects that cancel out perfectly. On the
other hand, surface terms combine into $S_{i}$ objects cast in the end.

\textbf{Standard tensor} - integrand%
\begin{equation}
\varepsilon _{\mu _{abc}\nu _{123}}t_{\mu _{d}}^{\left(
s_{1}s_{2}s_{3}\right) \nu _{123}}=\varepsilon _{\mu _{abc}\nu
_{123}}[K_{1\mu _{d}}K_{234}^{\nu _{123}}+s_{1}K_{2\mu _{d}}K_{134}^{\nu
_{123}}+s_{2}K_{3\mu _{d}}K_{124}^{\nu _{123}}+s_{3}K_{4\mu
_{d}}K_{123}^{\nu _{123}}]/D_{1234}
\end{equation}

\textbf{Standard tensor} - finite part%
\begin{eqnarray}
\varepsilon _{\mu _{abc}\nu _{123}}\hat{T}_{\mu _{d}}^{\left(
s_{1}s_{2}s_{3}\right) \nu _{123}} &=&\varepsilon _{\mu _{abc}\nu
_{123}}\left\{ \left( 1+s_{1}\right) p_{31}^{\nu _{2}}p_{41}^{\nu
_{3}}(J_{4\mu _{d}}^{\nu _{1}}+p_{21\mu _{d}}J_{4}^{\nu _{1}})\right.  \notag
\\
&&\left. -\left( 1-s_{2}\right) p_{21}^{\nu _{2}}p_{41}^{\nu _{3}}(J_{4\mu
_{d}}^{\nu _{1}}+p_{31\mu _{d}}J_{4}^{\nu _{1}})+\left( 1+s_{3}\right)
p_{21}^{\nu _{2}}p_{31}^{\nu _{3}}(J_{4\mu _{d}}^{\nu _{1}}+p_{41\mu
_{d}}J_{4}^{\nu _{1}})\right\}
\end{eqnarray}

\textbf{Subamplitude of the first and second versions} - finite part%
\begin{eqnarray}
-\varepsilon _{\mu _{1234}}^{\qquad \nu _{12}}\hat{T}_{\nu _{12}}^{\tilde{T}%
PPP} &=&\varepsilon _{\mu _{1234}\nu _{12}}\left\{ 16[\left( p_{31}\cdot
p_{43}\right) p_{21}^{\nu _{2}}-\left( p_{21}\cdot p_{42}\right) p_{31}^{\nu
_{2}}+\left( p_{21}\cdot p_{32}\right) p_{41}^{\nu _{2}}]J_{4}^{\nu
_{1}}\right.  \notag \\
&&+8(p_{21}^{\nu _{1}}p_{41}^{\nu _{2}}p_{31}^{2}-p_{31}^{\nu
_{1}}p_{41}^{\nu _{2}}p_{21}^{2}-p_{21}^{\nu _{1}}p_{31}^{\nu
_{2}}p_{41}^{2})J_{4}  \notag \\
&&+8[2p_{43}^{\nu _{2}}J_{3}^{\nu _{1}}\left( p_{31},p_{41}\right)
+p_{31}^{\nu _{1}}p_{41}^{\nu _{2}}J_{3}\left( p_{31},p_{41}\right)
]+8[2p_{21}^{\nu _{2}}J_{3}^{\nu _{1}}\left( p_{21},p_{41}\right) ]  \notag
\\
&&\left. +8[-p_{21}^{\nu _{1}}p_{41}^{\nu _{2}}J_{3}\left(
p_{21},p_{41}\right) +p_{32}^{\nu _{1}}p_{43}^{\nu _{2}}J_{3}\left(
p_{32},p_{42}\right) +p_{21}^{\nu _{1}}p_{31}^{\nu _{2}}J_{3}\left(
p_{21},p_{31}\right) ]\right\}
\end{eqnarray}%
\begin{eqnarray}
-\varepsilon _{\mu _{1234}}^{\qquad \nu _{12}}\hat{T}_{\nu _{1}\nu
_{2}}^{STPP} &=&\varepsilon _{\mu _{1234}\nu _{12}}\left\{ 16\left[ -\left(
p_{41}\cdot p_{43}\right) p_{21}^{\nu _{2}}+\left( p_{41}\cdot p_{42}\right)
p_{31}^{\nu _{2}}-\left( p_{31}\cdot p_{32}\right) p_{41}^{\nu _{2}}\right]
J_{4}^{\nu _{1}}\right.  \notag \\
&&+8\left[ p_{31}^{\nu _{1}}p_{43}^{\nu _{2}}p_{21}^{2}-p_{21}^{\nu
_{1}}p_{42}^{\nu _{2}}p_{31}^{2}+p_{21}^{\nu _{1}}p_{32}^{\nu
_{2}}p_{41}^{2}-4m^{2}p_{32}^{\nu _{1}}p_{42}^{\nu _{2}}\right] J_{4}  \notag
\\
&&+8\left[ 2p_{41}^{\nu _{2}}J_{3}^{\nu _{1}}\left( p_{21},p_{41}\right)
-2p_{32}^{\nu _{2}}J_{3}^{\nu _{1}}\left( p_{21},p_{31}\right) -p_{32}^{\nu
_{1}}p_{43}^{\nu _{2}}J_{3}\left( p_{42},p_{43}\right) \right]  \notag \\
&&+\left. 8\left[ -p_{31}^{\nu _{1}}p_{43}^{\nu _{2}}J_{3}\left(
p_{31},p_{41}\right) +p_{21}^{\nu _{1}}p_{42}^{\nu _{2}}J_{3}\left(
p_{21},p_{41}\right) -p_{21}^{\nu _{1}}p_{32}^{\nu _{2}}J_{3}\left(
p_{21},p_{31}\right) \right] \right\}
\end{eqnarray}

\textbf{Divergent contributions of the first and second versions}%
\begin{eqnarray}
3S_{1\mu _{1234}} &=&-8[\varepsilon _{\mu _{134}\nu _{123}}p_{32}^{\nu
_{2}}p_{42}^{\nu _{3}}\Delta _{4\mu _{2}}^{\nu _{1}}+\varepsilon _{\mu
_{124}\nu _{123}}p_{21}^{\nu _{2}}p_{43}^{\nu _{3}}\Delta _{4\mu _{3}}^{\nu
_{1}}+\varepsilon _{\mu _{123}\nu _{123}}p_{21}^{\nu _{2}}p_{31}^{\nu
_{3}}\Delta _{4\mu _{4}}^{\nu _{1}}]  \notag \\
&&-8\varepsilon _{\mu _{1234}\nu _{12}}(p_{43}^{\nu _{2}}P_{134}^{\nu
_{3}}+p_{21}^{\nu _{2}}P_{124}^{\nu _{3}})\Delta _{4\nu _{3}}^{\nu _{1}}
\end{eqnarray}%
\begin{eqnarray}
3S_{2\mu _{1234}} &=&+8[-\varepsilon _{\mu _{234}\nu _{123}}p_{32}^{\nu
_{2}}p_{43}^{\nu _{3}}\Delta _{4\mu _{1}}^{\nu _{1}}-\varepsilon _{\mu
_{124}\nu _{123}}p_{31}^{\nu _{2}}p_{41}^{\nu _{3}}\Delta _{4\mu _{3}}^{\nu
_{1}}+\varepsilon _{\mu _{123}\nu _{123}}p_{32}^{\nu _{2}}p_{41}^{\nu
_{3}}\Delta _{4\mu _{4}}^{\nu _{1}}]  \notag \\
&&+8\varepsilon _{\mu _{1234}\nu _{12}}(p_{32}^{\nu _{2}}P_{123}^{\nu
_{3}}-p_{41}^{\nu _{2}}P_{124}^{\nu _{3}})\Delta _{4\nu _{3}}^{\nu _{1}}
\end{eqnarray}

\begin{acknowledgments}
The authors express their gratitude to O. A. Battistel for valuable
discussions that contributed to their formation. To S. A. Dias for his
supportive demeanor throughout the research and writing process. To CAPES
and CNPQ for the provided financial~support.
\end{acknowledgments}

\bigskip

\bigskip

\bigskip

\bigskip

\bigskip

\end{document}